\documentclass{aa}
\usepackage{graphicx}
\usepackage{amssymb}
\usepackage{natbib}
\usepackage{aalongtable}
\bibpunct{(}{)}{;}{a}{}{,}

\def\vsini{$V\!\sin i$}
\def\teff{$T_{\rm eff}$}

\def\logg{$\log~g$}

\def\omc{$\Omega/\Omega_{\rm{c}}$}

\def\kms{km~s$^{-1}$}

\def\rv{RV}

\begin{document}
%
\title{ Be stars and binaries in the field of the SMC open cluster NGC330 with VLT-FLAMES}

\titlerunning{Be stars and binaries in the SMC}
\author{
C. Martayan \inst{1}
\and  M. Floquet \inst{1}
\and  A.M. Hubert  \inst{1}
\and J. Guti\'errez-Soto \inst{1,2}
\and J. Fabregat \inst{2}
\and  C. Neiner  \inst{1}
\and  M. Mekkas  \inst{1}
}
\offprints {C. Martayan}
\mail{christophe.martayan@obspm.fr}
\institute{GEPI, Observatoire de Paris, CNRS, Universit\'e Paris Diderot; place Jules Janssen 92195 Meudon Cedex, France 
\and Observatorio Astron\'omico de Valencia, edifici Instituts d'investigaci\'o, Poligon la Coma, 46980 Paterna Valencia, Spain}
\date{Received /Accepted}
\abstract
{}
{Observations of hot stars belonging to the young cluster SMC-NGC330  and its surrounding region  were
obtained with the VLT-GIRAFFE facilities in MEDUSA mode. We  investigated the B and Be star properties
and proportions in this environment of low metallicity. We also  searched for rapid variability in Be stars using
photometric databases.}
{With spectroscopic measurements we  characterized the emission and properties of Be stars. By cross-correlation
with photometric databases such as MACHO and OGLE, we  searched for binaries in our sample of hot stars, as well 
as for short-term variability in Be stars.}
{ We report on the global characteristics of the Be star sample (131 objects). We find that the proportion of early
Be stars with a large equivalent width of the H$\alpha$ emission line is higher in the SMC than in the LMC and MW. We
find a slight increase in the proportion of Be stars  compared to B-type stars with decreasing metallicity. We also
discovered spectroscopic and photometric binaries, and for the latter  we give their orbital period. We identify 13
Be stars  with short-term variability.  We determine their period(s) and find that  9 Be stars are multiperiodic. }
{}


\maketitle
%

\section{Introduction}
%

The Magellanic Clouds (MC), which contain a huge number of early-type stars, are particularly appropriate
to investigate the effect of low metallicity on the B and Be stars  populations, comparatively to the ones in the
Milky Way (MW). The FLAMES-GIRAFFE instrumentation  \citep{pasquini02} installed at the VLT  allowed us to obtain
significant samples of B and Be stars spectra in the Large and Small Magellanic Clouds (LMC and SMC) which are needed
to achieve our goal. In \citet{marta06a}, we presented an overview of spectroscopic results for 176 early-type stars
observed in the field of the LMC open cluster NGC2004. In \citet{marta06b} and \citet{marta07} (hereafter Papers I
and II) we searched for the effects of metallicity in the LMC and SMC, respectively. We showed that the lower the
metallicity, the higher the rotational velocities. These observational results support theoretical predictions by
\citet{meynet00,meynet02}  and \citet{maeder01}  for massive stars. 
 Therefore the percentage of Be stars seems to be higher in lower metallicity environments such as the SMC (Z$<$0.001). 
In this fourth paper we present an overview
of spectroscopic results for 346 early-type stars observed in the field of the SMC open cluster NGC330 with
VLT-FLAMES. Note that the determination of their fundamental parameters (\teff, \logg, \vsini, and \rv) has already
been reported in Paper II. We also search for pulsators among our Be stars sample through an analysis of their
 MACHO\footnote{http://wwwmacho.mcmaster.ca/} data. Theoretically, the pulsational  instability of hot stars has a
great dependence on metallicity. \citet{pam99}  showed that the instability  strip for \object{$\beta$ Cephei} and Slow
Pulsating B stars (SPB) vanishes at Z$<$0.01 and Z$<$0.006, respectively. Thus, Be stars, which show the same
pulsational characteristics as those of classical B-type pulsators \citep[e.g.][]{neiner05,walker05a,walker05b}  in the
MW, are among the best objects in the SMC to test  their theoretical predictions.\\
In the present paper, the observations and the reduction process are described in Sect.~\ref{reduc}. In
Sect.~\ref{result} we present  the characteristics of the H$\alpha$ emission line
and the proportion of Be stars in the field as well as in clusters and OB associations. We perform  a comparison
with Be stars in the MW (Sect.~\ref{propBe}).  In Sect.~\ref{var} we describe the variability detected in our Be stars 
sample thanks to an investigation of the MACHO and  OGLE databases \citep{ogleI}. We report on the discovery of spectroscopic and
photometric binaries (Sect.~\ref{bin}) and  the identification of multi-periods in the light curves of several objects
(Sect.~\ref{varbesh}) which pleads in  favour of pulsations. 
 We discuss the impact of metallicity on the proportion of Be stars and the presence of pulsations in Sect.~\ref{discussion}. 
 Finally, we give a detailed study of 3 peculiar emission line objects that are not classical Be stars
(Appendix~\ref{pec}), of binary systems (Appendix~\ref{appB}), and of short-term periodic Be stars (Appendix~\ref{indBe}).

\section{Observations}
\label{reduc}

Spectra of a significant sample of the B star population in
the young cluster SMC NGC330 and its surrounding field have been obtained with the ESO 
VLT-FLAMES facilities, as part of the Guaranteed Time Observation
programs of the Paris Observatory (P.I.: F. Hammer). The multi-fibre
spectrograph VLT-FLAMES has been used in MEDUSA mode (132 fibres) 
at medium resolution. \\
As shown in Paper I and II the use of the setup LR02 (396.4--456.7 nm, hereafter blue spectra) is
adequate for the determination of fundamental parameters, while the LR06 setup
(643.8--718.4 nm, hereafter red spectra) is used to identify Be stars 
and to study the H$\alpha$ emission line characteristics. The spectral
resolution is 6400 for LR02 and 8600 for LR06. 
The respective instrumental broadenings are $\simeq$ 50 \kms~and 35 \kms.
Observations (ESO runs 72.D-0245A and 72.D-0245C) were carried out  on  October 21, 22 and 23, 2003 (blue 
and red spectra) and on September 9 (blue spectra) and 10 (red spectra), 2004. The observational seeing ranged from 0.4 to 2$\arcsec$. \\
The observed fields are centred at $\alpha$(2000) = 00h 55mn 15s and $\delta$(2000) = $-$72$^{\circ}$ 20$\arcmin$ 00$\arcsec$ 
for the observations of 2003 and at $\alpha$(2000) = 00h 55mn 25s and $\delta$(2000) = $-$72$^{\circ}$ 23$\arcmin$ 30$\arcsec$ 
for the run of 2004. Besides the young cluster NGC330,  these fields  contain several high-density 
groups of stars. The position of all our stellar and sky fibre targets is plotted in Fig.~\ref{figure0} (online).\\
A sample of 346 stars  among the 5370 B-type star candidates located in the selected fields has been observed during
the two observing runs (see Paper II, Sect. 2). 
It contains 131 Be stars,  202 B-type stars, 4 O-type stars, 6 A-type stars, and 3 other types of stars, 
which are discussed in this paper.\\
The data reduction  was performed as described in Paper I and II. The S/N ratio of spectra obtained in the blue region varies from
$\sim$ 15 for the fainter objects to $\sim$ 135 for the brighter ones (see Table 3 in Paper II).\\

\section{Results}
\label{result}

After subtraction of the sky line contribution, it appeared that more than 80\% of the 346 stars are contaminated by
nebular lines. This contamination is particularly detectable in the H$\alpha$ line. Depending of the intensity level
of the nebular H$\alpha$ line, a weak nebular contribution can also be detected in forbidden lines of [NII]  at 6548
and 6583 {\AA}  and [SII] at 6717 and 6731 {\AA}  in the LR06 setup. When the nebular H$\alpha$ line is strong, the
H$\gamma$ and H$\delta$ line profiles are also affected by a nebular component. With the same technique as the one used
to identify stars with the Be phenomenon in the LMC \citep{marta06a}, we tried to disentangle the circumstellar 
line emission (CS) component from emission produced by the nebular line in polluted spectra. 

\subsection{Stellar and nebular radial velocities}
\label{rv}

 For each star, the radial velocity (\rv) of nebular lines (H$\alpha$, [NII], and [SII]) has been measured and compared
to the stellar \rv. The mean accuracy is $\pm$ 10 \kms. The statistical distribution of stellar and nebular RVs is
mono-modal with a maximum at +155 \kms~and +165 \kms, respectively.  The \rv difference between stellar and nebular
lines peaks around -10 \kms~and does not seem to indicate any clear link between stars and the structures giving rise to
the nebular lines. However, due to the weak difference between the stellar and nebular RVs, it has not been easy to
correctly estimate the nebulosity contribution for Be stars which present a single-peaked H$\alpha$ emission line
profile in their spectrum.

\addtocounter{figure}{+1}

\subsection{Be stars}
\label{Be}

Our sample contains 131 Be stars (see Table~\ref{tablebeobs}): 41 known Be stars including 39 ones from \citet{keller99b}
and  2 others from  \citet{grebel92b}, and 90 Be stars discovered in Paper II (see Table 3 therein). Note that among this second
group of Be stars, 28 ones are suspected to be emission line stars from a slit-less ESO-WFI survey \citep{marta06c}.


\subsubsection{Emission-line characteristics}
\label{emhalpha}

The equivalent width (EW), maximum intensity (Imax for a single peak, and I(V) and I(R) for the violet and red
peaks respectively in a double-peak emission), and the Full Width at Half Maximum (FWHM) of the circumstellar
H$\alpha$ emission measured for each Be star are given in Table~\ref{tablebeobs}. The FWHM of the CS H$\alpha$
emission line ranges from 146 to 622 \kms. The telluric lines as well as the nebular lines are resolved and the FWHM
of the H$\alpha$ nebular line is $\sim$ 60 \kms.\\
Not correcting for the nebular H$\alpha$ emission line contributes to overestimate  I$_{max}$ and underestimate 
the FWHM of the CS emission line. To a lesser degree it also leads to overestimate of
EW${\alpha}$. We thus determined and subtracted this contribution from the H$\alpha$ emission line profile.
As in \citet{marta06a}, we used the nebular ratio [\ion{S}{ii}]/H$\alpha$ to estimate the
nebular contribution in the CS H$\alpha$ line in Be star spectra. From the fibres located
on sky positions (without stars in the background) we derived a mean intensity ratio 
[\ion{S}{ii}]/H$\alpha$=0.2 $\pm$ 0.1. The intensity and RV of the [\ion{S}{ii}] 6717 line can be
measured in each Be star spectrum, if the line is present, and we assumed that the nebular
line ratio [SII]/H$\alpha$ is the same for sky and star spectra. 
Consequently, we can generally state whether the H$\alpha$ CS emission is single, double or affected by these narrow absorption
core of circumstellar origin. 

Since we used the mean intensity ratio [SII]/H$\alpha$, the
nebular contribution can be underestimated locally in some cases, which may lead to an overestimate of the number
of Be stars with a single-peaked H$\alpha$, even more so because the \rv~of the nebular line is close to the stellar
\rv. Moreover, the lowest value of the V and R emission peak
separation that can be measured is about 70-80 \kms. This limitation is chiefly found in 
Be stars with a low or moderate \vsini~and a large EW in H$\alpha$.\\ 
Early Be stars, which represent the quasi-totality of our sample, show a CS
emission  contribution in the H$\gamma$ line profile when  EW${\alpha}$ is $\ge$~15 \AA, and
in the H$\delta$ line profile when EW${\alpha}$ is  $\ge$  30-40 \AA. The detection limit 
depends on the S/N value. 
\ion{Fe}{ii} emission lines are often detected, however some spectra are too noisy to investigate their presence 
even if EW${\alpha}$ is large.

The H$\gamma$ and H$\delta$ CS emission contribution is
double-peaked except for Be stars with a low \vsini~and a large EW${\alpha}$. From
Table~\ref{tablebeobs}, we find that the H$\alpha$ emission line is single-peaked for 60 objects after
correction of the nebular contribution, and double-peaked  for 71 objects. Moreover, about 30\% of Be 
stars show an asymmetric or double emission H$\alpha$ profile with V/R or R/V $\geq 1.05$, 
consistent with the percentage in the MW \citep{hummel01}. 
Shell features observed in H lines  and very often in metallic lines, are present in 22 Be
stars.

\subsubsection{Envelopes of Be stars}
\label{envBe}

\begin{figure}[htp!]
    \centering
    \resizebox{\hsize}{!}{\includegraphics[angle=0]{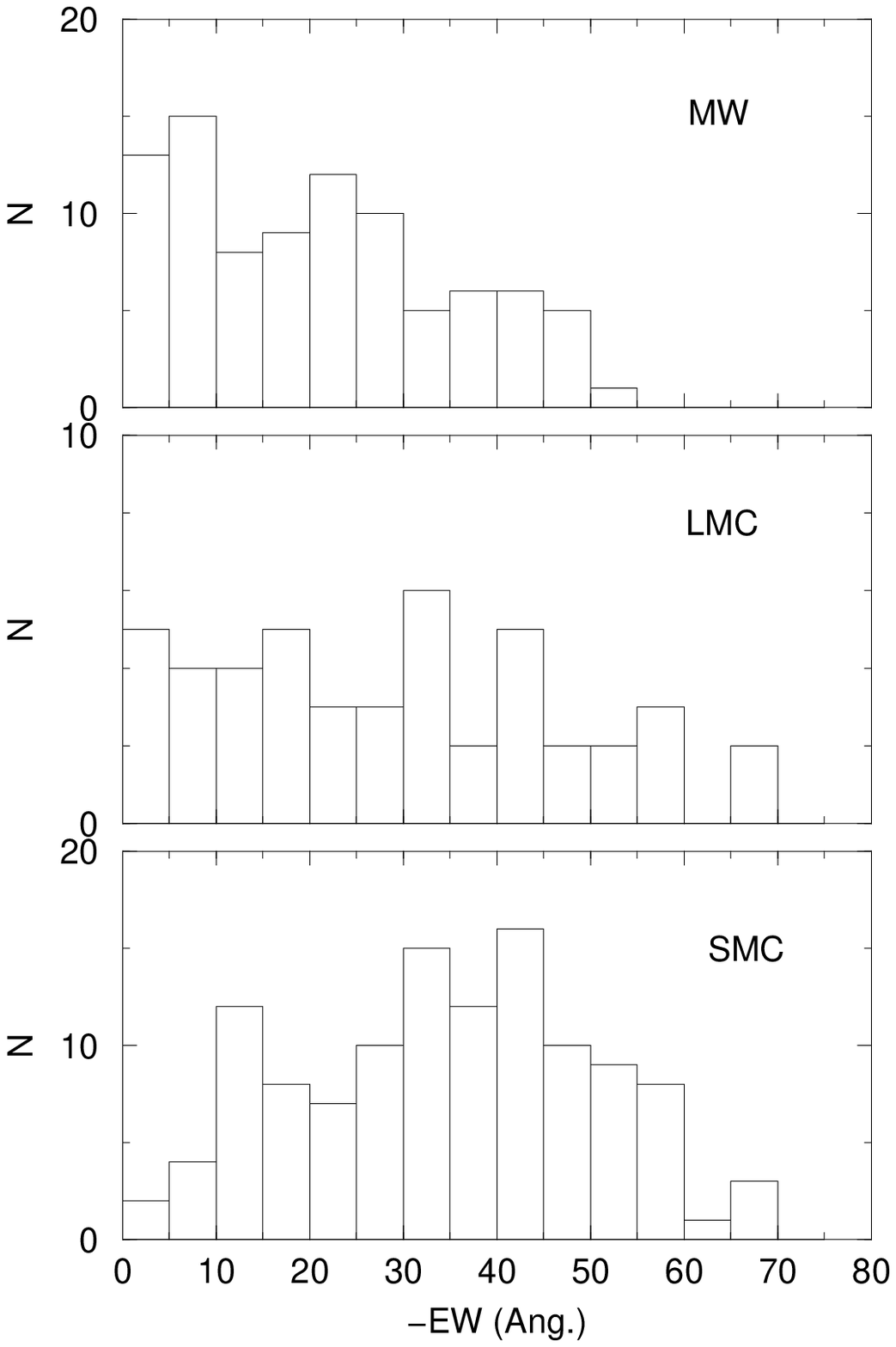}}
    \caption{Distributions of Be stars versus the EW of the H$\alpha$ emission line.
    Top: in the MW from \citet{dachs92}, \citet{andrillat82} and \citet{andrillat83};
    middle: in the LMC from Paper I;
    bottom: in the SMC (this study).
   }
    \label{ew3gal}
\end{figure}

\begin{figure}[htp!]
    \centering
    \resizebox{\hsize}{!}{\includegraphics[angle=0]{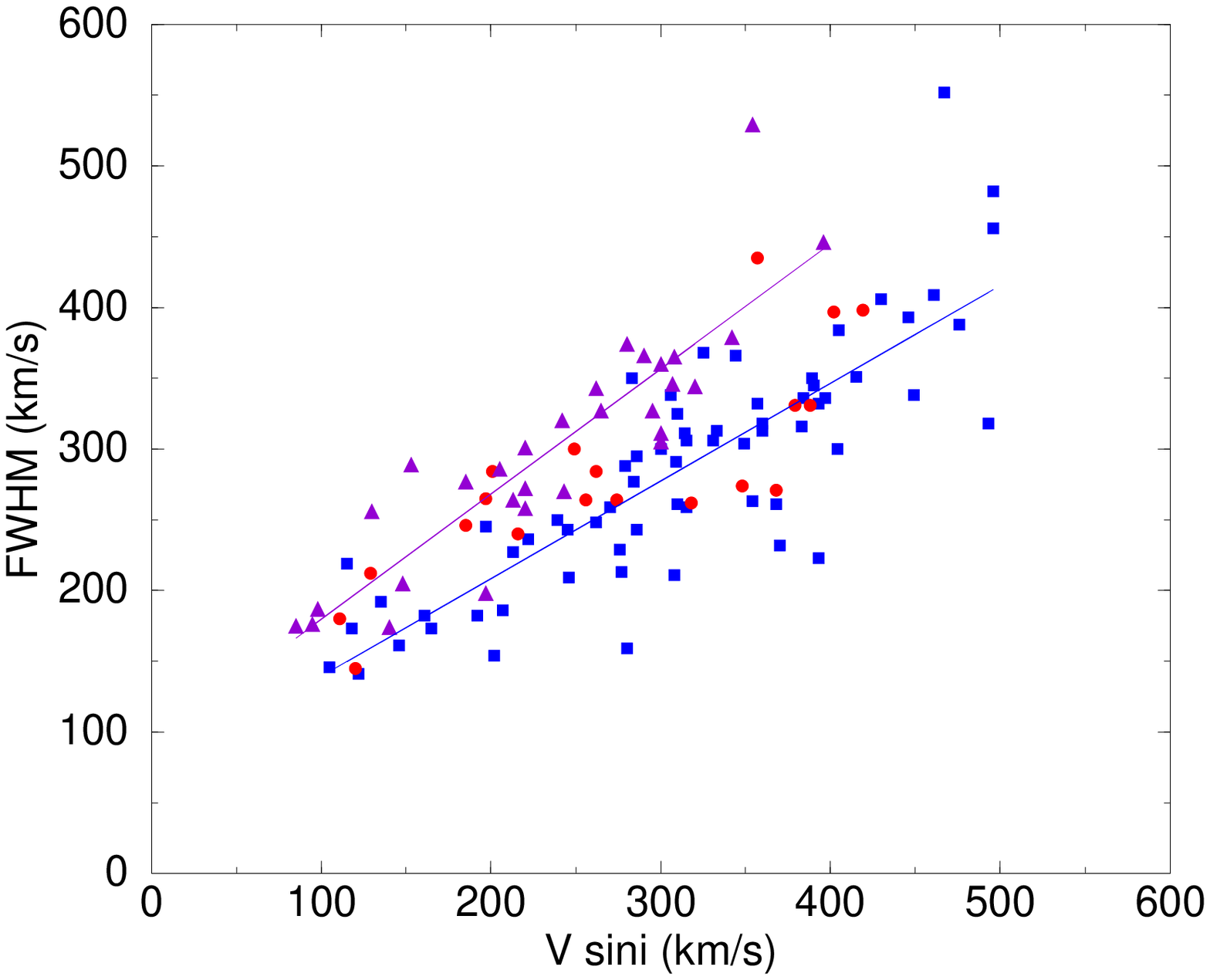}}
    \caption{FWHM of the H$\alpha$ emission line as a function of  \vsini~for Be stars with
    symmetric H$\alpha$ profiles and EW$\alpha > $20 \AA. 
    Squares correspond to Be stars in the SMC, circles to Be stars in the LMC from Paper I, 
     and triangles to Be stars in the MW from  \citet{dachs92}, \citet{andrillat82} and \citet{andrillat83}.   
     The linear regression for samples in the SMC and MW is shown.}
    \label{envall}
\end{figure}

\begin{figure}[htp!]
    \centering
    \resizebox{\hsize}{!}{\includegraphics[angle=-90]{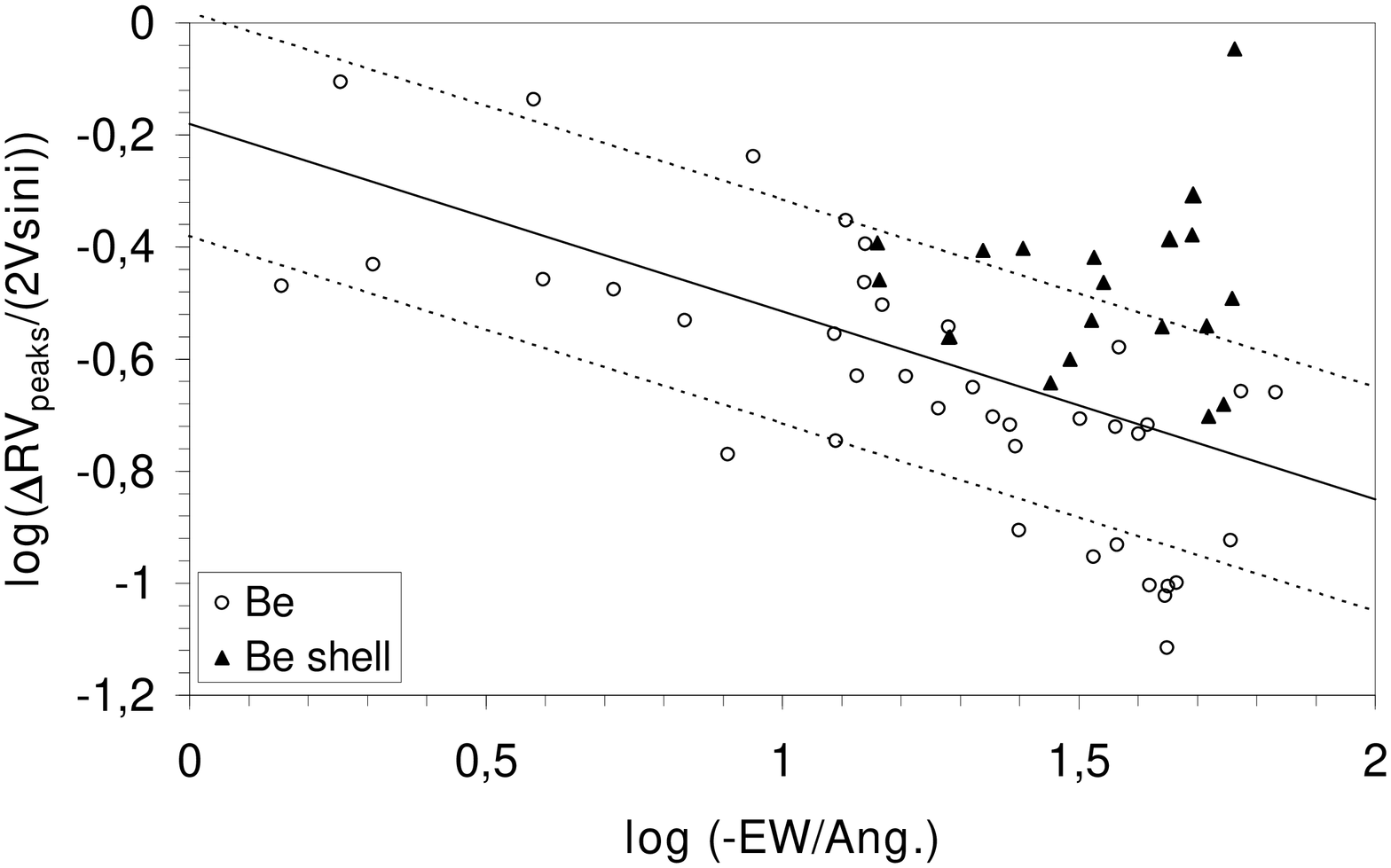}}
    \caption{Peak separation velocity of the H$\alpha$ CS emission line as
    a function of its equivalent width. Filled triangles correspond to Be-shell stars 
    and open circles to Be stars.  
    The solid line represents the mean correlation 
    from \citet{zamanov01},  with dashed lines showing the dispersion.}
    \label{zamaSMC}
\end{figure}

We examine the EW$\alpha$ distribution of the H$\alpha$ emission line for Be stars in the
SMC  and we search for  correlations between \vsini,  FWHM$\alpha$, EW$\alpha$ and 
$\Delta RV{\rm peaks}$ (the peak separation). Then, we compare them to the ones in the LMC and MW, respectively. \\
To study the EW$\alpha$ distribution we only consider Be stars in the MW with spectral
types ranging from  B0 to B4, in order to compare homogeneous samples, since the
quasi-totality of Be stars we observed  in the SMC and LMC  are early-type B stars.
Spectral types for Be stars in the SMC are taken from Paper II, as well as their \vsini~obtained by fitting the observed
spectra by theoretical spectra. 
For  Be stars in the LMC we use data (EW$\alpha$, spectral types and \vsini) reported in
Paper I, and for Be stars in the MW  the ones summarized in  \citet{dachs92}, \citet{andrillat82}  and  \citet{andrillat83}.\\
The EW$\alpha$ distributions for the SMC, LMC and MW are shown in Fig.~\ref{ew3gal}. 
The rate of early Be stars with large EW$\alpha$ ($> $20 \AA) is higher in the SMC than in the LMC and MW: 74\% 
in the SMC to be compared with 62\% in the LMC and 50\% in the MW.  Note that the shape of the distribution of EW is 
different in the SMC and peaks around -30 to -40 \AA. It remains the same when we subtract the Be stars selected from 
the WFI survey \citep{marta06c}.\\
We also examine the FWHM$\alpha$ distribution for Be stars in the SMC versus \vsini.
We obtain a linear correlation slightly different for stars with weak and strong EW$\alpha$. 
As in the MW \citep[e.g.][]{dachs92}, FWHM$\alpha$ at given EW$\alpha$ of emission increase as \vsini.
However, points corresponding to Be stars with an asymmetric H$\alpha$  emission line
profile  or with  a V/R ratio $\mbox{$\not=$ 1}$  are scattered in the graph. 
For more clarity we thus focus our study on Be stars with a symmetric single and double-peaked
H$\alpha$ line profile and with EW$\alpha$ $>$ 20 \AA. In Fig.~\ref{envall} we plot the linear regressions for 
this sample in the SMC and its counterpart in MW. \\
From Figs.~\ref{ew3gal} and \ref{envall} we can conclude that on
average Be stars in the SMC display lower FWHM$\alpha$ and larger EW$\alpha$
than their counterparts in the MW and the LMC.  In the frame of  a
differentially rotating Keplerian  disk model in which there is a direct
relation between EW$\alpha$ and the radius of the emitting disk expressed in
stellar units \citep{dachs92,grundstromgies06} our result supports larger disk radii in the SMC compared to the MW
and the LMC.\\
We find a correlation between  ($\Delta RV{\rm peaks}$)/(2\vsini) 
and EW$\alpha$, as found for Galactic Be stars by \citet{dachs86} 
and \citet{hanuschik88}, and refined by \citet{zamanov01}, and for the LMC by \citet{marta06c}. 
This provides evidence for a rotationally supported CS envelope, as confirmed with the VLTI by \citet{meilland06} 
for the Galactic Be star  \object{$\alpha$ Arae}. \\
Taking into account the fact that the dispersion of points for Galactic  Be stars is
similar to that for SMC Be stars \citep[see Fig. 2a of ][]{zamanov01}, 
the average slope of the peak separation versus the equivalent widths of  Galactic Be stars determined by
\citet{zamanov01}  is comparable to ours in the SMC (see Fig.~\ref{zamaSMC}). \\
Note that as in the LMC, the SMC Be stars with a shell (triangles in Fig.~\ref{zamaSMC}) are
located in the right part of the diagram, above the mean correlation. The increase of the peak
separation in those objects is an opacity effect \citep[see eq.(7) in][]{hanuschik88}. 
It indicates that SMC Be stars in shell phases have denser disks than stars without shell.\\

In addition to these distributions and graphs, we find the expected correlation between
the intensity and the equivalent widths of the H$\alpha$ emission line  for Be stars in the
LMC and SMC.


\subsubsection{Proportion of Be stars}
\label{propBe}

\addtocounter{table}{+1}
\begin{table}[tbph!]
\tiny{
\centering
\caption[]{ Percentage of Be stars. Top: in open clusters. 
Bottom:  in the field. In field A, we exclude NGC330 and the area covered by the study of \citet{keller99b}, 
and in field B, we exclude only the clusters instead of the entire area studied by \citet{keller99b}. 
}
\centering
\begin{tabular}{lcccc}
\hline
\hline	
Cluster & log(t) & Nstars & NBe & N(Be)/N(B+Be)\% \\
\hline	
NGC306 & 7.9$\pm$0.2 & 8 & 3 & 38 $\pm$ 12 \\
OGLE-SMC99 & 7.8$\pm$0.2 & 10 & 4 & 40$\pm$10 \\
Range & & & & 20-40\\
\hline
\hline
Area & & Nstars & NBe & N(Be)/N(B+Be)\% \\ 
\hline
Field A & & 178 & 51 & $\simeq$22 \\
Field B & & 180 & 78 & $\simeq$30  \\
Average field & & & & 26 $\pm$ 4 \\
\hline
\end{tabular}
\label{propBetab}
}
\end{table}

As in the LMC \citep{marta06a},  
we determine the percentage of Be stars in the SMC found on 
the one hand in the observed clusters and OB concentrations and on the other hand in the field close
to NGC330. ``Be stars'' are all non-supergiant O, B and early A-type stars that have shown some
emission in their Balmer lines at least once \citep{jaschek82}.   
This definition is also used here.
We have removed from the statistics all the Be stars intentionally observed from \citet{keller99b}, or 
\citet{grebel92b}   or preselected from our  ESO-WFI  survey  \citep{marta06c}, to remove any bias.\\
The ratios of Be stars in the clusters and in the field surrounding NGC330, but
excluding NGC330 itself, are given in Table~\ref{propBetab}.

Due to the characteristics of the instrumentation, which do not allow to position many fibres
in a small angular size region (typically $\le 1 \arcmin$),  and due to the characteristics of our survey and
the variable nature of the Be phenomenon, the above results should be considered as a rough estimate
of the frequency of Be stars.  \\
With FLAMES observations, the use of field stars is suited  because  the number of
observed stars is  higher and ages can be averaged out. The proportion of Be stars, which is 26 $\pm$ 4\%  in
fields in the  SMC, has to be compared to the one obtained for the LMC with the same methods : 17.5 $\pm$ 2.5\% 
(Paper I).

\section{Variability}
\label{var}

\subsection{Binaries}
\label{bin}

We search for spectroscopic binaries (SB) among our sample. 
As the observations have been carried on the same day or within a 1-day interval and as  only two spectra (one blue and one
red)  for each star are generally available, it is not possible to perform a similar search as in  the LMC (Paper I).  
However, 8 spectroscopic binaries with 2 spectra (SB2) have been identified.  Moreover, we searched for photometric binaries 
in the light curves of the 280 stars of the GIRAFFE survey identified in the MACHO and OGLE databases from  their coordinates
within a radius smaller than 1\arcsec. Among the 13 photometric binaries found, 8 were not known as binaries before and 4 of
them are SB2, 2 have a Be component. All the systems are compounds with 1 or 2 B-type stars,  excepted SMC5\_002807, which
has a cold supergiant component (see Sect.~\ref{pec2807} for this latter object).\\
The orbital period of the systems is determined using 3 different methods: the CLEAN algorithm \citep{roberts87},    the Least Squares
method and the PDM method \citep{stell78}  with an  accuracy of $10^{-6}$d. Results  are given in
Table~\ref{tablebinPNM}. Illustrations of the most characteristic systems, associated with their MACHO light curve when
available, are shown in Figs~\ref{bin977} to \ref{bin49816-74928-84353}. The light curve magnitudes are expressed in
instrumental units. \\
In Table~\ref{tablebinPNM2}, we present additional parameters on the binaries.
In this table, the ephemeris have been calculated with the primary eclipse at phase 0. 
The phases at which the blue and red spectra were obtained are also indicated.
In the case of rather equal minima it has been difficult to pinpoint the one corresponding to the primary eclipse. \\
\begin{table*}[tbph!]
\tiny{
\centering
\caption[]{Binaries stars in the SMC.
The orbital period  is given in days. The accuracy on the radial velocities of blue \ion{He}{i} lines 
is $\pm$ 10 \kms.
The last column provides some additional information:
``cl8'' is for the star projected onto the open cluster OGLE-SMC99.
EB means Eclipsing Binary, SB2 is for spectroscopic binary with two spectra, ``ell'' means ellipsoidal binary.
Note that the star SMC5\_016461 was observed twice within 11 months.
}
\centering
\begin{tabular}{lllllllll}
\hline
\hline	
Star & MACHO & $\alpha$ & $\delta$ & Vmag  & P(d)  &  RV1  &  RV2  & Comments \\
     &       & (2000)   & (2000)   &       &       & (\kms) & (\kms) & \\
\hline	
SMC5\_000977 & 207.16205.63 & 00 54 06.550 & -72 14 47.31 & 16.7 & 3.128 & +8  & + 253  &  SB2, EB \\
SMC5\_002807 & 207.16373.18 & 00 56 09.420 & -72 28 09.30 & 14.6 & 454.959 & +160    &   & cool Sg, EB $^{1}$ \\
SMC5\_003789 & 207.16203.202 & 00 53 26.690 & -72 22 07.40 & 17.6 & 2.087 &  +42   &   & Be, EB \\
SMC5\_004477 & 207.16375.7 & 00 56 11.620 & -72 18 23.70 & 14.7 & 4.480 &  -42  &   & EB, SB2, eccentric $^{2}$ \\
SMC5\_004534 & 207.16318.41 & 00 55 40.340 & -72 17 50.90 & 16.4 & 4.051 & +167   &  & EB \\
SMC5\_013723 & 207.16259.277 & 00 54 53.330 & -72 28 12.20 & 17.5 & 2.059 & +134  &   & cl8, EB $^{1}$ \\
SMC5\_016461 & 207.16316.21 & 00 55 49.619 & -72 25 27.43 & 14.9 & 54.317 &   +95  &   & Be, ell?, EB, SB1-2? \\
             &              &              &              &      &        &   +83  &   &              \\
SMC5\_020391 & 207.16374.9 & 00 56 23.580 & -72 21 23.60 & 15.0 & 2.320 &   +85  & +216  & SB2, EB $^{2}$\\
SMC5\_023571 & 207.16375.29 & 00 56 51.090 & -72 17 37.70 & 15.9 & 3.534 &   +63  &  +288 & ell, EB, SB2 \\
SMC5\_023641 & 207.16375.38 & 00 56 34.250 & -72 17 37.30 & 15.9 & 2.010 &  -67   &  +293 & SB2, EB\\
SMC5\_049816 & 207.16374.65 & 00 56 19.270 & -72 21 03.50 & 16.6 & 0.664 &  -59  & +380  & SB2, EB $^{3}$ \\
SMC5\_052516 & 207.16376.67 & 00 56 50.579 & -72 16 50.85 & 16.7 &      &  +17   &  +253 & SB2\\
SMC5\_052663 & 207.16319.70 & 00 55 57.859 & -72 16 36.95 & 17.1 &      &   +53  &  +278 & SB2\\
SMC5\_074928 & 207.16318.129 & 00 55 09.195 & -72 17 13.65 & 17.3 & 2.137 &  +200   &   & ell, EB \\
SMC5\_084353 & 207.16204.47 & 00 53 54.750 & -72 18 29.50 & 16.2 & 1.557 &  +128  &   &  EB \\
\hline
\end{tabular}
\label{tablebinPNM}
}

1: Also observed by \citet{wyrz04} with OGLE, P(SMC5\_002807)=452.295d, P(SMC5\_013723)=2.059d.\\ 
2: Also observed by \citet{bayne02}, catalogue MOA, P(SMC5\_004477)=4.482d, P(SMC5\_020391)=2.320d.\\
3: Also observed by \citet{samus04}, star found as a W~UMa with P=0.6d. 

\end{table*}

\begin{table*}[tbph!]
\centering
\tiny{
\caption[]{ Binaries  stars in the SMC.
For each eclipsing binary, the ephemeris is listed in col. 2 (phase 0 at the primary eclipse),  
the phases corresponding to the spectra are given in cols. 3 and 4, 
the phase of the secondary eclipse in col. 5,  the intensity ratio of the 2 minima in  col. 6 
and  the number of the corresponding figure in col. 7.}
\centering
\begin{tabular}{ccccccc}
\hline
\hline	
Star & Ephemeris & phase1 & phase2 &  phase 2nd eclipse & Imini2/Imini1 &  Figure \\
\hline	
SMC5\_000977 & MJD48857.0852 + 3.127654E & 0.87 & 0.19 &0.48 & 0.85 & \ref{bin977}\\
SMC5\_003789 & MJD48855.3943 + 2.087258E & 0.13 & 0.65 & 0.49 & $\sim$0.46 & \ref{bin4477-2807-3789}\\
SMC5\_004477 & MJD48856.3937 + 4.479996E & 0.19 & 0.20 & 0.33 & 0.66 & \ref{bin4477-2807-3789}\\
SMC5\_004534 & MJD48858.1849 + 4.051494E & 0.01 & 0.03 &0.50 & 0.38 & \ref{bin4534-13723} \\
SMC5\_013723 & MJD48856.0050 + 2.059270E & 0.40 & 0.92 &0.48 & 0.55 & \ref{bin4534-13723} \\
SMC5\_016461 & MJD48865.8675 + 54.337490E & 0.87 & 0.87 & 0.48 & $\sim$0.86 & \ref{bin16461}\\
	     &                            & 0.82 & 0.84 &      &            &                \\
SMC5\_020391 & MJD49145.5595+ 2.320051E & 0.92 & 0.96 &0.50 & 0.95 &\ref{bin20391-23571}\\
SMC5\_023571 & MJD48857.8867+ 3.534556E & 0.22 & 0.25 &0.52 & 0.56 & \ref{bin20391-23571}\\
SMC5\_023641 & MJD48855.0948+ 2.009626E & 0.69 & 0.74 &0.50 & 0.85 & \ref{bin23641} \\
SMC5\_049816 & MJD49144.3831+ 0.663033E & 0.85 & 0.46 &0.50 & 0.98 & \ref{bin49816-74928-84353} \\
SMC5\_074928 & MJD49144.1543+ 2.137252E & 0.50 & 0.96 & 0.50 & 0.82 & \ref{bin49816-74928-84353} \\
SMC5\_084353 & MJD49144.4066+ 1.557259E & 0.52 & 0.57 & 0.50 & 0.57 & \ref{bin49816-74928-84353} \\
\hline
\end{tabular}
\label{tablebinPNM2}
}
\end{table*}

\subsection{Short-term variability in Be stars}
\label{varbesh}

In a preliminary analysis of the MACHO light curves \citet{marta05} discovered 13 short-period variables 
among Be stars in the SMC.
We reanalysed the same time series  in an attempt to refine these periods and to detect additional ones. 
The tools used for this analysis are Period04 \citep{lenz05},  
the CLEAN algorithm \citep{roberts87}   and the Least Squares method.  In order to know if the detected 
frequencies are significant or not, we follow
the signal to noise ratio (SNR) criterion described in \citet{breger93}.  
All the detected frequencies presented in this study fulfill the SNR requirement, i.e.
have a SNR greater than 4.\\
The time span of the observations is $\sim$ 2690 d in the majority of the stars, and 
thus, the resolution in frequency is $\sim$~0.00037 $\mbox{c d}^{-1}$. The determination of the error
in frequency has been obtained with Period04, which follows the equations derived by 
\citet{montgomery99}. 
In our case, the estimate of the error in frequency, 
accounting for the correlation in the residuals \citep{schwa91}, 
is of the order of $1-5 \times 10^{-5}$ $\mbox{c d}^{-1}$.\\
We found 9 multi-periodic and 4 mono-periodic stars. 
The results of the spectral analysis are reported in Table~\ref{tablevarBePNM}.
Amplitudes are given in this table only for the b filter, although the same 
analysis was applied to the dataset for the r filter. 
Only frequencies detected in both filters were considered as certain.
Phase diagrams of the 13 Be stars folded with the detected frequencies  
are displayed in Figs.~\ref{varbe1} to~\ref{varbe7}. Additional comments on 
those Be stars are given in Appendix~\ref{indBe}. 

\begin{table*}[tbph!]
\centering
\tiny{
\caption[]{ Short-term photometric variability of Be stars in the SMC.
The EIS names and MACHO numbers of the studied stars are given in cols. 1 and 2, 
the spectral types taken from \citet{marta05}  in col. 3,
the detected frequencies (in $\mbox{c d}^{-1}$)  in cols. 4 and 5, 
 the amplitudes (in mmag) of the corresponding frequencies for the b filter 
in col. 5 and 6 and  additional information in col. 7.
Comments:
'**' indicates that the star was pre-selected from ESO/ WFI survey \citep{marta06c}.  
``limit'' indicates that the difference between the two detected frequencies is
similar to the resolution in frequency and thus, the second frequency is at the 
limit of detectability.
f$_{Balona}$ stands for frequencies (in $\mbox{c d}^{-1}$) previously obtained by \citet{balona92}.}
\begin{tabular}{ccccccc}
\hline
\hline	
Star & MACHO  & Sp. T. &f1  & f2  & Amp. f1, f2 & Comments \\
\hline	
SMC5\_002232 & 207.16372.22 & B2III &1.32661 & 1.32616 & 24,  15 &  **, limit\\
SMC5\_003296 & 207.16373.5496& B2IV&2.00320 &         & 8&    \\
SMC5\_013978 & 207.16373.58  & B3III&1.37946 & 0.59335 & 16, 14 &  f$_{\mathrm{Balona}, \mathrm{N964}}$=1.361\\
SMC5\_014212 & 207.16259.29  & B2III&1.27691 &         & 24 & **\\ 
SMC5\_014727 & 207.16373.63  & B2IV&1.12991 &         & 17 &  f$_{\mathrm{Balona}, \mathrm{N585}}$=1.120\\
SMC5\_016523 & 207.16316.30  & B2III&1.29297 & 1.29344 & 40, 24 &  limit\\
SMC5\_016544 & 207.16373.129 & B2IV&1.70774 & 1.64993 & 30, 25 &  \\
SMC5\_021152 & 207.16147.14  & B2III&0.98514 & 1.00443 & 19, 19 &  \\
SMC5\_037013 & 207.16315.26  & B2III&1.18153 & 1.21709 & 15, 9 &  **\\
SMC5\_037162 & 207.16259.57  & B2III&0.88531 & 0.90612 & 37, 19 &  \tiny{OGLE005440.73-722752.4 (f=0.443 $\mbox{cd}^{-1}$)} \\
SMC5\_043413&  207.16315.41  & B2IV&2.00716 &         & 7 &  \\
SMC5\_082042&  207.16375.41  & B3III&2.48834 & 1.16625 & 19, 17 &  \\
SMC5\_082941&  207.16203.47  & B3III&0.62483 & 0.62525 & 51, 12 &    f3=0.15324 not a pulsation \\
\hline
\end{tabular}
\label{tablevarBePNM}
}
\end{table*}

\section{Discussion}
\label{discussion}

\subsection{Proportion of Be stars versus metallicity}
The comparison of the proportions of Be
stars in the SMC and  LMC to the MW is a difficult issue. 

In clusters and OB associations of the Magellanic Clouds, it is not easy to give any firm  
conclusions from our study about the dependence of the 
rates of Be stars on the metallicity because the number of stars observed in each cluster is low ($\le$15) 
and the dispersion in the rates of Be stars is high. The  only estimates for clusters in these galaxies have 
been derived from photometry. \citet{wisn06} find an increase of the proportion of Be stars to B stars with 
decreasing metallicity in open clusters.\\
On the contrary in the field of the Magellanic Clouds for which our samples are highly significant our results 
on proportion of Be stars are robust.
For field stars in the MW  \citet{zorec97}  obtained a 20\% value  considering several decades of
Galactic searches for Be stars, and several corrections for different
sources of incompleteness of the currently known Be star samples.
Conversely, the proportions we have obtained for the SMC and LMC are
made from a single epoch survey, in which many Be stars are likely to be
missed due to the non-permanent nature of the Be phenomenon. 
Therefore,  both kind of values are not directly comparable.
It is difficult to estimate what would be the result of a single epoch
survey for Be stars in the Galaxy, but it will certainly be lower than
20\%, probably between 10 and 15\%. \\
If we consider that the probable value
in the MW for a single epoch survey is lower than 15\%, our estimates 
in the field of the  MC (17.5\% in the LMC and 26\% in the SMC) indicate an increase 
of the proportion of Be stars with decreasing metallicity.
Note that \citet{keller99b} did  not find such an increase but the fields  observed by these authors 
were smaller and closer to the clusters than ours while \citet{evans06}, from a restricted sample 
also obtained with the FLAMES-GIRAFFE  instrumentation, found a proportion of Be stars similar to 
ours in the LMC and in the SMC.\\
The  higher proportion of Be stars in the SMC could be explained by the higher ratio of angular velocity
to breakup velocity \omc~found in this galaxy (\omc=95\%, Paper II) compared to the ones in the LMC and
MW (\omc=85\% and 80\% respectively, Paper II). 
Our observational results support predictions by \citet{MM2000}, which link the occurence of Be stars to enhanced stellar
rotation for B-type stars in low metallicity environments.


\subsection{Be stars as pulsators}
Several authors have previously investigated the pulsational behavior of
Be stars in the SMC, by means of photometric and spectroscopic techniques,
with negative results. \citet{balona92} conducted a photometric
variability study of Be stars in the region of the young open cluster NGC330 
in the SMC. He found many of them to be short-period variables, but
he was able to find only one period in each star. He concluded that the
short-term variability is produced by rotational modulation of the light
curve caused by inhomogeneities in the stellar photosphere. \citet{baade02} 
obtained time-resolved high-resolution spectroscopy of two bright
Be stars close to NGC330. They failed to find line-profile variability,
and their analysis cast doubts on the presence of pulsations in these two
stars.\\
Our finding of photometric multi-periodicity in 9 SMC Be stars
clearly shows that (1) they are variable, and (2) they are pulsating stars. 
In particular, the detected periods fall in the range of 
SPB-type pulsating modes (from 0.40 to 1.60 days). \\
The current theoretical models do not predict the presence of pulsational instabilities in
massive stars at metallicities much lower than the Galactic one \citep{pam99}. 
However, these models have recently been challenged by
several observational results. \citet{kola04}  detected some $\beta$ Cephei and SPB stars in
the LMC. Recently, \citet{fab05}  and \citet{sch04}  have also detected 
multiple periods in high mass X-ray binaries with a Be primary component 
(Be/X system) in the SMC. These results point towards the necessity of new modeling or 
improved determination of the opacities used in current theoretical models.\\
In an attempt to investigate  the discrepancies 
between theory and observations, \citet{miglio06}  have analyzed the effect 
of uncertainties in the opacity computations on the excitation of pulsation 
modes in B-type stars.
Computations for a low metallicity (Z=0.005) showed that none of the 
different opacities with different mixtures predicts $\beta$ Cep-type pulsations, 
whereas they found excited SPB-type modes  when considering the recently 
updated OP opacity  with the new metal mixture with Fe enhancement. 
The instability strip is compatible with the position of the studied 
stars in our work \citetext{M.A. Dupret, priv. \ comm.}.
However, the metallicity of NGC330 and its surrounding field is Z=0.002 
\citep[see][and references therein]{maeder99},
which is lower than the lowest  metallicity used by  \citet{miglio06}.

\section{Conclusions}

Medium-resolution spectroscopic observations of a large
sample of B and Be stars in the SMC-NGC330 region are
presented. 131 Be stars were identified in the field 
as well as in small clusters and OB associations.\\
Characteristics of the H$\alpha$ emission line were
investigated through  spectral parameters (FWHM$\alpha$, EW$\alpha$, peak separation). The 
CS envelope of Be stars seems to be rotationally supported.
Higher EW$\alpha$ and lower FWHM$\alpha$ values are found for a large fraction of Be 
stars in the SMC compared to the LMC and MW.
It could indicate the presence of more extended CS envelopes around a fraction of Be stars in the SMC.\\
The proportion of field Be stars compared to B-type stars is found to be slightly higher in the 
SMC (26\%) than in the LMC (17.5\%) 
and MW  (estimated $<$15\% for a single epoch survey). This result  could be explained by the ratios of angular velocities
\omc~of Be stars, which are close to the critical value in the SMC (see Paper II).
Thus, the decrease in  metallicity over \omc~seems to influence the occurrence of the Be phenomenon.\\
We have also found 13 photometric binaries by cross-correlation with the MACHO and OGLE databases. 
Among them, 8 are newly discovered, 2 include a Be star, 4 are SB2. For each of these stars, the period of the system is 
given. In addition, we discovered 2 SB2 in the GIRAFFE spectra, without any photometric variations in 
the MACHO database.
Finally, we  studied the short-term variability of Be stars in the SMC and we  found  a periodicity for
13 of the observed Be stars. Moreover, for the first time, we  found evidence for 
multi-periodicity (2 frequencies) for 9 of them. Such multi-periodicity, which is the signature of stellar pulsations,  
had never been found so far in Be stars in such a low metallicity environment. 
Our result points toward the necessity of modeling of pulsation instabilities in B-type stars at the metallicity of the SMC.

\tiny{
\begin{acknowledgements}

We would like to thank Dr D. Baade for very fruitful discussions. This research has made use of the 
Simbad database and Vizier database maintained at CDS, Strasbourg,
France.
This paper utilizes public domain data originally obtained by MACHO Project,
whose work was performed under the joint auspices of the U.S. Department of
Energy, National Security Administration by the University of California,
Lawrence Livermore National Laboratory under contract No. W-7405-Eng-48, the
National Science Foundation through the Center for Particle Astrophysics of
the University of California under cooperative agreement AST-8809616, the
Mount Stromlo and Siding Spring Observatory, part of the Australian National
University.

\end{acknowledgements}
%


\bibliographystyle{aa}
\bibliography{article3d006}

\begin{thebibliography}{62}
\expandafter\ifx\csname natexlab\endcsname\relax\def\natexlab#1{#1}\fi

\bibitem[{{Andrillat}(1983)}]{andrillat83}
{Andrillat}, Y. 1983, \aaps, 53, 319

\bibitem[{{Andrillat} \& {Fehrenbach}(1982)}]{andrillat82}
{Andrillat}, Y. \& {Fehrenbach}, C. 1982, \aaps, 48, 93

\bibitem[{{Arp}(1959)}]{arp59}
{Arp}, B.~H. 1959, \aj, 64, 254

\bibitem[{{Baade} {et~al.}(2002){Baade}, {Rivinius}, {{\v S}tefl}, \&
  {Kaufer}}]{baade02}
{Baade}, D., {Rivinius}, T., {{\v S}tefl}, S., \& {Kaufer}, A. 2002, \aap, 383,
  L31

\bibitem[{{Balona}(1992)}]{balona92}
{Balona}, L.~A. 1992, \mnras, 256, 425

\bibitem[{{Bayne} {et~al.}(2002){Bayne}, {Tobin}, {Pritchard}, {Bond},
  {Pollard}, {Besier}, {Noda}, {Sumi}, {Yanagisawa}, {Sekiguchi}, {Honda},
  {Muraki}, {Takeuti}, {Hearnshaw}, {Kilmartin}, {Dodd}, {Sullivan}, \&
  {Yock}}]{bayne02}
{Bayne}, G., {Tobin}, W., {Pritchard}, J.~D., {et~al.} 2002, \mnras, 331, 609

\bibitem[{{Breger} {et~al.}(1993){Breger}, {Stich}, {Garrido}, {Martin},
  {Jiang}, {Li}, {Hube}, {Ostermann}, {Paparo}, \& {Scheck}}]{breger93}
{Breger}, M., {Stich}, J., {Garrido}, R., {et~al.} 1993, \aap, 271, 482

\bibitem[{{Dachs} {et~al.}(1986){Dachs}, {Hanuschik}, {Kaiser}, {Ballereau}, \&
  {Bouchet}}]{dachs86}
{Dachs}, J., {Hanuschik}, R., {Kaiser}, D., {Ballereau}, D., \& {Bouchet}, P.
  1986, \aaps, 63, 87

\bibitem[{{Dachs} {et~al.}(1992){Dachs}, {Hummel}, \& {Hanuschik}}]{dachs92}
{Dachs}, J., {Hummel}, W., \& {Hanuschik}, R.~W. 1992, \aaps, 95, 437

\bibitem[{{Evans} {et~al.}(2006){Evans}, {Lennon}, {Smartt}, \&
  {Trundle}}]{evans06}
{Evans}, C.~J., {Lennon}, D.~J., {Smartt}, S.~J., \& {Trundle}, C. 2006, \aap,
  456, 623

\bibitem[{{Fabrycky}(2005)}]{fab05}
{Fabrycky}, D. 2005, \mnras, 359, 117

\bibitem[{{Grebel} \& {Richtler}(1992)}]{grebel92a}
{Grebel}, E.~K. \& {Richtler}, T. 1992, \aap, 253, 359

\bibitem[{{Grebel} {et~al.}(1992){Grebel}, {Richtler}, \& {de
  Boer}}]{grebel92b}
{Grebel}, E.~K., {Richtler}, T., \& {de Boer}, K.~S. 1992, \aap, 254, L5

\bibitem[{{Grundstrom} \& {Gies}(2006)}]{grundstromgies06}
{Grundstrom}, E.~D. \& {Gies}, D.~R. 2006, \apjl, 651, L53

\bibitem[{{Hanuschik} {et~al.}(1988){Hanuschik}, {Kozok}, \&
  {Kaiser}}]{hanuschik88}
{Hanuschik}, R.~W., {Kozok}, J.~R., \& {Kaiser}, D. 1988, \aap, 189, 147

\bibitem[{{Hummel} {et~al.}(2001){Hummel}, {G{\"a}ssler}, {Muschielok},
  {Schink}, {Nicklas}, {Conti}, {Mattaini}, {Keller}, {Mantel}, {Appenzeller},
  {Rupprecht}, {Seifert}, {Stahl}, \& {Tarantik}}]{hummel01}
{Hummel}, W., {G{\"a}ssler}, W., {Muschielok}, B., {et~al.} 2001, \aap, 371,
  932

\bibitem[{{Hummel} {et~al.}(1999){Hummel}, {Szeifert}, {G{\"a}ssler},
  {Muschielok}, {Seifert}, {Appenzeller}, \& {Rupprecht}}]{hummel99}
{Hummel}, W., {Szeifert}, T., {G{\"a}ssler}, W., {et~al.} 1999, \aap, 352, L31

\bibitem[{{Jaschek} \& {Egret}(1982)}]{jaschek82}
{Jaschek}, M. \& {Egret}, D. 1982, in IAU Symp. 98: Be Stars, ed. M.~{Jaschek}
  \& H.-G. {Groth}, 261

\bibitem[{{Keller}(1999)}]{keller99a}
{Keller}, S.~C. 1999, \aj, 118, 889

\bibitem[{{Keller} {et~al.}(1999){Keller}, {Wood}, \& {Bessell}}]{keller99b}
{Keller}, S.~C., {Wood}, P.~R., \& {Bessell}, M.~S. 1999, \aaps, 134, 489

\bibitem[{{Ko{\l}aczkowski} {et~al.}(2004){Ko{\l}aczkowski}, {Pigulski},
  {Soszy{\'n}ski}, {Udalski}, {Szyma{\'n}ski}, {Kubiak}, {{\.Z}ebru{\'n}},
  {Pietrzy{\'n}ski}, {Wo{\'z}niak}, {Szewczyk}, {Wyrzykowski}, \& {The Ogle
  Team}}]{kola04}
{Ko{\l}aczkowski}, Z., {Pigulski}, A., {Soszy{\'n}ski}, I., {et~al.} 2004, in
  ASP Conf. Ser. 310: IAU Colloq. 193: Variable Stars in the Local Group, ed.
  D.~W. {Kurtz} \& K.~R. {Pollard}, 225

\bibitem[{{Ku{\v c}inskas} {et~al.}(2000){Ku{\v c}inskas}, {Vansevi{\v c}ius},
  {Sauvage}, \& {Tanab{\'e}}}]{kuc00}
{Ku{\v c}inskas}, A., {Vansevi{\v c}ius}, V., {Sauvage}, M., \& {Tanab{\'e}},
  T. 2000, \aap, 353, L21

\bibitem[{{Lamers} {et~al.}(1998){Lamers}, {Zickgraf}, {de Winter}, {Houziaux},
  \& {Zorec}}]{lamers98}
{Lamers}, H.~J.~G.~L.~M., {Zickgraf}, F.-J., {de Winter}, D., {Houziaux}, L.,
  \& {Zorec}, J. 1998, \aap, 340, 117

\bibitem[{{Leisy} \& {Dennefeld}(1996)}]{leisy96}
{Leisy}, P. \& {Dennefeld}, M. 1996, \aaps, 116, 95

\bibitem[{{Leisy} \& {Dennefeld}(2006)}]{leisy06}
{Leisy}, P. \& {Dennefeld}, M. 2006, \aap, 456, 451

\bibitem[{{Lenz} \& {Breger}(2005)}]{lenz05}
{Lenz}, P. \& {Breger}, M. 2005, Communications in Asteroseismology, 146, 53

\bibitem[{{Lindsay}(1961)}]{lindsay61}
{Lindsay}, E.~M. 1961, \aj, 66, 169

\bibitem[{{Maeder} {et~al.}(1999){Maeder}, {Grebel}, \&
  {Mermilliod}}]{maeder99}
{Maeder}, A., {Grebel}, E.~K., \& {Mermilliod}, J.-C. 1999, \aap, 346, 459

\bibitem[{{Maeder} \& {Meynet}(2000)}]{MM2000}
{Maeder}, A. \& {Meynet}, G. 2000, \aap, 361, 159

\bibitem[{{Maeder} \& {Meynet}(2001)}]{maeder01}
{Maeder}, A. \& {Meynet}, G. 2001, \aap, 373, 555

\bibitem[{{Martayan} {et~al.}(2006{\natexlab{a}}){Martayan}, {Baade}, {Hubert},
  {Floquet}, {Fabregat}, {Bertin}, \& {Neiner}}]{marta06c}
{Martayan}, C., {Baade}, D., {Hubert}, A.-M., {et~al.} 2006{\natexlab{a}},
  SF2A-2006, 481

\bibitem[{{Martayan} {et~al.}(2007{\natexlab{a}}){Martayan}, {Floquet},
  {Hubert}, \& {Mekkas}}]{marta05}
{Martayan}, C., {Floquet}, M., {Hubert}, A.-M., \& {Mekkas}, M.
  2007{\natexlab{a}}, in Astronomical Society of the Pacific Conference Series,
  Vol. 361, Active OB-Stars: Laboratories for Stellare and Circumstellar
  Physics, ed. A.~T. {Okazaki}, S.~P. {Owocki}, \& S.~{Stefl}, 460

\bibitem[{{Martayan} {et~al.}(2006{\natexlab{b}}){Martayan}, {Fr{\'e}mat},
  {Hubert}, {Floquet}, {Zorec}, \& {Neiner}}]{marta06b}
{Martayan}, C., {Fr{\'e}mat}, Y., {Hubert}, A.-M., {et~al.} 2006{\natexlab{b}},
  \aap, 452, 273

\bibitem[{{Martayan} {et~al.}(2007{\natexlab{b}}){Martayan}, {Fr{\'e}mat},
  {Hubert}, {Floquet}, {Zorec}, \& {Neiner}}]{marta07}
{Martayan}, C., {Fr{\'e}mat}, Y., {Hubert}, A.-M., {et~al.} 2007{\natexlab{b}},
  \aap, 462, 683

\bibitem[{{Martayan} {et~al.}(2006{\natexlab{c}}){Martayan}, {Hubert},
  {Floquet}, {Fabregat}, {Fr{\'e}mat}, {Neiner}, {Stee}, \& {Zorec}}]{marta06a}
{Martayan}, C., {Hubert}, A.~M., {Floquet}, M., {et~al.} 2006{\natexlab{c}},
  \aap, 445, 931

\bibitem[{{Meilland} {et~al.}(2007){Meilland}, {Stee}, {Vannier}, {Millour},
  {Domiciano de Souza}, {Malbet}, {Martayan}, {Paresce}, {Petrov}, {Richichi},
  \& {Spang}}]{meilland06}
{Meilland}, A., {Stee}, P., {Vannier}, M., {et~al.} 2007, \aap, 464, 59

\bibitem[{{Meynet} \& {Maeder}(2000)}]{meynet00}
{Meynet}, G. \& {Maeder}, A. 2000, \aap, 361, 101

\bibitem[{{Meynet} \& {Maeder}(2002)}]{meynet02}
{Meynet}, G. \& {Maeder}, A. 2002, \aap, 390, 561

\bibitem[{{Meyssonnier} \& {Azzopardi}(1993)}]{meyss93}
{Meyssonnier}, N. \& {Azzopardi}, M. 1993, \aaps, 102, 451

\bibitem[{{Miglio} {et~al.}(2006){Miglio}, {Bourge}, {Montalban}, \&
  {Dupret}}]{miglio06}
{Miglio}, A., {Bourge}, P.~., {Montalban}, J., \& {Dupret}, M.~. 2006, ArXiv
  Astrophysics e-prints 0611944

\bibitem[{{Momany} {et~al.}(2001){Momany}, {Vandame}, {Zaggia}, {Mignani}, {da
  Costa}, {Arnouts}, {Groenewegen}, {Hatziminaoglou}, {Madejsky}, {Rit{\'e}},
  {Schirmer}, \& {Slijkhuis}}]{eis01}
{Momany}, Y., {Vandame}, B., {Zaggia}, S., {et~al.} 2001, \aap, 379, 436

\bibitem[{{Montgomery} \& {O'Donoghue}(1999)}]{montgomery99}
{Montgomery}, M. \& {O'Donoghue}, D. 1999, Delta Scuti Newsletter, 13, 28

\bibitem[{{Neiner} {et~al.}(2005){Neiner}, {Floquet}, {Hubert}, {Fr{\'e}mat},
  {Hirata}, {Masuda}, {Gies}, {Buil}, \& {Martayan}}]{neiner05}
{Neiner}, C., {Floquet}, M., {Hubert}, A.~M., {et~al.} 2005, \aap, 437, 257

\bibitem[{{Nussbaumer} {et~al.}(1989){Nussbaumer}, {Schmid}, \&
  {Vogel}}]{nuss89}
{Nussbaumer}, H., {Schmid}, H.~M., \& {Vogel}, M. 1989, \aap, 211, L27

\bibitem[{{Pamyatnykh}(1999)}]{pam99}
{Pamyatnykh}, A.~A. 1999, Acta Astronomica, 49, 119

\bibitem[{{Pasquini} {et~al.}(2002){Pasquini}, {Avila}, {Blecha}, {Cacciari},
  {Cayatte}, {Colless}, {Damiani}, {de Propris}, {Dekker}, {di Marcantonio},
  {Farrell}, {Gillingham}, {Guinouard}, {Hammer}, {Kaufer}, {Hill}, {Marteaud},
  {Modigliani}, {Mulas}, {North}, {Popovic}, {Rossetti}, {Royer}, {Santin},
  {Schmutzer}, {Simond}, {Vola}, {Waller}, \& {Zoccali}}]{pasquini02}
{Pasquini}, L., {Avila}, G., {Blecha}, A., {et~al.} 2002, The Messenger, 110, 1

\bibitem[{{Roberts} {et~al.}(1987){Roberts}, {Lehar}, \& {Dreher}}]{roberts87}
{Roberts}, D.~H., {Lehar}, J., \& {Dreher}, J.~W. 1987, \aj, 93, 968

\bibitem[{{Robertson}(1974)}]{rob74}
{Robertson}, J.~W. 1974, \aaps, 15, 261

\bibitem[{{Samus} \& {Durlevich}(2004)}]{samus04}
{Samus}, N.~N. \& {Durlevich}, O.~V. 2004, VizieR Online Data Catalog, 2250

\bibitem[{{Schmid}(1989)}]{schmid89}
{Schmid}, H.~M. 1989, \aap, 211, L31

\bibitem[{{Schmidtke} {et~al.}(2004){Schmidtke}, {Cowley}, {Levenson}, \&
  {Sweet}}]{sch04}
{Schmidtke}, P.~C., {Cowley}, A.~P., {Levenson}, L., \& {Sweet}, K. 2004, \aj,
  127, 3388

\bibitem[{{Schwarzenberg-Czerny}(1991)}]{schwa91}
{Schwarzenberg-Czerny}, A. 1991, \mnras, 253, 198

\bibitem[{{Sebo} \& {Wood}(1994)}]{sebo94}
{Sebo}, K.~M. \& {Wood}, P.~R. 1994, \aj, 108, 932

\bibitem[{{Stellingwerf}(1978)}]{stell78}
{Stellingwerf}, R.~F. 1978, \apj, 224, 953

\bibitem[{{Sterken} {et~al.}(1994){Sterken}, {Vogt}, \&
  {Mennickent}}]{sterken94}
{Sterken}, C., {Vogt}, N., \& {Mennickent}, R. 1994, \aap, 291, 473

\bibitem[{{Szymanski}(2005)}]{ogleI}
{Szymanski}, M.~K. 2005, Acta Astronomica, 55, 43

\bibitem[{{Walker} {et~al.}(2005{\natexlab{a}}){Walker}, {Kuschnig},
  {Matthews}, {Cameron}, {Saio}, {Lee}, {Kambe}, {Masuda}, {Guenther},
  {Moffat}, {Rucinski}, {Sasselov}, \& {Weiss}}]{walker05a}
{Walker}, G.~A.~H., {Kuschnig}, R., {Matthews}, J.~M., {et~al.}
  2005{\natexlab{a}}, \apjl, 635, L77

\bibitem[{{Walker} {et~al.}(2005{\natexlab{b}}){Walker}, {Kuschnig},
  {Matthews}, {Reegen}, {Kallinger}, {Kambe}, {Saio}, {Harmanec}, {Guenther},
  {Moffat}, {Rucinski}, {Sasselov}, {Weiss}, {Bohlender}, {Bo{\v z}i{\'c}},
  {Hashimoto}, {Koubsk{\'y}}, {Mann}, {Ru{\v z}djak}, {{\v S}koda}, {{\v
  S}lechta}, {Sudar}, {Wolf}, \& {Yang}}]{walker05b}
{Walker}, G.~A.~H., {Kuschnig}, R., {Matthews}, J.~M., {et~al.}
  2005{\natexlab{b}}, \apjl, 623, L145

\bibitem[{{Wisniewski} \& {Bjorkman}(2006)}]{wisn06}
{Wisniewski}, J.~P. \& {Bjorkman}, K.~S. 2006, \apj, 652, 458

\bibitem[{{Wyrzykowski} {et~al.}(2004){Wyrzykowski}, {Udalski}, {Kubiak},
  {Szymanski}, {Zebrun}, {Soszynski}, {Wozniak}, {Pietrzynski}, \&
  {Szewczyk}}]{wyrz04}
{Wyrzykowski}, L., {Udalski}, A., {Kubiak}, M., {et~al.} 2004, Acta
  Astronomica, 54, 1

\bibitem[{{Zamanov} {et~al.}(2001){Zamanov}, {Reig}, {Mart{\'{\i}}}, {Coe},
  {Fabregat}, {Tomov}, \& {Valchev}}]{zamanov01}
{Zamanov}, R.~K., {Reig}, P., {Mart{\'{\i}}}, J., {et~al.} 2001, \aap, 367, 884

\bibitem[{{Zorec} \& {Briot}(1997)}]{zorec97}
{Zorec}, J. \& {Briot}, D. 1997, \aap, 318, 443

\end{thebibliography}

}

\tiny{
\begin{appendix}

\section{Peculiar objects}
\label{pec}
In this section, we present the 3 peculiar emission-line objects (non Be stars)
observed on October 21, 22 and 23, 2003.

\subsection{SMC5\_037102}

\begin{figure}[htp!]
    \centering
    \resizebox{\hsize}{!}{\includegraphics[angle=0]{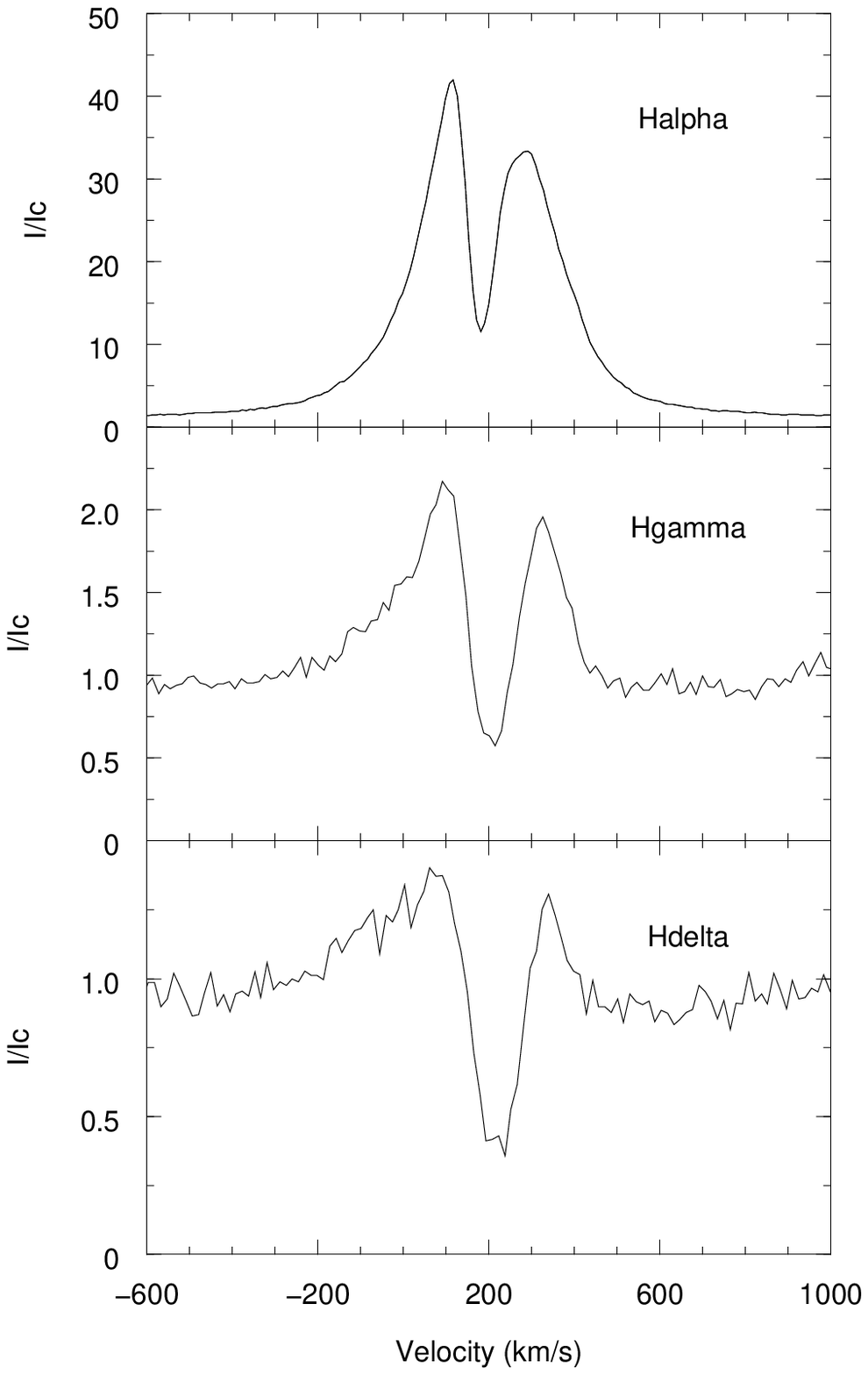}}
    \resizebox{\hsize}{!}{\includegraphics[angle=-90]{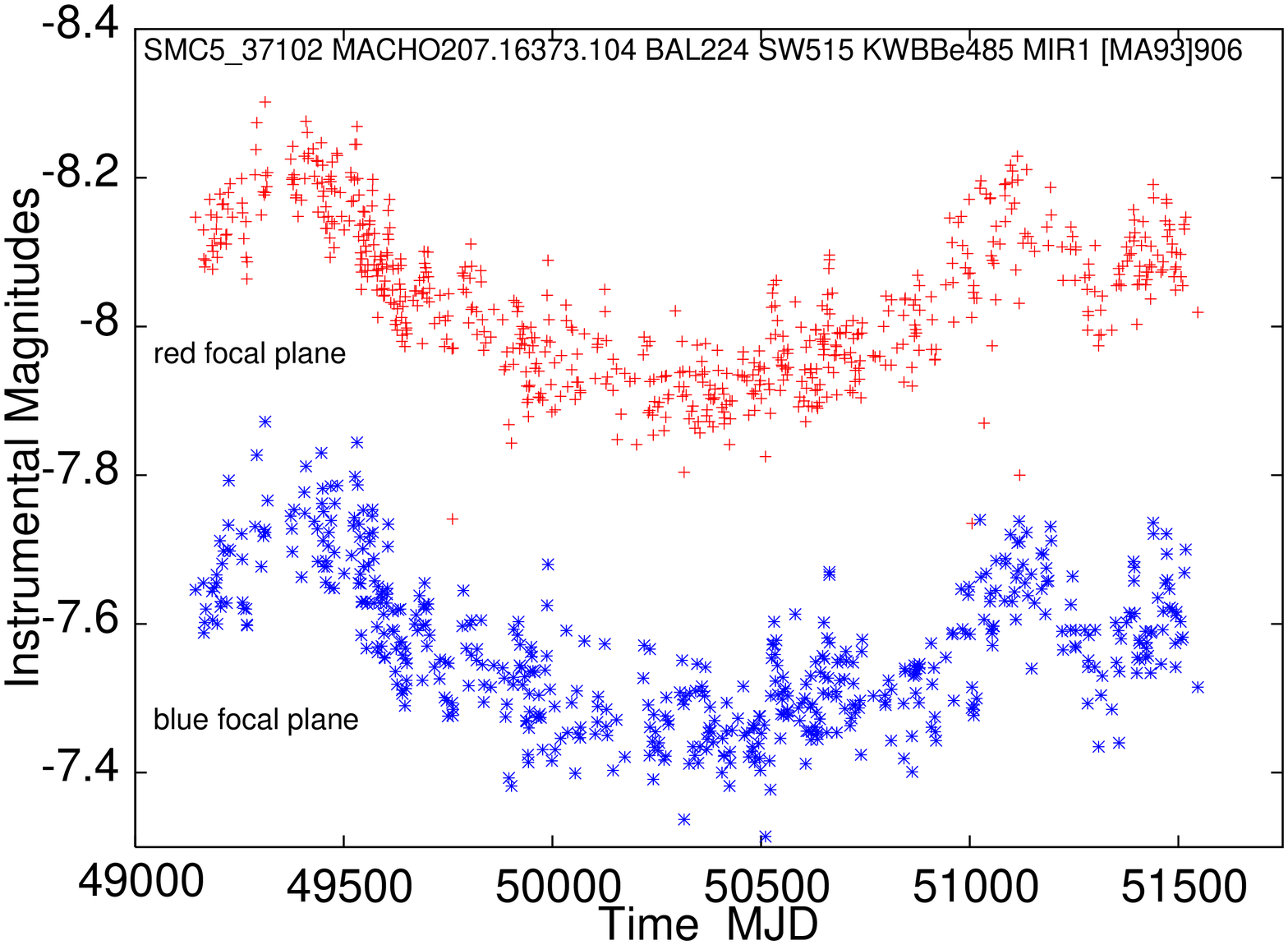}}
    \caption{SMC5\_037102. Top : Balmer emission lines H$\alpha$, H$\gamma$ and H$\delta$.
    Bottom: MACHO light-curves in the blue and red bands.}
    \label{spectrelon37102}
\end{figure}

The star is also known as  KWBBe485 \citep{keller99b},  BAL 224  \citep{balona92},  SW 515  \citep{sebo94},    
and [MA93] 906 \citep{meyss93}.  It has been reported as an IR object (MIR1) in \citet{kuc00}.   
From VLT-FORS1 low resolution spectroscopic observations \citet{hummel99}  
suggested that the absence of emission in \ion{He}{i}
lines and the strong Balmer decrement could indicate that this star has a shell
of  gas cooler than 5000 K. Thanks to ISOCAM observations \citet{kuc00}  
found a prominent mid-IR excess consistent with a dust shell of low
temperature (360 K). Without discarding the possibility that the object may be
an isolated Ae/Be star, they,  nevertheless, favoured a supergiant Be or an AGB star
to  explain visual and mid-IR photometry at once. The star also varied
photometrically. According to \citet{balona92}  it displayed a fading of 0.2 mag in 1991 and a
rapid variability close to 1 day but without any period satisfactorily fitting 
the data. We investigate the star's variability thanks to the MACHO and OGLE
databases; 2 strong outbursts, separated by smaller ones without any periodic
recurrence, occurred on a time-scale of about 3100 days and with a 0.4 mag
amplitude (see Fig.~\ref{spectrelon37102}). No short period can fit the data in 
spite of rapid variability close to a 1-d time-scale. \\
The GIRAFFE red spectrum is dominated by a very strong
asymmetric double-peaked H$\alpha$ emission line (EW = 360 \AA, V/R = 1.24)
affected by a shell feature. The R component seems to be larger than the blue one. 
In the blue spectrum H$\gamma$ and H$\delta$ also display  V/R $> 1$,  but the blue wing of 
their emission is highly complex compared to the red one. The mean value of the \rv ~of the shell
component of  H$\alpha$, H$\gamma$ and H$\delta$ (see
Table~\ref{vitradS37102}) is +193 \kms; it is red-shifted with respect to the
emission (Fig.~\ref{spectrelon37102}). Spectral parameters of the H lines are reported 
in Table~\ref{vitradS37102} and compared
to the ones obtained by \citet{hummel99}. We confirm the strong Balmer
decrement.\\
Table~\ref{S37102raies} gives the identification of emission and absorption
lines. 
Thanks to the resolution used in this study, it is possible, for the first
time, to identify \ion{Fe}{ii} and [\ion{Fe}{ii}] emission lines as well as the [SII] nebular
lines 4068, 6717 and 6731; however, the nebular  [NII] lines 6548 and 6583 are
not detected in the spectrum. The mean radial velocity  (\rv) of emission lines
is +165 $\pm$10  \kms~indicating that this object belongs to the SMC. The FWHM
of metallic emission lines is about 80  \kms. We do not notice any significant
difference between RVs and FWHMs of permitted and forbidden emission lines.  
In spite of the low S/N ratio (S/N$\simeq$20 in the continuum)
and the presence of numerous metallic emission lines it has been possible to
measure RVs of absorption lines identified as  \ion{He}{i} lines 4026, 4388, 4471 and
6678 and \ion{O}{ii} (see Table~\ref{S37102raies}), however \ion{Mg}{ii} 4481 line is absent. All
these identifications indicate that the central object is  a hot star. All the
helium lines are strongly red-shifted except \ion{He}{i} 4388. By comparison with
corresponding profiles of B stars having a similar radial velocity we can conclude
that this red-shift is in fact  due to  \ion{He}{i} emission which fills in the blue part
of the \ion{He}{i} photospheric lines. \ion{He}{i} line profiles as well as Balmer emission line profiles
may thus be explained by an accretion disk \citep{hummel99}. \\
In conclusion this star shows common characteristics with the pre-main sequence B[e] stars following 
\citet{lamers98}:  presence of \ion{Fe}{ii} and [\ion{Fe}{ii}]  emission lines with FWHM lower than 100 \kms, 
spectroscopic evidence of accretion or in-fall, large irregular photometric variations on time scale from 1 to
1000 days.

\begin{table}[tbph]
\tiny{
\caption{Spectral parameters of Balmer lines in  SMC5\_037102.
Values between brackets are taken from \citet{hummel99}. }
\centering
\begin{tabular}{cccc}
\hline
\hline	
	& H$\alpha$ & H$\gamma$ & H$\delta$ \\
\hline	
RV$_{V}$ ($\pm$20) \kms & 104 [140 $\pm$50] & 86 & 62 \\
RV$_{shell}$ ($\pm$20) \kms & 173 & 201 & 207 \\
RV$_{R}$ ($\pm$20)  \kms & 276 [301 $\pm$50] & 317 & 327 \\
FWHM ($\pm$20) \kms   &  320 [443 $\pm$50] & 410 & 600 \\
I$_{V}$ & 41.8 [18.6]& 2.2 & 1.4 \\
I$_{R}$ & 33.4 [15.2]& 1.9 & 1.3 \\
Mean I & 37.6 [16.9]& 2.1 & 1.4 \\
EW ($\pm$20) \AA & 360 [202 $\pm$20] & & \\
\hline
\end{tabular}
\label{vitradS37102}
}
\end{table}

\subsection{SMC5\_081994}

This star, also known as KWBBe4154, is the planetary nebula  SMC SMP 21 in \citet{lindsay61}.
From its spectral characteristics it has been classified as a type I PNe by \citet{leisy96,leisy06}.  
These authors claimed that the central star could belong to a binary system. We present the
first spectroscopic observations of this object obtained at medium resolution. The identification of
emission lines and their corresponding flux  are given in Table~\ref{S81994raies}.  The flux have been
obtained by fitting a gaussian function. The blue and red spectrum are displayed in Fig.~\ref{spec81994}.
The mean FWHM of nebular and emission lines is 1.18 \AA.\\
Thanks to the good quality of the spectra it
is possible to distinguish a triple structure at the foot of the most intense emission lines H$\alpha$, H$\gamma$, 
H$\delta$ and nebular lines of [NII], [SII] and [OIII] (see
Fig.~\ref{spec81994f}). Each central emission peak is flanked by 2 weak emission peaks, which are
about symmetric around the central one ($\pm$ 150 \kms~in the red domain and $\pm$ 120 \kms~in the blue domain). These
secondary emission peaks could be the signature of matter previously  ejected from the central star.
In the case of Balmer lines the blue component is about 5.4 times stronger than the red one whereas in the
case of nebular lines the two components seem to be of equal intensity.\\
It has to be noted that we observed broad emission patterns at 6834.74 and 7092.52 \AA, which can be identified 
as Raman scattering by neutral hydrogen of the OVI resonance doublet at 1032 and 1038 \AA~\citep{nuss89,schmid89}.  

\begin{figure}[htp!]
    \centering
    \resizebox{\hsize}{!}{\includegraphics[angle=-90]{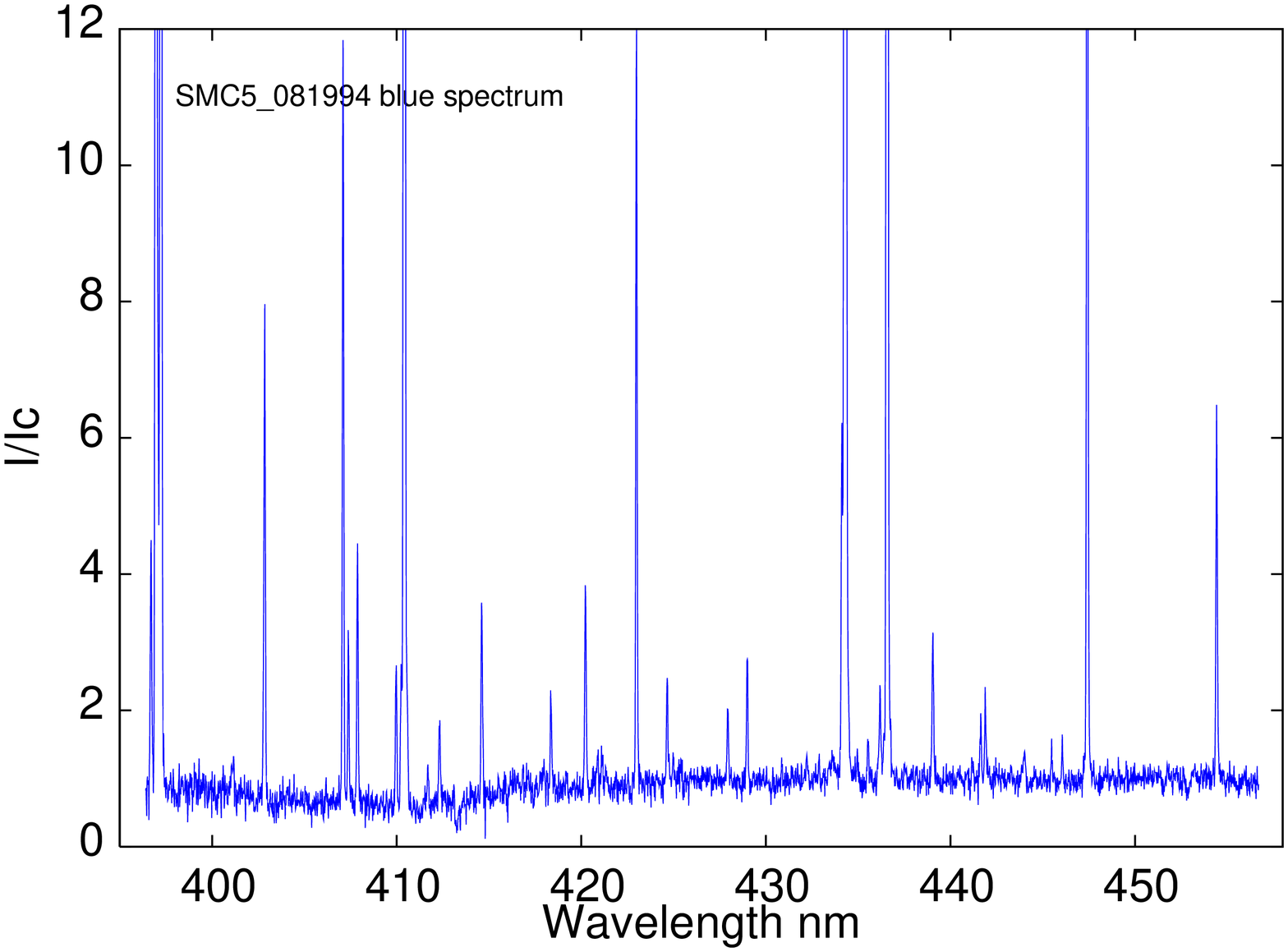}}
    \resizebox{\hsize}{!}{\includegraphics[angle=-90]{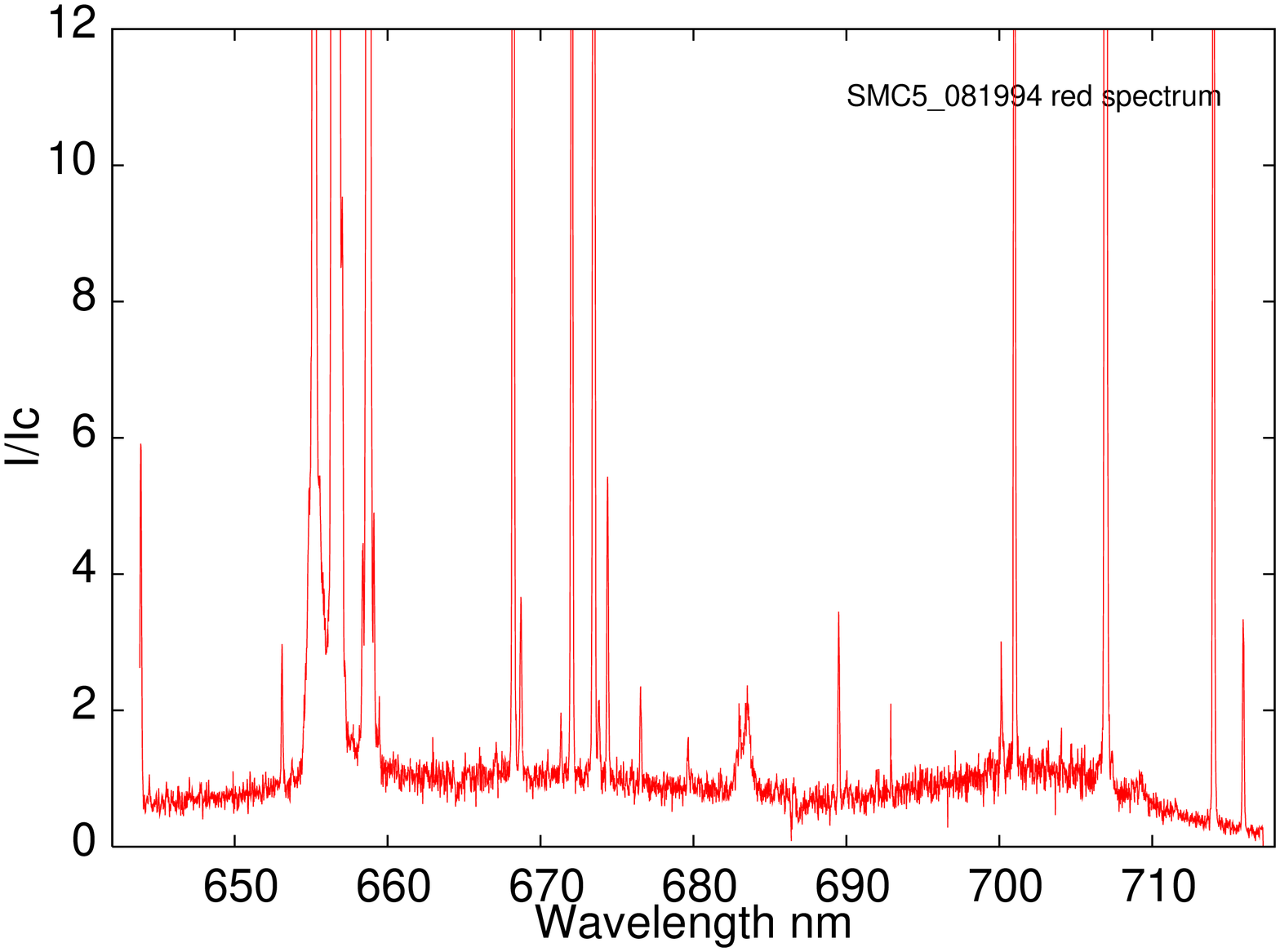}}    
    \caption{Spectrum of the PN SMC5\_081994. Top: blue domain. Bottom: red domain.}
    \label{spec81994}
\end{figure}
\begin{figure}[htp!]
    \centering
    \resizebox{\hsize}{!}{\includegraphics[angle=0]{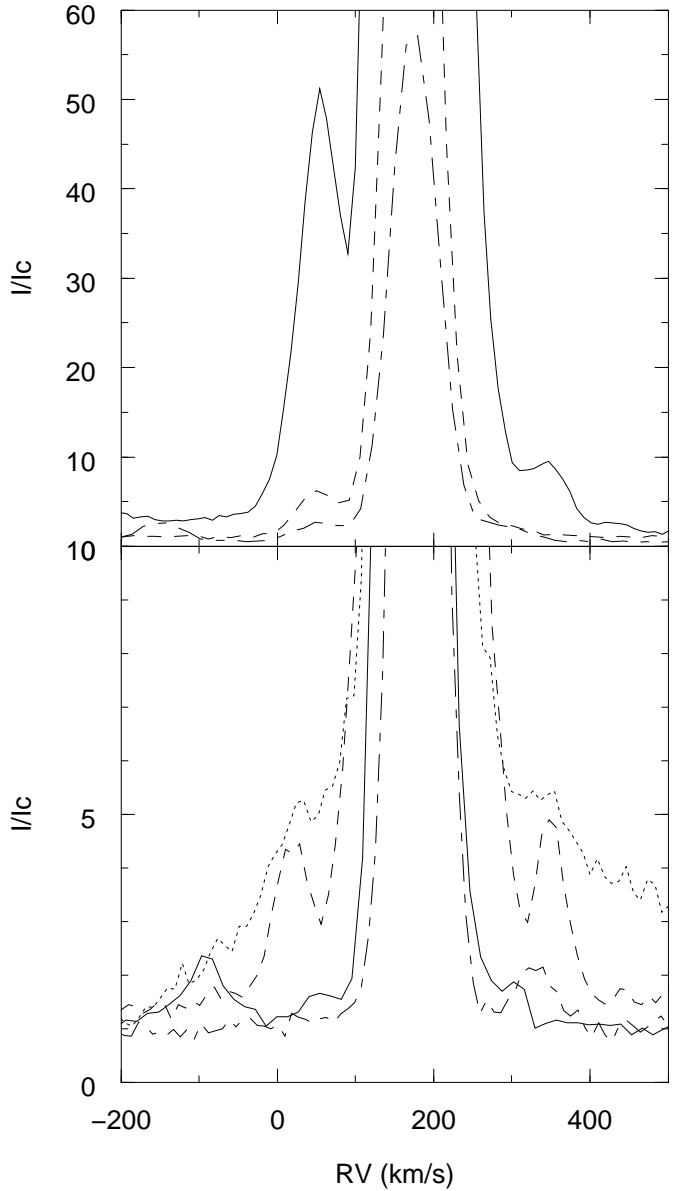}}    
    \caption{Triple structure at the foot of the most intense emission lines in
	     the spectrum of the PN SMC5\_081994. Top: Balmer lines. The 
	     solid line is H$\alpha$, the dashed line is H$\gamma$ and the dot-dashed line is H$\delta$.
	     Bottom: nebular lines. The
	     solid line is [OIII] 4363, the dotted line is [NII] 6548, the dashed line is [NII] 6583 and the 
	     dot-dashed line is [SII] 6730. }
    \label{spec81994f}
\end{figure}

\subsection{SMC5\_2807}
\label{pec2807}

\begin{table*}[htbp!]
\tiny{
\caption{Characteristics of the H$\alpha$ emission line of SMC5\_2807 in the spectra from this study
 and \citet{hummel01}.}
\centering
\begin{tabular}{cccccccc}
\hline
Date & EW~\AA & I(V)/Ic & I(R)/Ic & RV(V)~\kms & RV(R)~\kms & FWHM~\kms & Source\\
\hline	
2003/10/22 & 20.7 & 4.1 & 3.0 & +8  & +188 & 374 & this study\\
1999/11/12 &  8   & 1.6 & 2.1 & -70 & +240 & 460 & \citet{hummel01}\\
\hline	
\end{tabular}
\label{S2807Ha}
}
\end{table*}

This star is also known as KWBBe44  \citep{keller99b},  ROBB31 \citep{rob74},  BAL317  \citep{balona92},  
ARP2  \citep{arp59}  and SW77 \citep{sebo94}. These studies are mainly based on photometric surveys and/or  
low-resolution spectroscopic data.  From CCD Str\"omgren photometry \citet{grebel92a}  
 found that this star has a very low metallicity: [M/H] = -2.37 dex. From a 4-year monitoring in V 
 and I bands with CCD images to search for Cepheids and Long-Period Variables (LPV)  \citet{sebo94}   
 detected a 227d-period. 
 Then, from the light curve and a H$\alpha$ spectrum, they investigated
the nature of the star and concluded that the star is a foreground binary system
consisting of a red supergiant  or upper AGB star and  a compact star
surrounded by a disk where the H$\alpha$ emission line is formed. They estimated
the mass  M $\simeq$ 13 M$_{\odot}$ and the stellar radius R $\simeq$ 171~R$_{\odot}$
for the red star, the companion would have M = 1 M$_{\odot}$. From broadband IR
photometry, \citet{keller99a}  showed that the star is a red supergiant with Teff
$\sim$ 4355 K and log(L/L$_{\odot}$) =  3.971.\\
The star is included in the OGLE catalogue of eclipsing binary stars \citep{wyrz04}   
which gives a 452.29480-day period. Using MACHO data we find a
454.959-day period (see Sect.~\ref{bin}, Table~\ref{tablebinPNM}), close to the one derived from the
OGLE survey.\\
The GIRAFFE observations in LR02 and LR06 have been obtained at phase 0.07 i.e. out of 
eclipses (Fig.~\ref{spec2807}). We observe a cool star spectrum with
RV$_{*}$ = +160~\kms. Thus this star seems to be a SMC member. The H$\alpha$
emission line shows an asymmetric double profile, with V$<$R and a deep central
absorption, similar to the one shown in \citet{sebo94}  and \citet{hummel01}. 
However, the spectral parameters of the H$\alpha$ emission line are
variable (see Table~\ref{S2807Ha}) and changed between our study and \citet{hummel01} who then observed this
system in a stage of lower emission.
The RV of the narrow deep absorption
is +98~\kms~for H$\alpha$ and +120~\kms~for H$\gamma$  and H$\delta$ in
GIRAFFE spectra. Those features,  as well as the double  emission in H$\alpha$,
are probably formed in a disk which surrounds the secondary star or in a
circumbinary disk.

\begin{figure}[htp!]
    \centering
    \resizebox{\hsize}{!}{\includegraphics[angle=-90]{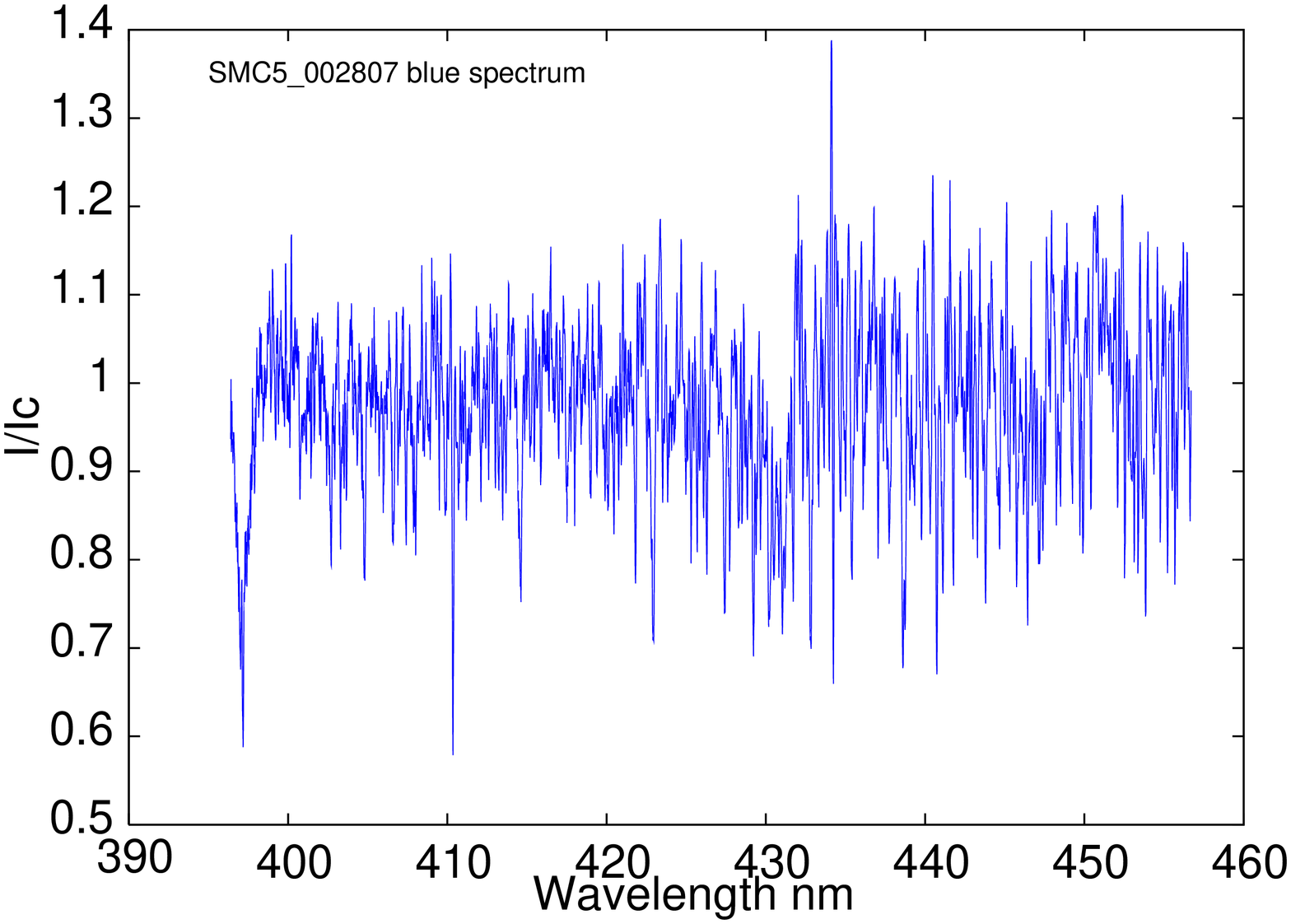}}
      \resizebox{\hsize}{!}{\includegraphics[angle=0]{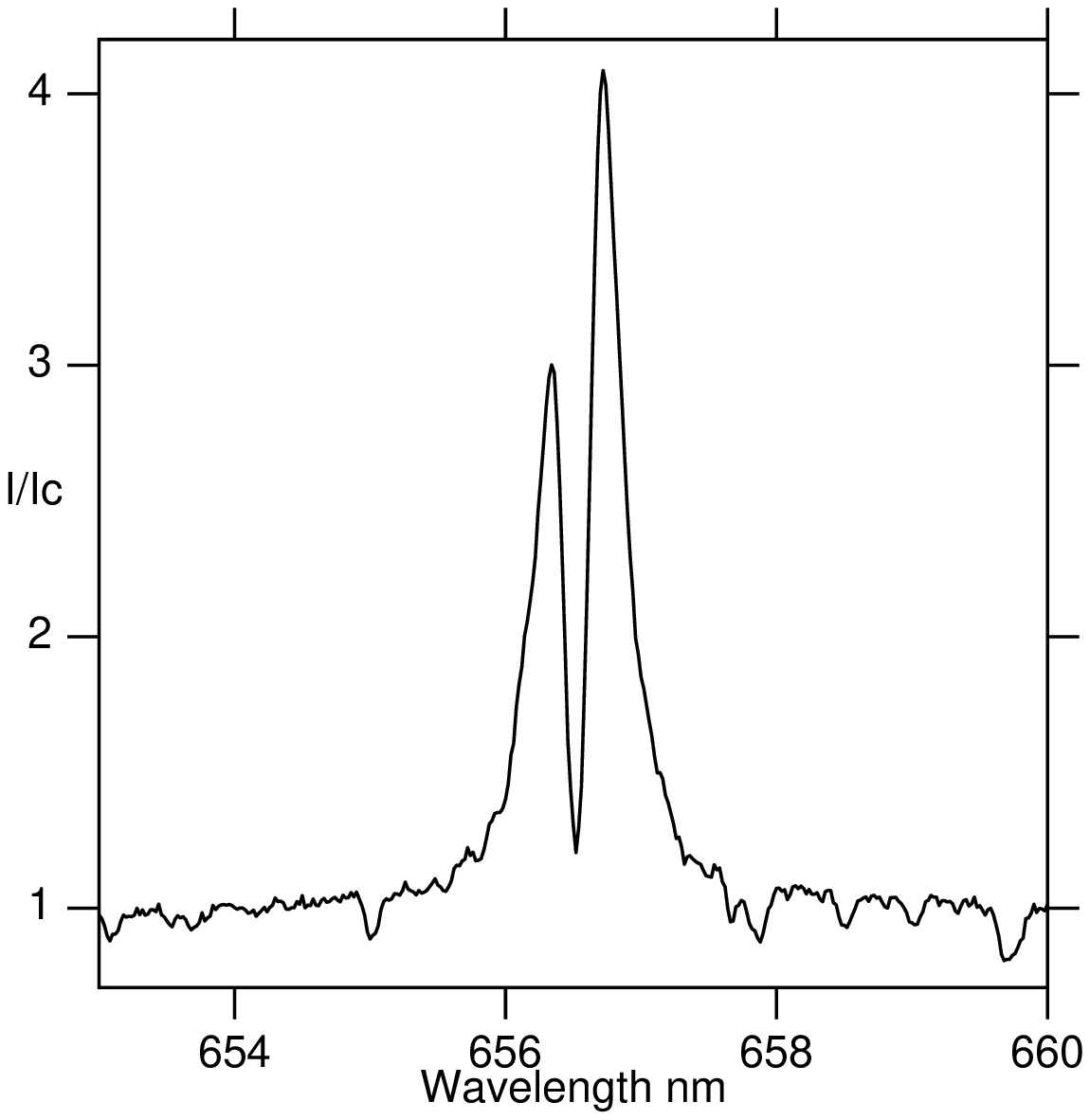}}
    \caption{Spectra of the cool supergiant SMC5\_002807, top: blue spectrum; bottom: zoom on H$\alpha$.}
    \label{spec2807}
\end{figure}
\end{appendix}

\begin{appendix}
\section{Additional comments on binaries}
\label{appB}

\textbf{SMC5\_016461:}
This object is a Be-shell star with a strong double H$\alpha$ emission line  (EW$\alpha$ = 52\AA, Imax $\leq$ 9.5 with V$<$ R).  This
star was observed twice within an interval of 11 months (2003 and 2004).
The blue spectrum displays numerous metallic shell lines of  \ion{Fe}{ii}, \ion{Ti}{ii}, \ion{Cr}{ii} and \ion{Si}{ii};  the stronger
\ion{Fe}{ii} lines are flanked by V and R emission components. In 2004 we observed 2 absorption components in \ion{Fe}{ii} and
\ion{Ti}{ii} lines (RV = +81 and +115 \kms) but only one in  \ion{He}{i}, \ion{Cr}{ii}, \ion{Mg}{ii} and \ion{Si}{ii}  (on average RV
= +83~\kms). In the 2003 spectra,  the lines are deeper and simple and their average RV is  +108 \kms. The radial velocity of
the shell components of the Balmer  lines   is +126 \kms~in  2003 and +130 \kms~in 2004. 
The accuracy on the RV measurements is $\pm$2 \kms.\\
The MACHO and OGLE data show a periodic variability with P = 27.1687 d with a peak to peak  amplitude $\sim$ 0.2 mag, but we cannot
exclude that the true period could be twice the detected period  i.e. P = 54.3374 d (see Fig.~\ref{bin16461}). In this latter case the
two minima are not very different from each other. In this picture, this object would belong to the group of \object{$\beta$ Lyrae} type
eclipsing variables with long period in which the light curve does not show any plateau and varies continuously. 
This star has a behavior similar to the one of \object{HD50123} \citep{sterken94}: similar light curve, long period, a Be star as 
primary, relative intensity of the 2 minima. As HD50123, this object could thus be an interacting binary with a Be star as primary, and
the companion displaying ellipsoidal variations with a period of 54.3374d.   More spectroscopic observations are needed to  validate our
proposition.\\ 
\textbf{SMC5\_052516:}
The MACHO data do not show any evidence of periodicity. However, the star could be a SB2. The \ion{He}{i} lines display either a
splitting or an asymmetry (red-winged profile).\\
\textbf{SMC5\_052663:}
The MACHO data do not show any evidence of periodicity. However, the star is clearly a SB2 system with a separation in \rv~of 225
\kms~(see Fig.~\ref{bin52663}).

\end{appendix}


\begin{appendix}
\section{Comments on the individual Be stars}
\label{indBe}

\textbf{SMC5\_013978:}
\citet{balona92} has previously observed this star (number 964 in his list) and 
found it variable with a frequency of 1.361 $\mbox{c d}^{-1}$. 
The large time baseline of the MACHO light curves ($\sim$ 2700 d) allows us to refine his 
frequency and, furthermore, to find an additional frequency (Fig.~\ref{varbe2}).\\
\textbf{SMC5\_014727:}
\citet{balona92} has previously observed this star (number 585 in his list) and found it variable with
a frequency at 1.120 $\mbox{c d}^{-1}$.
Thanks to the larger time span of the MACHO observations, we are able to 
refine this frequency (Fig.~\ref{varbe3}).\\
\textbf{SMC5\_016544:}
To confirm the two frequencies found, we folded the light curve 
with the frequency produced by the beating of the two frequencies,
i.e. f$_{\mathrm{beating}}$ = 1.70774-1.64993 = 0.05781 $\mbox{c d}^{-1}$ (Fig.~\ref{varbe4}). This plot 
shows evidence of beating phenomenon.\\
\textbf{SMC5\_037013:}
After prewhitening for the two first frequencies (Fig.~\ref{varbe5}), 
we find a significant frequency at f3 = 1.18113 $\mbox{c d}^{-1}$ (SNR $\sim$ 6).
As the difference between f1 and f3
is very close to our resolution in frequency and the phase diagram folded with f3 is
very scattered, we consider it as uncertain.  \\
\textbf{SMC5\_037162:}
OGLE \citep{wyrz04} found a frequency f = 0.443 $\mbox{c d}^{-1}$.
From our analysis, we obtained the frequency f1, which is twice the OGLE's frequency, and an additional frequency.\\
\textbf{SMC5\_082941:}
Note that the frequency f3, due to its time-scale (period of 6.526 d), is probably not due to 
pulsation, but to surface activity or a possible binarity. 
We also detected an additional frequency at 0.62525 $\mbox{c d}^{-1}$ which fulfills the SNR requirement.

\end{appendix}
}


\Online

%
\begin{figure*}[htp]
    \centering
    \resizebox{\hsize}{!}{\includegraphics[angle=-90]{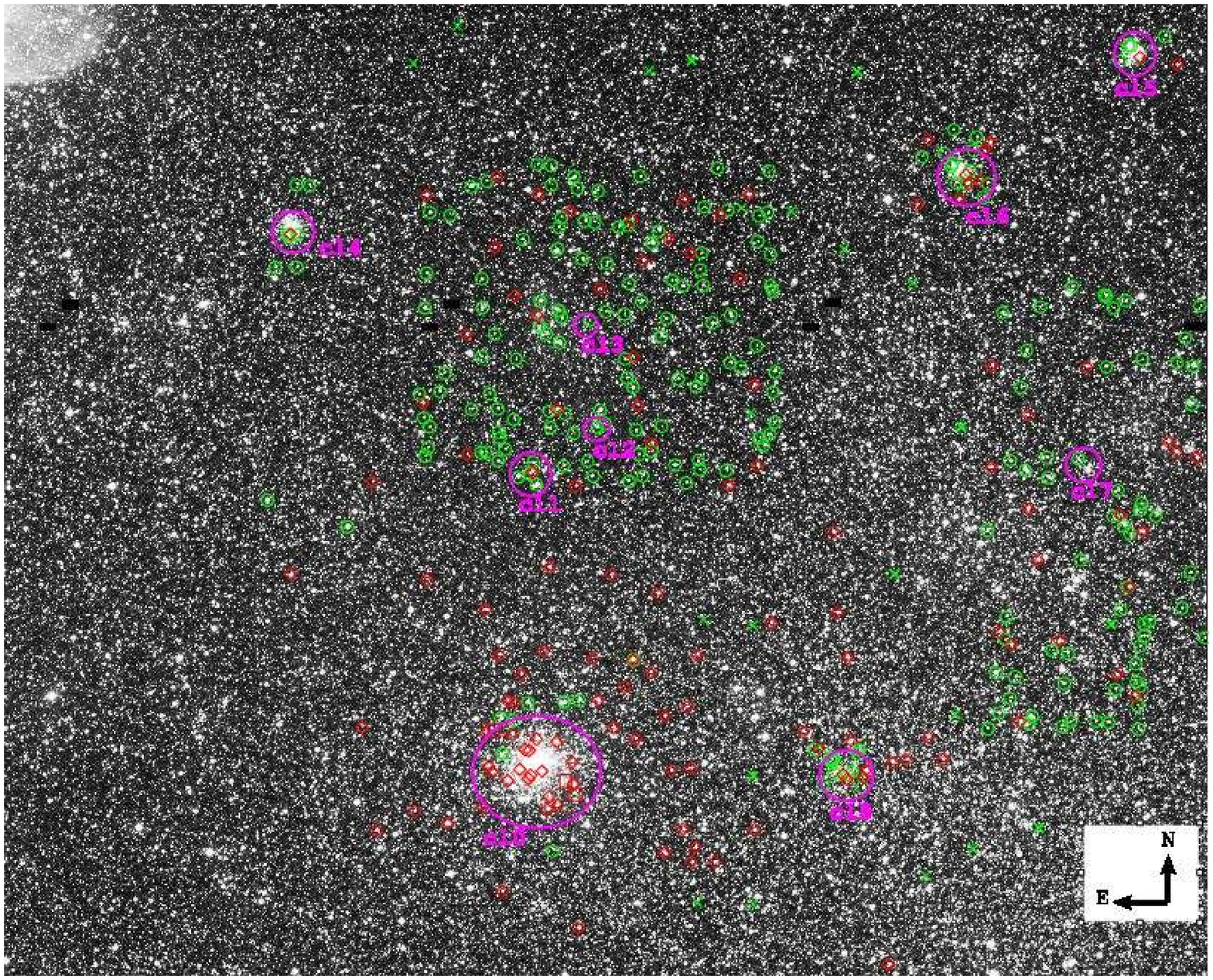}}
    \caption{Part of the SMC5 field from EIS pre-FLAMES survey \citep{eis01}. 
    Red diamonds are for Be stars, red squares for the other emission line stars, small green circles are for non-emission line stars
    (mainly B-type stars), green crosses for the sky fibre positions. The large pink circles indicate the clusters and associations:
    cl0 for NGC330, cl1 for H86 170, cl2 for [BS95]78, cl3 for SMC ASS39, cl4 for OGLE SMC109, cl5 for NGC299, cl6 for NGC306, cl7 for
    H86 145, and cl8 for OGLE SMC99.}
    \label{figure0}
\end{figure*}
\tiny{
\begin{longtable}{llllllllll}
\caption{\label{tablebeobs} Observational characteristics of Be stars in the SMC. The first column
gives the name of the Be star in the EIS catalogue or following our catalogue if the star is 
not found in the EIS catalogue. In the last column some complementary 
indications on the spectrum are given: `cl' the star belongs to a cluster, `bin' the star 
is a binary; `sh' the star shows  shell lines; 
`\ion{Fe}{ii}' (red or blue lines) and `\ion{He}{i}' (red lines) indicate that
the corresponding lines present emission components, 'dble' and 'w' when the corresponding emission  
is a double peak one and  weak respectively; `noneb' no nebular lines have been
detected; `$\alpha$(CS+neb)' the nebular line is present at H$\alpha$ but
cannot be disentangled from the CS emission and `$\alpha$CS+$\alpha$neb' the 
H$\alpha$ CS and nebular lines can be disentangled; the mention kXXX is the number of the star in 
 \citet{keller99b}. '**' indicates that the star was pre-selected from the ESO-WFI slit-less study  \citet{marta06c}.
Note that the star SMC5\_016461 was observed twice within 11 months interval.}\\
\hline
\hline
Name  & $\alpha$  & $\delta$  & Vmag  & EW$\alpha$  & I(V)  & I(R)  &  Imax  & FWHM$\alpha$ & Remarks \\
      & 2000      &  2000     &       & \AA         &       &       &        & \kms &  \\
\hline 
\endfirsthead
\caption{continued.}\\
\hline
\hline
Name  & $\alpha$  & $\delta$  & Vmag  & EW$\alpha$  & I(V)  & I(R)  &  Imax  & FWHM$\alpha$ & Remarks \\
      & 2000      &  2000     &       & \AA         &       &       &        & \kms &  \\
\hline 
\endhead
\hline
\endfoot
MHF[S9]47315 & 00 54 49.559 & -72 24 22.35 & \_ & 52.73  &   &   & 8.78  & 232  & **, $\alpha$(CS+neb), \ion{He}{i} dble, \ion{Fe}{ii}\\
MHF[S9]51066 & 00 54 50.936 & -72 22 34.63 & \_ & 52.08  &   &   & 6.76  & 351  & **, $\alpha$(CS+neb), \ion{Fe}{ii} dble\\
SMC5\_000476 & 00 53 23.700 & -72 23 43.80 & 16.36 & 19.52  &   &   & 5.15  & 291  & $\alpha$CS+$\alpha$neb\\
SMC5\_000643 & 00 55 44.490 & -72 20 38.00 & 16.30 & 39.49  &   &   &  8.94:  & 159  & $\alpha$CS+$\alpha$neb\\
SMC5\_002232 & 00 56 05.560 & -72 31 25.68 & 15.60 & 57.53  &   &   & 10.89  & 209  & **, $\alpha$(CS+neb), \ion{Fe}{ii}\\
SMC5\_002483 & 00 55 32.170 & -72 29 56.70 & 16.80 & 39.27  &   &   & 6.73  & 263  & k1064, $\alpha$CS+$\alpha$neb \\
SMC5\_002751 & 00 56 14.260 & -72 28 30.10 & 15.99 & 60.54  &   &   & 9.69  & 259  & cl, $\alpha$(CS+neb), \ion{Fe}{ii}\\
SMC5\_002825 & 00 54 41.373 & -72 28 02.42 & 17.43 & 25.32  &   &   & 5.79  & 213  & cl, $\alpha$(CS+neb)\\
SMC5\_002957 & 00 54 45.158 & -72 27 13.68 & 15.28 & 33.92  &   &   & 6.58  & 250  &  **, $\alpha$(CS+neb), \ion{Fe}{ii} w\\
SMC5\_002984 & 00 55 48.780 & -72 27 12.70 & 17.40 & 14.47  & 2.67  & 2.12  &   & 438  & sh, $\alpha$CS+$\alpha$neb, \ion{Fe}{ii} \\
SMC5\_003119 & 00 55 59.900 & -72 26 21.30 & 15.62 & 39.34  &   &   & 6.93  & 247  & k258, $\alpha$(CS+neb), \ion{Fe}{ii} w \\
SMC5\_003296 & 00 56 15.964 & -72 25 15.98 & 16.83 & 15.50  & 3.95  & 4.03  &  &  216  & k916, $\alpha$CS+$\alpha$neb \\
SMC5\_003315 & 00 53 58.250 & -72 25 02.60 & 16.06 & 19.06  & 3.12  & 3.14  &   & 393  & sh, $\alpha$CS+$\alpha$neb\\
SMC5\_003389 & 00 55 08.999 & -72 24 37.24 & 16.24 & 56.98  & 7.50  & 7.28  &   & 336  & **, $\alpha$CS+$\alpha$neb, \ion{Fe}{ii} dble\\
SMC5\_003537 & 00 56 50.484 & -72 23 40.08 & 16.27 & 27.73  &   &   & 5.58  & 245  & **, $\alpha$CS+$\alpha$neb, \ion{Fe}{ii}?\\
SMC5\_003789 & 00 53 26.690 & -72 22 07.40 & 17.62 & 36.45  & 3.95  & 3.82?  &  & 490  & bin, $\alpha$CS+$\alpha$neb\\
SMC5\_003919 & 00 57 06.510 & -72 21 29.55 & 16.95 & 46.02  &   &   & 7.00:  & 311  & **, $\alpha$CS+$\alpha$neb\\
SMC5\_004026 & 00 56 38.660 & -72 20 52.80 & 17.53 & 28.31  & 3.81  & 3.69  &   & 452  & sh, noneb, \ion{Fe}{ii} dble \\
SMC5\_004201 & 00 56 12.270 & -72 19 52.90 & 16.33 & 69.04  &   &   & 6.13  & 454  & $\alpha$CS+$\alpha$neb, \ion{Fe}{ii} dble+w\\
SMC5\_004509 & 00 56 38.767 & -72 18 12.67 & 16.52 & 12.12  &   &   & 4.07  & 159  & $\alpha$(CS+neb)\\
SMC5\_004685 & 00 55 59.850 & -72 17 11.40 & 16.36 & 25.41  &   &   & 4.91  & 243  & $\alpha$(CS+neb), \ion{Fe}{ii}?\\
SMC5\_004982 & 00 55 24.880 & -72 15 30.60 & 16.29 & 36.38  & 6.48:  & 6.41  &   & 259  & $\alpha$CS+$\alpha$neb\\
SMC5\_005045 & 00 54 15.010 & -72 15 08.80 & 15.80 & 6.84  & 1.99  & 1.93  &   & 327  & $\alpha$CS+$\alpha$neb\\
SMC5\_008231 & 00 56 04.638 & -72 33 41.61 & 16.26 & 51.15  &   &   & 6.43  & 368  & **, $\alpha$(CS+neb), \ion{Fe}{ii}\\
SMC5\_009378 & 00 57 08.105 & -72 32 40.18 & 16.51 & 51.67  & 6.94  & 7.29:  &   & 304  & **, $\alpha$CS+$\alpha$neb, \ion{Fe}{ii} dble?\\
SMC5\_011371 & 00 56 28.170 & -72 30 36.80 & 16.94 & 41.30  & 3.81  & 4.31  &  &502  & $\alpha$CS+$\alpha$neb, \ion{Fe}{ii}?\\
SMC5\_011991 & 00 55 25.244 & -72 29 56.50 & 16.28 & 42.03  &   &   & 9.64  &  182  & **, $\alpha$(CS+neb)\\
SMC5\_012717 & 00 57 05.450 & -72 29 16.20 & 16.48 & 39.87  & 5.66  & 6.28  &  &316  & $\alpha$CS+$\alpha$neb, \ion{Fe}{ii}?\\
SMC5\_012767 & 00 55 34.670 & -72 29 13.10 & 16.66 & 33.56  & 3.20  & 3.66  &  &536  & k1054, sh, $\alpha$CS+$\alpha$neb, \ion{Fe}{ii} dble, \ion{He}{i}  \\
SMC5\_013233 & 00 56 33.898 & -72 28 43.92 & 18.15 & 8.09  & 2.26  & 2.75:  &   & 270 &  cl, $\alpha$CS+$\alpha$neb, \ion{Fe}{ii}?\\
SMC5\_013978 & 00 56 31.140 & -72 27 57.80 & 15.60 & 50.98  &   &   & 8.08  & 277  & cl, k206, $\alpha$(CS+neb), \ion{Fe}{ii}? \\
SMC5\_014052 & 00 56 23.010 & -72 27 53.90 & 15.35 & 44.96  & 10.38  & 10.07  &   & 170  & cl, k215, sh, noneb, \ion{Fe}{ii} \\
SMC5\_014114 & 00 56 32.261 & -72 27 50.17 & 15.52 & 40.36  &   &   & 7.20  & 219  & cl, k203, noneb, \ion{He}{i} w, \ion{Fe}{ii} \\
SMC5\_014212 & 00 54 32.960 & -72 27 41.94 & 15.42 & 55.28  &   &   & 10.40  & 227  &  **, $\alpha$(CS+neb), \ion{Fe}{ii} dble\\
SMC5\_014271 & 00 54 18.116 & -72 27 37.21 & 15.56 & 46.28  & 5.27  & 5.59  &   &  388  & **, $\alpha$CS+$\alpha$neb, \ion{Fe}{ii} dble, \ion{He}{i}\\
SMC5\_014637 & 00 56 12.130 & -72 27 16.90 & 15.42 & 23.33  &   &   & 5.80  & 182  & cl, k242, $\alpha$(CS+neb) \\ 
SMC5\_014727 & 00 56 18.130 & -72 27 13.50 & 15.65 & 23.72  &   &   & 4.20:  & 306  & cl, k228, $\alpha$CS+$\alpha$neb, \ion{Fe}{ii} dble? \\
SMC5\_014864 & 00 56 33.110 & -72 27 04.99 & 16.12 & 25.09  &   &   & 7.65  & 146  & cl, k419, $\alpha$(CS+neb), \ion{Fe}{ii} \\
SMC5\_014878 & 00 54 59.326 & -72 27 02.12 & 15.71 & 40.49  &   &   & 5.82  & 300  & **, $\alpha$(CS+neb), \ion{Fe}{ii} dble\\
SMC5\_015429 & 00 55 33.650 & -72 26 29.90 & 17.63 & 20.97  & 2.57  & 2.58  &   & 552  & k2299, $\alpha$CS+$\alpha$neb \\
SMC5\_015509 & 00 56 24.620 & -72 26 24.70 & 16.98 & 44.58  &   &   & 5.43  & 388  & k857, $\alpha$CS+$\alpha$neb \\
SMC5\_015867 & 00 55 52.279 & -72 26 03.77 & 17.23 & 31.79  &    &   & 7.45  & 186  & k991, $\alpha$(CS+neb) \\
SMC5\_016177 & 00 55 44.521 & -72 25 43.91 & 16.99 & 30.26  &    &   & 5.85  &  243  & k1017 $\alpha$CS+$\alpha$neb, \ion{Fe}{ii} dble+w\\
SMC5\_016461 & 00 55 49.619 & -72 25 27.43 & 14.90 & 52.33  & 9.08  & 9.68  &  & 245  & k137, bin, sh,**, noneb, \ion{Fe}{ii} dble, \ion{He}{i}\\
	                    & 00 55 49.630 & -72 25 27.30 & 14.90 & 50.99  & 8.70  & 8.99  &   & 259  & k137, bin, sh,**, noneb, \ion{Fe}{ii} dble, \ion{He}{i}\\
SMC5\_016477 & 00 56 01.510 & -72 25 25.78 & 18.76 & 13.74  & 2.65  & 2.36  &   &  418  & $\alpha$CS+$\alpha$neb \\
SMC5\_016486 & 00 54 46.371 & -72 25 22.73 & 15.58 & 48.80  &   &   & 5.45  & 482 & **, $\alpha$(CS+neb), \ion{Fe}{ii} dble\\
SMC5\_016523 & 00 55 30.790 & -72 25 20.30 & 15.70 & 34.46  &   &   & 4.11  & 406  & k278, $\alpha$CS+$\alpha$neb, \ion{He}{i}, \ion{Fe}{ii} \\
SMC5\_016544 & 00 56 29.100 & -72 25 21.50 & 16.84 & 44.70  & 5.69  & 5.72  &   &  365  & k837, $\alpha$CS+$\alpha$neb, \ion{Fe}{ii} dble \\
SMC5\_016824 & 00 53 44.010 & -72 24 56.30 & 15.06 & 5.47  &   &  & 1.66  & 328 & $\alpha$CS+$\alpha$neb\\
SMC5\_017596 & 00 56 33.330 & -72 24 19.80 & 17.16 & 29.20  &    &    & 7.78  & 154  & k818, $\alpha$(CS+neb), \ion{Fe}{ii}? \\
SMC5\_018501 & 00 56 14.450 & -72 23 23.60 & 15.20 & 26.46  &   &  & 4.00  & 331  & k128, $\alpha$CS+$\alpha$neb, \ion{Fe}{ii} dble+w, \ion{He}{i}? \\
SMC5\_020211 & 00 56 06.798 & -72 21 35.34 & 16.86 & 34.83  & 4.90  & 3.63  &   & 393  & sh, noneb, \ion{Fe}{ii} dble \\
SMC5\_021152 & 00 53 12.660 & -72 20 29.50 & 15.30 & 8.93  & 1.81  & 1.94  &    & 390   & $\alpha$CS+$\alpha$neb\\
SMC5\_021886 & 00 55 48.566 & -72 19 46.88 & 17.50 & 19.16  & 3.21  & 3.10  &    & 388  & sh, $\alpha$CS+$\alpha$neb, \ion{Fe}{ii} dble  \\
SMC5\_022295 & 00 55 14.500 & -72 19 18.60 & 15.92 & 49.87  & ?  & ?  &   &    &  bad pixels in H$\alpha$, \ion{Fe}{ii}?\\
SMC5\_022628 & 00 53 37.080 & -72 18 50.60 & 15.86 & 22.67  &    &    & 3.39  & 398  & $\alpha$CS+$\alpha$neb, \ion{He}{i} asym., \ion{Fe}{ii}?\\
SMC5\_022842 & 00 55 49.880 & -72 18 42.10 & 17.77 & 58.91  & 8.20:  & 6.83  &   & 318  & $\alpha$CS+$\alpha$neb, \ion{Fe}{ii} dble+w \\
SMC5\_023931 & 00 56 24.635 & -72 17 20.79 & 17.36 & 14.59  & 3.21  & 3.16  &   & 379  & sh, $\alpha$CS+$\alpha$neb, \ion{Fe}{ii}  \\
SMC5\_025052 & 00 55 39.810 & -72 16 04.20 & 17.72 & 1.43  & 1.28  & 1.28  &   &  427  & H$\alpha$w, $\alpha$CS+$\alpha$neb \\
SMC5\_025589 & 00 56 08.450 & -72 15 28.00 & 17.97 & 16.16  & 2.90  & 2.97  &    &  347  & $\alpha$CS+$\alpha$neb \\
SMC5\_025718 & 00 54 27.140 & -72 15 15.90 & 16.87 & 49.15  & 6.54  & 5.27  &   & 393  & sh, noneb, \ion{Fe}{ii } \\
SMC5\_025816 & 00 55 16.580 & -72 15 04.70 & 15.62 & 48.43  &    &    & 8.74  & 236  & $\alpha$(CS+neb), \ion{Fe}{ii}\\
SMC5\_025829 & 00 56 17.880 & -72 15 05.90 & 16.23 &        &   &    &   &   & no red spectrum, $\alpha$CS+$\alpha$neb \\
SMC5\_026182 & 00 54 08.940 & -72 14 42.50 & 17.98 & 23.37  &   &   & 3.94  & 318  &  cl, $\alpha$CS+$\alpha$neb\\
SMC5\_026689 & 00 54 07.970 & -72 14 03.90 & 17.52 & 30.79  &   &   & 4.85  & 335  & noneb  \\
SMC5\_028368 & 00 53 11.930 & -72 12 04.60 & 16.61 & 12.30  & 2.02  & 2.03  &   & 518  & $\alpha$CS+$\alpha$neb\\ 
SMC5\_036967 & 00 55 40.100 & -72 29 44.70 & 16.36 & 51.96  & 8.89  & 5.31  &   & 270  & sh, k528, noneb, \ion{Fe}{ii} dble \\
SMC5\_037013 & 00 55 13.616 & -72 29 13.87 & 15.50 & 49.45  &   &   & 9.70  & 192  &  **, $\alpha$(CS+neb), \ion{Fe}{ii}\\ 
SMC5\_037137 & 00 56 26.602 & -72 28 09.40 & 15.84 & 75.48  &   &   & 10.86  & 295  & cl, k211, noneb, \ion{Fe}{ii} dble \\
SMC5\_037158 & 00 55 38.260 & -72 27 54.60 & 17.23 & 36.97  & 3.44  & 3.61  &   & 561  & k1041, noneb, \ion{Fe}{ii}? \\
SMC5\_037162 & 00 54 40.790 & -72 27 52.50 & 16.02 & 47.38  &   &   & 7.55  & 261  &  cl, $\alpha$(CS+neb), \ion{Fe}{ii}?\\
SMC5\_038007 & 00 53 54.170 & -72 19 55.42 & 17.27 & 13.07  & 3.15:  & 2.99:  &   &  286  & $\alpha$CS+$\alpha$neb\\
SMC5\_038312 & 00 55 18.950 & -72 16 56.60 & 17.72 & 23.27  &   &   & 4.47  & 280  &  $\alpha$CS+$\alpha$neb\\
SMC5\_038363 & 00 55 33.327 & -72 16 22.70 & 16.54 & 43.73  & 4.45  & 5.63  &   & 396  & sh, noneb, \ion{Fe}{ii} dble\\
SMC5\_041410 & 00 54 33.353 & -72 32 11.79 & 16.37 & 36.90  &   &   & 4.42:  & 418  & **, $\alpha$CS+$\alpha$neb, \ion{He}{i}, \ion{Fe}{ii} dble\\
SMC5\_043413 & 00 55 30.950 & -72 29 36.70 & 15.84 & 49.29  &   &   & 8.66  &  229  & k27, $\alpha$(CS+neb) \\
SMC5\_044117 & 00 56 11.660 & -72 28 41.80 & 16.34 & 41.62  & 5.73  & 5.78  &   & 313  & cl, k471, $\alpha$CS+$\alpha$neb \\
SMC5\_044693 & 00 56 19.851 & -72 28 01.57 & 15.96 & 18.33  & 3.24  & 3.17  &   & 307  & cl, k222, $\alpha$CS+$\alpha$neb \\
SMC5\_044898 & 00 56 07.514 & -72 27 43.74 & 16.85 & 67.52  &   &   & 8.03  & 345  & cl, k480, $\alpha$(CS+neb), \ion{Fe}{ii} dble \\
SMC5\_045353 & 00 54 22.312 & -72 27 07.72 & 15.53 & 55.51  & 6.88  & 6.35  &   & 356  & sh, **, noneb, \ion{Fe}{ii} dble\\
SMC5\_045747 & 00 55 40.150 & -72 26 41.70 & 16.58 & 42.39  &   &   & 6.52  & 288  & k529, $\alpha$(CS+neb), \ion{Fe}{ii}? \\
SMC5\_046388 & 00 53 26.980 & -72 25 41.40 & 16.47 & 3.50   & 1.4  & 1.3  &   & 412 & H$\alpha$w, $\alpha$CS+$\alpha$neb\\
SMC5\_046462 & 00 56 22.590 & -72 25 47.20 & 16.76 & 32.52  & 4.88  & 4.28  &   & 334  & k874, $\alpha$CS+$\alpha$neb \\
SMC5\_047763 & 00 55 42.620 & -72 23 58.20 & 16.04 & 10.60  &    &   & 3.20  & 406  & k522, $\alpha$CS+$\alpha$neb \\
SMC5\_048045 & 00 55 56.320 & -72 23 33.30 & 16.59 & 44.47  &   &    & 6.51  & 306  & k509, $\alpha$(CS+neb) \\ 
SMC5\_048047 & 00 57 30.565 & -72 23 32.73 & 15.90 & 31.78  & 5.35  & 4.82:  &   & 300  & **, $\alpha$CS+$\alpha$neb, \ion{Fe}{ii}?\\ 
SMC5\_048289 & 00 53 50.650 & -72 23 09.20 & 15.87 & 18.71  & 3.79  & 3.33  &   & 266  & $\alpha$CS+$\alpha$neb\\ 
SMC5\_049651 & 00 56 19.491 & -72 21 17.05 & 17.86 & 3.95  & 1.72  & 1.61  &    & 370  &  cl, $\alpha$CS+$\alpha$neb\\
SMC5\_049746 & 00 54 04.690 & -72 21 06.30 & 16.49 & 43.13  &   &   & 6.58  & 245  & $\alpha$(CS+neb)\\
SMC5\_049780 & 00 55 13.492 & -72 21 06.75 & 16.71 & 43.19  & 5.26  & 5.18  &   & 393  & $\alpha$CS+$\alpha$neb, \ion{Fe}{ii} dble, \ion{He}{i} dble?\\  
SMC5\_049996 & 00 53 10.350 & -72 20 42.30 & 15.88 & 36.69  & 4.43  & 4.01  &  &  409  & $\alpha$CS+$\alpha$neb, \ion{Fe}{ii} dble \\
SMC5\_051315 & 00 54 04.820 & -72 18 50.70 & 17.96 & 3.81  & 1.86  & 1.71  &  & 336  &  $\alpha$CS+$\alpha$neb \\
SMC5\_052688 & 00 55 47.140 & -72 16 34.00 & 15.49 & 0.10  & 1.37  & 1.23  &   & 175  & $\alpha$CS+$\alpha$neb \\
SMC5\_053267 & 00 55 50.490 & -72 15 39.90 & 17.04 & 33.52  & 4.96  & 4.76  &   & 338  & $\alpha$CS+$\alpha$neb, \ion{Fe}{ii}  \\
SMC5\_053756 & 00 54 12.372 & -72 14 48.00 & 16.53 & 34.17  &   &   & 8.88  & 161  &  cl, $\alpha$(CS+neb), \ion{Fe}{ii} dble\\
SMC5\_055592 & 00 53 22.720 & -72 11 55.70 & 16.68 & 43.56  &   &   & 10.51  & 173  &  cl, $\alpha$CS+$\alpha$neb, \ion{Fe}{ii}\\
SMC5\_061950 & 00 56 29.960 & -72 14 41.70 & 17.67 & 2.04  & 1.39  & 1.33  &    & 431  & $\alpha$CS+$\alpha$neb\\  
SMC5\_064327 & 00 56 14.900 & -72 28 47.50 & 15.41 & 67.91  & 7.86  & 7.92  &   & 350  & cl, k238, $\alpha$CS+$\alpha$neb, \ion{Fe}{ii} dble \\ 
SMC5\_064576 & 00 54 46.290 & -72 28 05.00 & 17.38 & 31.25  &   &   & 6.77:  & 211  &  cl, $\alpha$CS+$\alpha$neb\\
SMC5\_064745 & 00 54 28.887 & -72 27 38.20 & 15.69 & 24.74  & 3.75  & 2.80  &   & 377  &  **, $\alpha$CS+$\alpha$neb, \ion{He}{i} complex\\
SMC5\_064832 & 00 54 54.577 & -72 27 23.64 & 16.79 & 38.44  &   &   & 6.07  & 300  & $\alpha$(CS+neb) \\
SMC5\_065055 & 00 53 55.340 & -72 26 45.30 & 14.72 & 24.23  & 3.47  & 3.22  &   & 390  & $\alpha$CS+$\alpha$neb, \ion{He}{i} dble\\ 
SMC5\_065746 & 00 54 01.887 & -72 24 45.57 & 17.42 & 30.76  & 3.65  & 3.77  &   & 413  & $\alpha$CS+$\alpha$neb \\
SMC5\_066754 & 00 55 21.820 & -72 21 33.70 & 16.09 & 36.90  &   &   & 8.85  & 172  & $\alpha$(CS+neb) \\
SMC5\_067333 & 00 56 51.700 & -72 19 45.20 & 16.21 & 14.75  & 2.52  & 2.49  &   & 393 & $\alpha$CS+$\alpha$neb, \ion{He}{i}  \\
SMC5\_073581 & 00 56 26.600 & -72 26 23.00 & 16.22 & 13.81  & 2.45  & 2.65  &   & 365  & $\alpha$CS+$\alpha$neb \\
SMC5\_073594 & 00 53 21.410 & -72 26 08.90 & 16.13 & 12.51  & 3.20  & 2.98  &   & 227  & $\alpha$CS+$\alpha$neb \\
SMC5\_074402 & 00 53 04.530 & -72 20 49.30 & 15.83 & 35.22  &   &   & 4.90  & 331  & $\alpha$(CS+neb), \ion{He}{i}?, \ion{Fe}{ii}? \\
SMC5\_074471 & 00 53 26.610 & -72 20 18.60 & 14.97 & 19.08  & 3.19  & 3.12  &   & 330  & $\alpha$CS+$\alpha$neb\\ 
SMC5\_075061 & 00 56 30.580 & -72 16 16.20 & 17.58 & 13.18  &   &   & 4.83  & 141  & $\alpha$(CS+neb) \\
SMC5\_075360 & 00 54 05.990 & -72 13 51.60 & 15.78 & 57.33  & 6.27  & 5.86  &   & 418  & sh, noneb, \ion{Fe}{ii} dble \\ 
SMC5\_078338 & 00 56 25.450 & -72 27 07.00 & 15.49 & 49.32  & 11.99  & 10.68  &   & 152  & cl, sh, k213, noneb, \ion{Fe}{ii} \\ 
SMC5\_078440 & 00 56 50.560 & -72 15 07.30 & 15.63 & 12.23  & 2.34  & 1.89  &   & 409  & $\alpha$CS+$\alpha$neb, \ion{Fe}{ii} dble? \\
SMC5\_078928 & 00 57 09.690 & -72 26 57.50 & 17.48 & 57.96  & 6.97  & 4.03  &   & 322  & sh, $\alpha$CS+$\alpha$neb, \ion{Fe}{ii} strong, \ion{He}{i}\\
SMC5\_080910 & 00 54 24.272 & -72 13 49.41 & 16.74 & 33.25  & 4.19  & 3.80  &   & 436  & sh, noneb, \ion{Fe}{ii} dble \\
SMC5\_081260 & 00 54 13.120 & -72 14 35.60 & 17.77 & 1.80  & 1.14  & 1.15  &   & 497 & cl, H$\alpha$w, $\alpha$CS+$\alpha$neb \\
SMC5\_082042 & 00 56 18.260 & -72 17 46.80 & 16.39 & 65.27  &   &   & 7.40  & 384  & $\alpha$CS+$\alpha$neb, \ion{Fe}{ii} dble+w \\
SMC5\_082202 & 00 57 30.310 & -72 15 58.40 & 15.76 & 30.55  & 4.35  & 4.44  &   & 350  & cl, sh, noneb, \ion{Fe}{ii}?\\  
SMC5\_082543 & 00 56 54.005 & -72 28 50.18 & 17.05 & 50.70  &   &   & 7.32  & 313  & k754, $\alpha$(CS+neb) \\
SMC5\_082819 & 00 56 07.190 & -72 28 13.70 & 13.46 & 11.88  &   &   & 2.68  & 286  &  cl, sh, $\alpha$(CS+neb), \ion{Fe}{ii}\\
SMC5\_082941 & 00 53 53.660 & -72 22 01.40 & 15.75 & 59.43  & 7.33  & 6.15  &   & 368  & $\alpha$CS+$\alpha$neb, \ion{Fe}{ii} dble \\ 
SMC5\_083491 & 00 53 19.930 & -72 22 29.20 & 15.95 & 13.37  & 2.29  & 1.71  &   & 456  & $\alpha$CS+$\alpha$neb \\
SMC5\_085503 & 00 54 47.457 & -72 27 58.77 & 16.88 & 21.79  & 2.79  & 4.89  &   & 284  & cl, sh, $\alpha$CS+$\alpha$neb, \ion{Fe}{ii}  \\
SMC5\_086200 & 00 55 32.480 & -72 27 52.00 & 17.29 & 12.79  &  &    &   &  622  & k1062, $\alpha$CS+$\alpha$neb \\
SMC5\_086251 & 00 55 35.310 & -72 15 11.70 & 16.86 & 25.47  & 3.00  & 3.05  &    & 529  & sh, $\alpha$CS+$\alpha$neb\\ 
SMC5\_086581 & 00 55 55.498 & -72 26 58.34 & 17.53 & 25.07  & 4.17  & 4.36  &   & 325  & k2118, $\alpha$CS+$\alpha$neb \\
SMC5\_086890 & 00 56 16.439 & -72 27 56.38 & 16.88 & 50.96  &   &   & 7.16  & 316  & cl, k462, $\alpha$(CS+neb), \ion{Fe}{ii} dble \\
SMC5\_086983 & 00 56 20.410 & -72 28 06.40 & 16.24 & 33.07  &   &   & 4.45  & 350  & cl, k441, $\alpha$(CS+neb), \ion{Fe}{ii} dble \\
SMC5\_087004 & 00 56 21.394 & -72 27 27.89 & 17.18 & 44.24  & 6.07  & 5.90  &   & 336  & cl, k882, $\alpha$CS+$\alpha$neb, \ion{Fe}{ii} dble \\ 
SMC5\_090914 & 00 56 20.250 & -72 27 28.70 & 16.03 & 42.42  &   &   & 7.23  & 261  & cl, k442, $\alpha$(CS+neb) \\
SMC5\_190576 & 00 56 44.310 & -72 29 06.30 & 14.56 & 26.74  &   &   &  5.78  &  222  & $\alpha$(CS+neb), \ion{He}{i} dble \\
\hline
SMC5\_002807 & 00 56 09.420 & -72 28 09.30 & 14.62 & 20.7  &  3.00 &  4.07 &   & 374 & cl, bin, cool Sg, k44 \\
SMC5\_037102 & 00 56 06.450 & -72 28 27.70 & 17.32 & 360 & 41.84  &  33.29 &   & 337 &cl,  HB[e], k485 \\
SMC5\_081994 & 00 56 30.750 & -72 27 02.00 & 17.32 & 673 &   &   & 2019 & 64 &  cl, PNe, k4154\\
\end{longtable}
}

\addtocounter{table}{+4}

\tiny{
\begin{longtable}{ccc}
\caption{\label{S37102raies}Lines identification of SMC5\_037102.
E=Emission, A=Absorption,  sh=shell component of Balmer lines.}\\
\hline
\hline
Wave. (\AA) & ID  & E/A   \\
\hline 
\endfirsthead
\caption{continued.}\\
\hline
\hline
Wave. (\AA) & ID  & E/A   \\
\hline 
\endhead
\hline
\endfoot
3972.63	& sh H$\epsilon$  3970.074 & A     \\
3994.70	& [\ion{Cr}{ii}] (4F) 3992.080  & E    \\
4026.84	& \ion{Fe}{ii} (127) 4024.552  & E  \\
4028.45	& [\ion{Ni}{ii}] (4F)   4025.800  & E  \\
4030.34	& \ion{He}{i} (3) 4026.190  &  A \\
4056.15	& \ion{O}{ii} (50,98) 4054.100  & A \\
4062.82	& \ion{O}{ii}  (97) 4060.580  & A \\
4070.81	& [\ion{S}{ii}] (1NF) 4068.620  & E \\
4104.56	& sh H$\delta$ 4101.737 & A \\
4119.16	& [\ion{Ti}{ii}] (20F) 4116.600 & E \\
4134.11	& [\ion{Fe}{ii}] (24F) 4131.510 & E \\
4175.74	& \ion{Fe}{ii} (27) 4173.450 & E \\
4179.32	& [\ion{Fe}{ii}] (21F) 4177.210 & E \\
4180.66	& [\ion{Fe}{ii}] (23F)+ \ion{Fe}{ii} (28) 4178.855 & E \\
4187.10	& \ion{O}{ii} (36) 4185.456 & A \\
4207.78	& \ion{Fe}{ii} (P22) 4205.480 & E \\
4229.76	&	\ion{Al}{ii} (46) 4226.827  & A  \\
4232.35	& ? [\ion{Fe}{iv}] (1F) 4229.800 & E \\
4235.39	& \ion{Fe}{ii} (27) 4233.170 & E \\
4236.86	& [\ion{Fe}{ii}] (37F) 4234.810  & E \\
4246.49	&	[\ion{Fe}{ii}] (21F) 4243.980 & E \\
4271.75	& ? \ion{S}{ii} (49) 4269.760 & A \\
4279.30	&	\ion{O}{ii} (54,67) 4277.050  & E \\
4289.76	&	[\ion{Fe}{ii}] (7F) 4287.400  & E \\
4298.16	&	\ion{Fe}{ii} (28) 4296.567  & E \\
4302.16	&	 \ion{Ti}{ii} (41) 4300.052 & E \\
4305.25 &  \ion{Fe}{ii} (27) 4303.166  & E  \\
4321.86	&	[\ion{Fe}{ii}] (21F) 4319.620 & E \\
4343.38 & sh H$\gamma$ 4340.470 & A \\
4349.01	&	[\ion{Fe}{ii}] (21F) 4346.850  &	E  \\
4354.75	&	\ion{Fe}{ii} (27) 4351.640& E \\
4360.75	&	[\ion{Fe}{ii}] (21,6F) 4358.230& E \\
4361.78	&	[\ion{Fe}{ii}] (7F) 4359.340& E \\
4385.58	&	[\ion{Fe}{ii}] (6F) 4382.750& E \\
4390.62	& \ion{He}{i} (51) 4387.930& A \\
4416.11	&	[\ion{Fe}{ii}] (6F) 4414.450& E \\
4418.64	&	[\ion{Fe}{ii}] (6F) 4416.270& E \\
4419.65	&	? \ion{Fe}{ii} (27) 4416.817& E \\
4445.29	&	? \ion{Ti}{ii} (19) 4443.802    & E \\
4454.54	&	[\ion{Fe}{ii}] (7F) 4452.110& E \\
4460.43	&	[\ion{Fe}{ii}] (6F) 4457.950& E \\
4468.93	&	[\ion{Ni}{ii}] (10F) 4466.330& E \\
4472.66	&	[\ion{Fe}{ii}]  (6F) 4470.290& E \\
4475.51	&	\ion{He}{i} (14) 4471.477& A \\
4477.64	&	[\ion{Fe}{ii}] (7F) 4474.910& E \\
4479.59	&	\ion{O}{ii} (88) 4477.880& A \\
4491.19	&	[\ion{Fe}{ii}] (6F)+\ion{Fe}{ii} (37) 4489.465& E \\
4493.51	&	\ion{Fe}{ii} (37) 4491.401& E \\
4510.51	&	\ion{Fe}{ii} (38) 4508.280& E \\
4517.45	&	[\ion{Fe}{ii}] (6F)+ \ion{Fe}{ii} (37) 4515.118& E \\
4523.19	&	\ion{Fe}{ii} (37) 4520.225& E \\
4525.15	&	\ion{Fe}{ii} (38) 4522.630& E \\
4527.05	&	\ion{Ti}{ii} (60) 4524.732& E \\
4550.54	&	[\ion{Fe}{i}] (21F) 4548.320& E \\
4551.81	&	\ion{Fe}{ii} (38) 4549.467& E \\
4558.07	&	\ion{Fe}{ii} (37) 4555.890& E \\
6459.66	&	\ion{Fe}{ii} (74) 6456.380& E \\
6462.01	&	?? \ion{Fe}{iii} (3P) 6458.680& E \\
6494.99	&	\ion{Fe}{ii} 6491.280& E \\
6496.59	&	\ion{Fe}{ii} 6493.050 &  E  \\
6519.07	&	\ion{Fe}{ii} (40) 6516.050& E \\
6521.43	&	\ion{Fe}{ii} 6517.010& E \\
6566.60	& sh H$\alpha$ 6562.817& E \\
6685.45	& \ion{He}{i} (45) 6678.149& A \\
6720.12	&	[\ion{S}{ii}] (2F) 6717.000& E \\
6733.21	&	[\ion{Fe}{ii}] (31F) 6729.850& E \\
6734.50	&	[\ion{S}{ii}] (2F) 6731.300& E \\
7159.16	&	[\ion{Fe}{ii}] (14F) 7155.140 &  E \\
\end{longtable}
}
\tiny{
\begin{longtable}{cccc}
\caption{\label{S81994raies} Lines identification for SMC5\_081994. 
E=Emission, A=Absorption, Neb.=nebular, sat=satellite emission line. 
The flux was measured above the continuum with an accuracy of 20\%.
}\\
\hline
\hline
Wave. (\AA) & ID  & flux (\AA) & comment \\
\hline 
\endfirsthead
\caption{continued.}\\
\hline
\hline
Wave. (\AA) & ID  & flux \AA & comment \\
\hline 
\endhead
\hline
\endfoot
3966.78 & \ion{He}{i} (5) 3964.44  & 3.97   & E  \\
3969.50 & [\ion{Ne}{iii}] Neb (1F) 3967.51 & 62.47 &E \\
3972.12 & H$\epsilon$ 3970.074    & 47.01 &E \\
4010.62 & [\ion{Fe}{iii}] (4F) 4008.30          & 0.61    &E \\
4011.51 & \ion{He}{i} (55) 4009.27             & 0.72     &E \\
4028.29 & \ion{He}{i} (3) 4026.60 + \ion{He}{i} (18) 4026.36 & 6.81 & E \\
4061.58 & [\ion{F}{iv}] (1F) 4059.30            & 0.15    & E \\
4070.84 & [\ion{S}{ii}] Neb (1F) 4068.62     & 9.94   & E \\
4073.74 & \ion{O}{ii} (49) 4071.20               & 1.87    & E \\
4078.61 & [\ion{S}{ii}] Neb (1F) 4076.22     & 3.14   & E \\
4099.54 & \ion{O}{ii} (20,48) 4097.26          & 1.93   & E \\
4102.37 & \ion{He}{i} (3) 4100.04  + H$\delta$  sat    &  1.82  & E \\
4103.98 & H$\delta$ 4101.74            & 58.94 & E \\
4105.25 & H$\delta$ sat               & 1.68   & E \\
4116.73 & [\ion{Fe}{ii}] (23F) 4114.48          & 0.50   & E \\
4123.11 & \ion{He}{i} (16) 4120.89 + \ion{He}{i} (16) 4120.81 + \ion{S}{ii}  (2) 4121.0  & 1.15 & E \\
4130.77 & ?\ion{Fe}{ii} (27) 4128.73             & 0.39 & E \\
4132.42 & \ion{O}{ii} (19) 4129.34 + \ion{Si}{ii} (3) 4128.05  &          & A \\
4145.95 & \ion{He}{i} (53) 4143.76 + \ion{O}{ii}  (106) 4143.52+4143.77 & 2.78  & E \\
4183.42 & \ion{N}{ii} (49) 4181.17+4182.42   & 1.22 & E \\
4191.89 & \ion{O}{ii} (36) 4189.79               & 0.34 & E \\
4202.12 & \ion{He}{i} (3) 4199.83               & 3.09 & E \\
4229.79 & [\ion{Fe}{v}] (2F) 4227.44           & 9.63 & E \\
4246.40 & [\ion{Fe}{ii}] (21F) 4243.98           & 1.34 & E \\ 
4247.27 &  [\ion{Fe}{ii}] (21F) 4244.81        & 0.28 & E \\
4279.09 & \ion{O}{ii} (54, 67) 4276.71         & 1.00 & E \\
4289.73 & [\ion{Fe}{ii}] (7F) 4287.40          & 1.74 & E \\
4308.47 & [\ion{Fe}{ii}] (21F) 4305.90                  & 0.26 & E \\
4322.08 & [\ion{Fe}{ii}] (21F) 4319.62          & 0.32 & E \\
4328.72 & [\ion{Ni}{ii}] (3F) 4326.85              & 0.34 & E \\
4341.07 & \ion{He}{ii} (3) 4338.67+ H$\gamma$ sat      & 4.54 & E \\
4342.81 & H$\gamma$ 4340.47       & 148.64 & E \\
4344.37 & H$\gamma$ sat &    &  E \\
4349.39& \ion{O}{ii} (16) 4347.42  or [\ion{Fe}{ii}] (36F) 4347.35  & 0.33 & E \\
4355.18 & [\ion{Fe}{ii}] (21F) 4352.78          & 0.58 & E \\
4361.66 & [\ion{Fe}{ii}] (7F) 4359.34             & 1.66 & E \\
4363.86 & [\ion{O}{iii}] Neb sat &               &    E  \\
4365.55 & [\ion{O}{iii}] Neb (2F) 4363.21    & 76.57 &E \\
4367.04 & [\ion{O}{iii}] Neb sat &               &    E  \\
4390.34 & \ion{He}{i} (51) 4387.93                       & 2.21 & E \\
4411.74 & [\ion{Fe}{ii}] (22F) 4407.25+4410.75 & 0.36 & E \\
4416.15& [\ion{Fe}{ii}] (7F) 4413.78             & 0.88 & E \\
4418.69 & [\ion{Fe}{ii}] (6F) 4416.27             & 1.19 & E \\
4439.84 & \ion{He}{i} (50) 4437.55 + \ion{Mg}{ii} (19) 4436.48 & 0.68 & E \\
4454.54 & \ion{O}{ii} (5) 4452.38                   & 0.37 & E \\
4460.54 & [\ion{Fe}{ii}] (6F) 4457.95              & 0.45& E \\
4474.94 & \ion{He}{i} (14) 4471.47                & 19.84 & E \\
4477.25 & \ion{O}{iii} (37) 4474.95                 &         & E \\
4544.08 & \ion{He}{ii} (2) 4541.59 + \ion{Fe}{ii} (38) 4541.52   & 4.95 & E \\
6530.78 & [\ion{N}{ii}] (1F) 6527.40                 & 2.41 & E \\
6551.69 & [\ion{N}{ii}] Neb (1F) 6548.06         & 350.90 & E\\
6563.73 & \ion{He}{ii} (2) 6560.10    + H$\alpha$ sat       & 91.99 & E \\
6566.43 & H$\alpha$ 6562.82              & 2943.0 & E \\
6570.12 &  H$\alpha$ sat                     & 15.6 & E \\
6583.66 &  [\ion{N}{ii}] Neb sat                        & 4.82 & E \\
6587.11 & [\ion{N}{ii}] Neb (1F) 6583.37           & 1167.0 & E\\
6590.62 &  [\ion{N}{ii}] Neb sat                                  & 6.14 & E \\
6681.80 & \ion{He}{i} (46) 6678.15                      & 50.62 & E \\
6686.88 & \ion{He}{ii} (7) 6683.39                     & 3.78 & E \\
6704.58 & [\ion{Fe}{ii}] (43F) 6700.68 + [\ion{Ni}{ii} ] (8F) 6700.61  &  0.18         & E \\
6713.17 & [\ion{S}{ii}] Neb sat                             &    1.00 & E \\
6720.17 & [\ion{S}{ii}] Neb (2F) 6716.42             & 28.35 & E \\
6734.56 & [\ion{S}{ii}] Neb (2F) 6730.78              &    52.87 & E \\
6737.93 &  [\ion{S}{ii}] Neb sat                          & 1.85 & E \\
6895.00 & \ion{He}{ii} (7)   6890.88     &   5.37      &    E   \\
7009.46 & [\ion{Ar}{v}] (1F) 7005.70       &  28.43 & E \\
7055.02 & [\ion{Fe}{ii}] (17F) 7051.04 & & E \\
7065.41& \ion{He}{i}  sat    &    & E \\
7069.11 & \ion{He}{i} (10) 7065.28              &      141.50  & E \\
7073.16 & \ion{He}{i}  sat           &    0.79 & E \\
7139.84 & [\ion{Ar}{iii}] (1F) 7135.78     & 29.52 & E \\
7159.23 & [\ion{Fe}{ii}] (14F) 7155.14       &  4.01 & E \\
\end{longtable}
}
%

\twocolumn

\addtocounter{figure}{+3}

\begin{figure}[]
\centering
\resizebox{\hsize}{!}{\includegraphics[angle=-90]{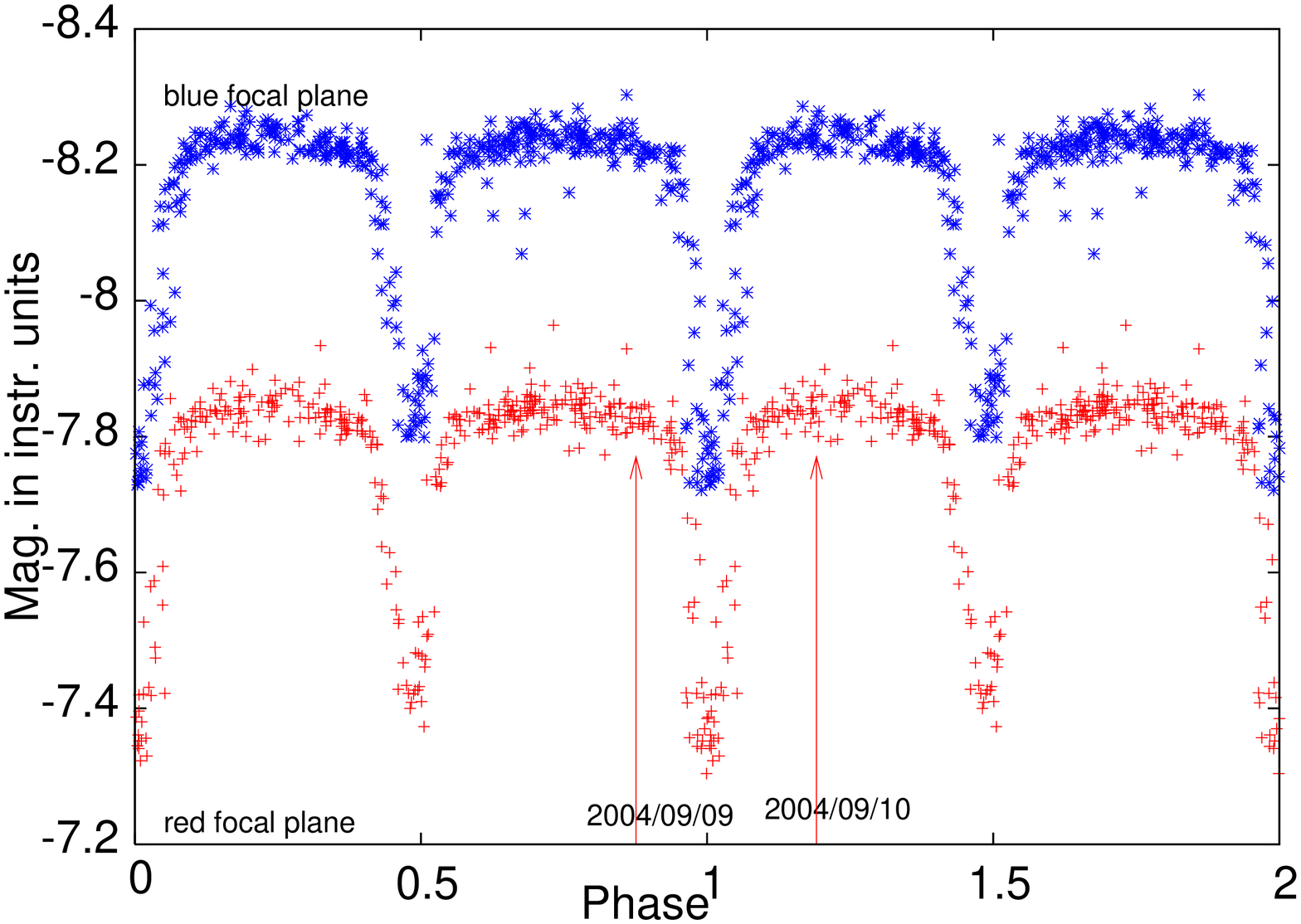}}
 \resizebox{\hsize}{!}{\includegraphics[]{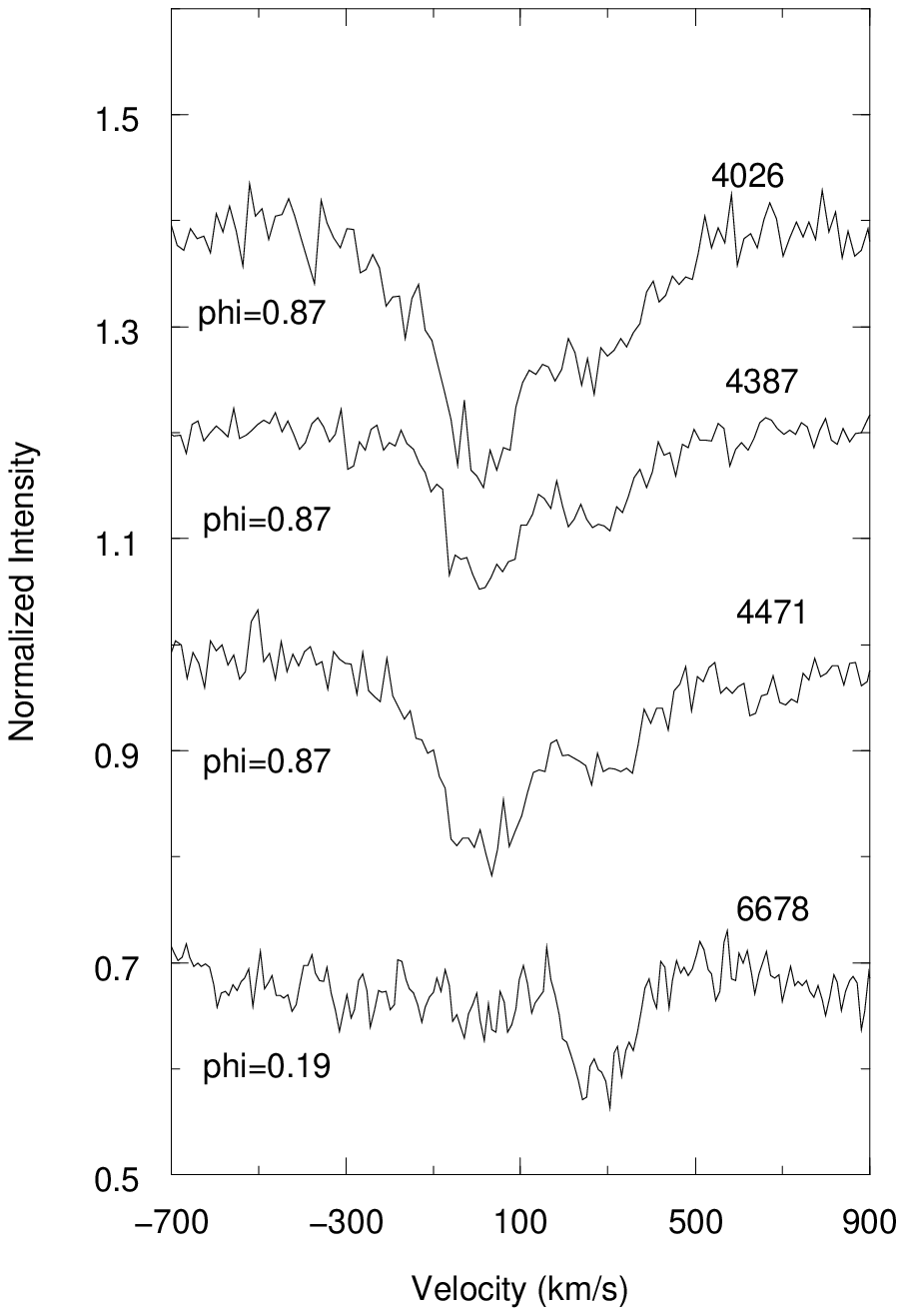}}
\caption{ SMC5\_000977. Top: light curve folded in phase with P=3.128d. Bottom: Helium lines  taken at
 $\phi$= 0.87 for the blue ones and at $\phi$= 0.19 for the red one.}  
\label{bin977}
\end{figure}

\begin{figure}[]
\centering
\resizebox{\hsize}{!}{\includegraphics[angle=-90]{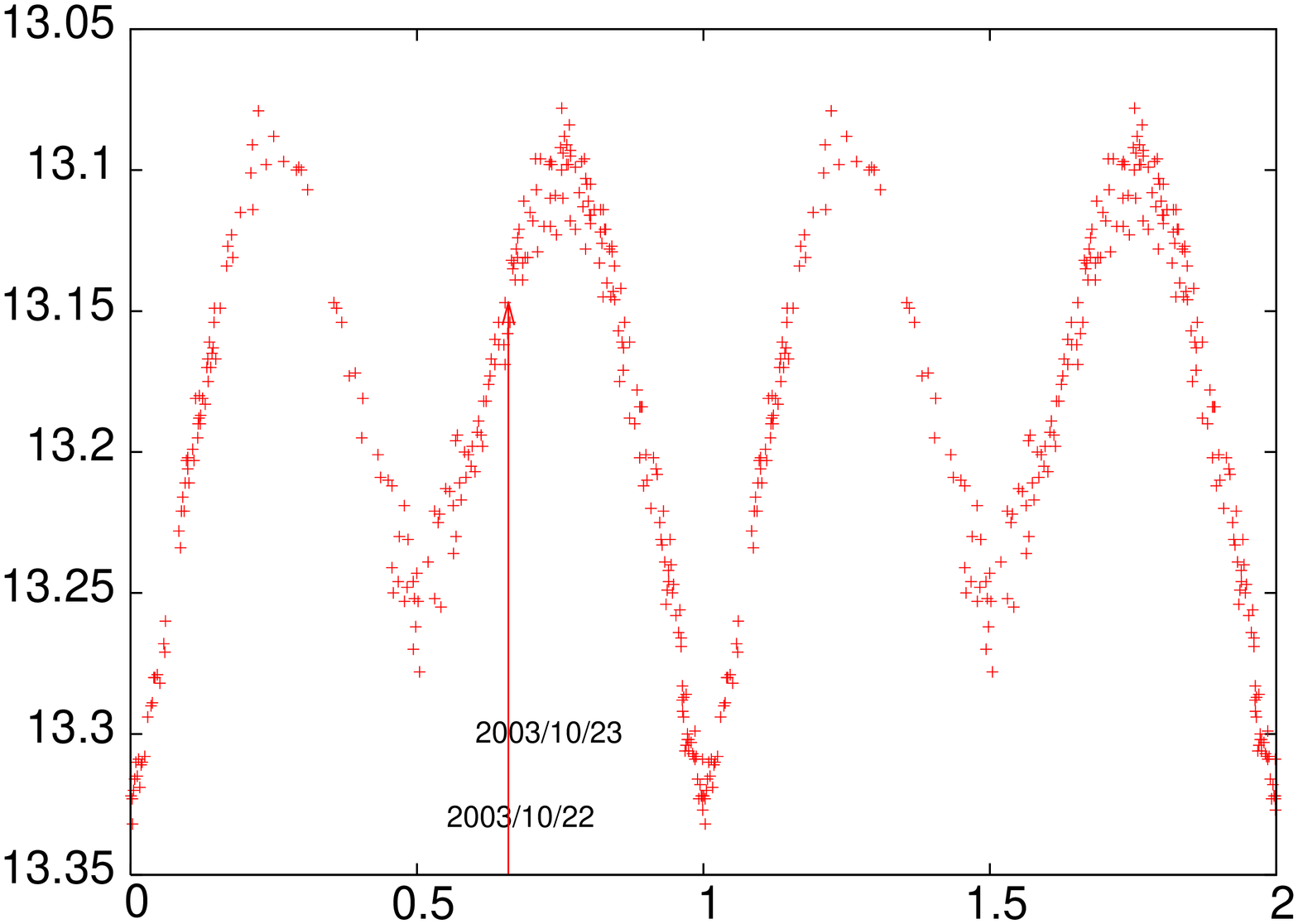}}
\resizebox{\hsize}{!}{\includegraphics[angle=-90]{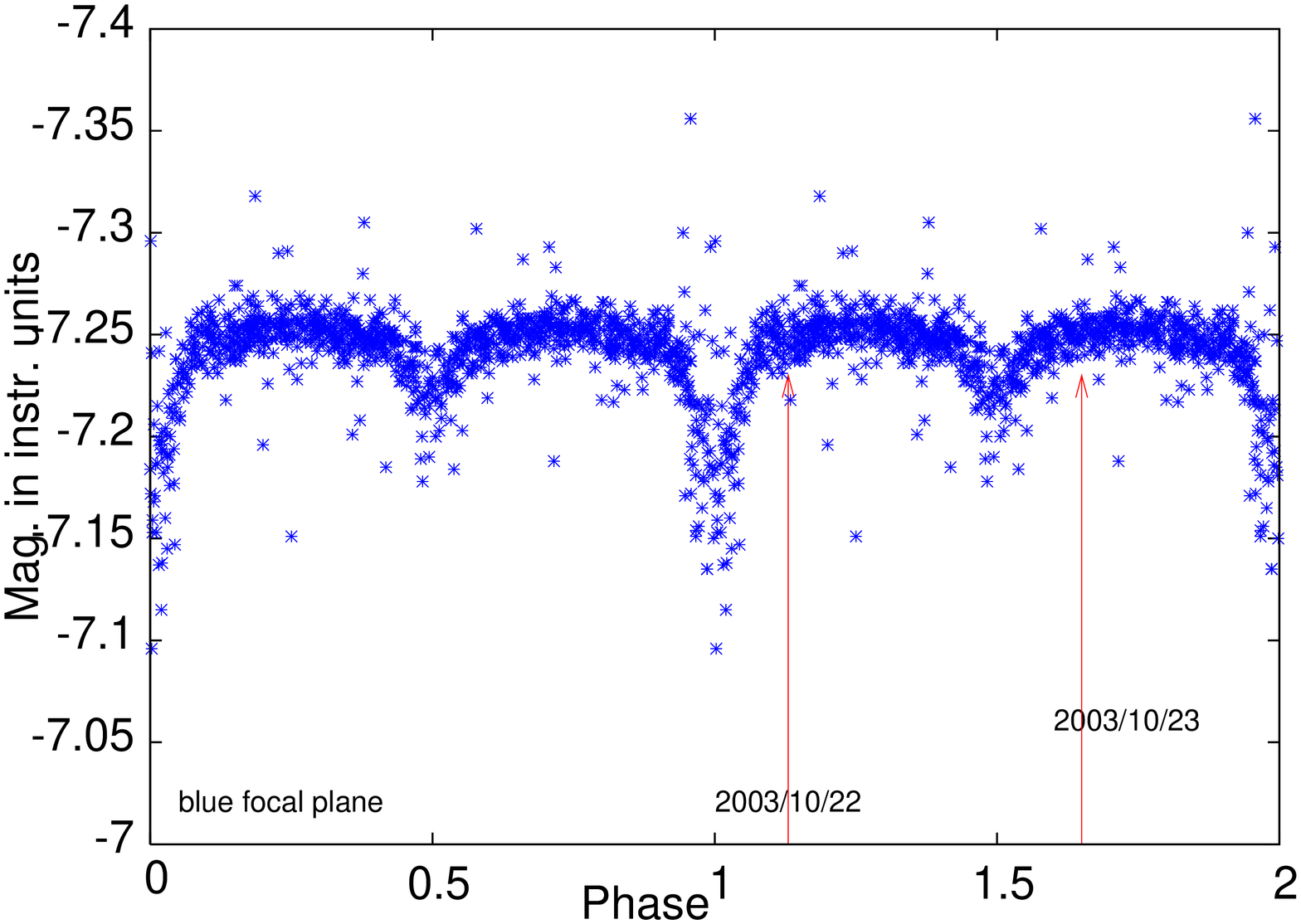}}
\resizebox{\hsize}{!}{\includegraphics[angle=-90]{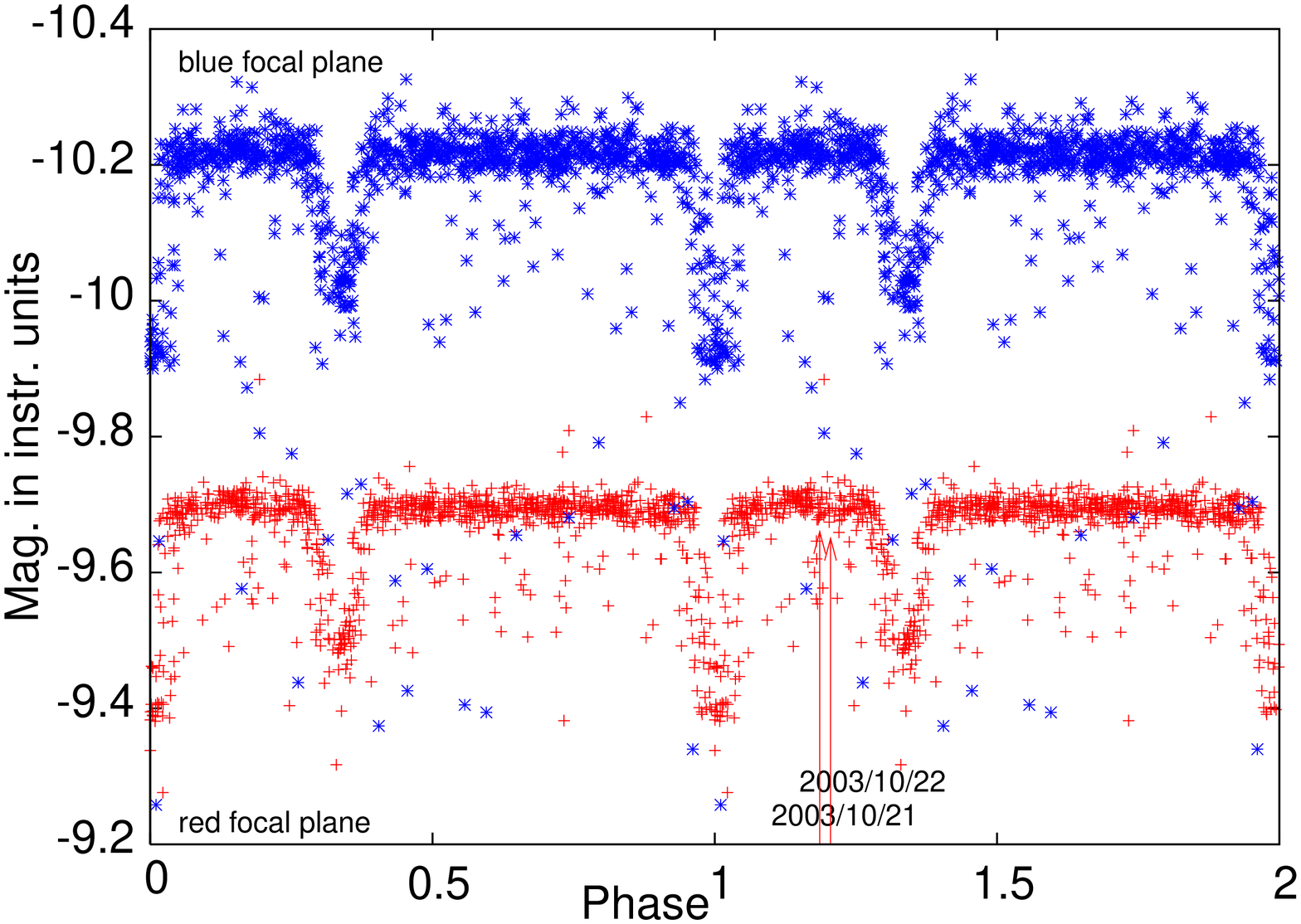}}
\caption{ Top: SMC5\_002807 folded in phase with
P=454.959d (OGLE data, Imag).
Middle: SMC5\_003789 Be star folded in phase with P=2.087d (MACHO data).
Bottom: SMC5\_0004477 folded in phase with P=4.480d (MACHO data).}  
\label{bin4477-2807-3789}
\end{figure}

\begin{figure}[]
\centering
\resizebox{\hsize}{!}{\includegraphics[angle=-90]{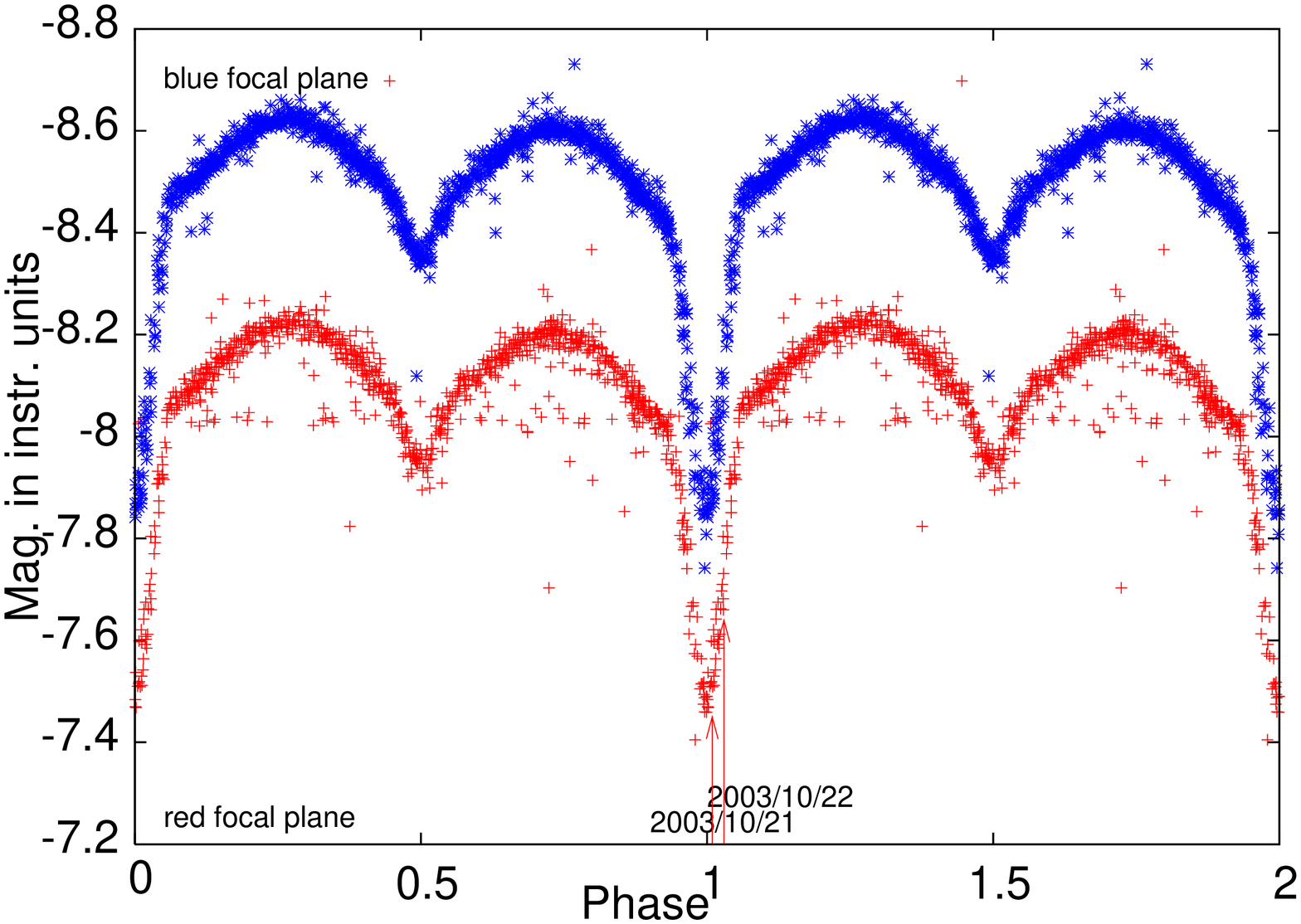}}
\resizebox{\hsize}{!}{\includegraphics[angle=-90]{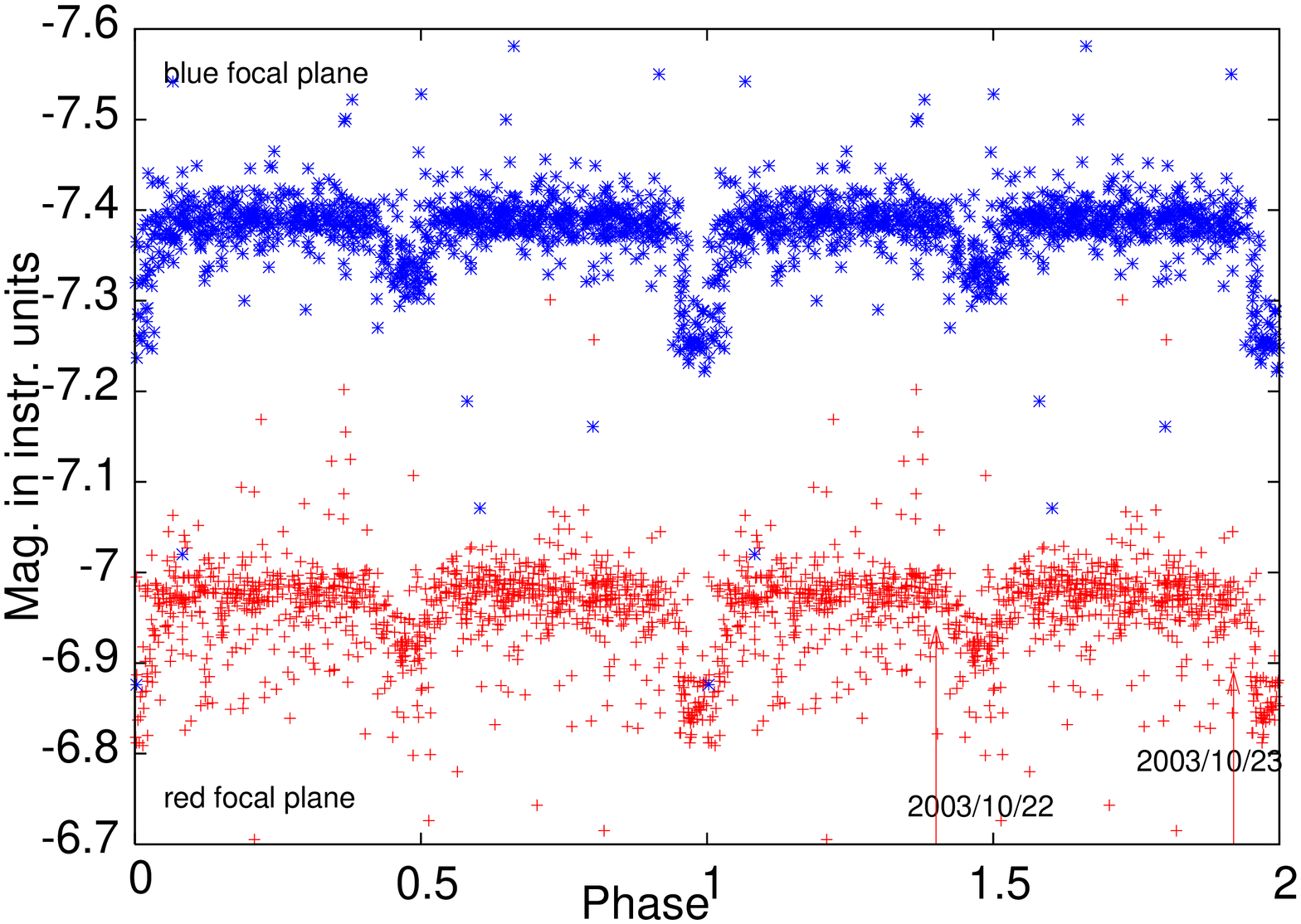}}
\caption{Top: SMC5\_004534 folded in phase with P=4.051d.
Bottom: SMC5\_013723 folded in phase with P=2.059d.}  
\label{bin4534-13723}
\end{figure}

\begin{figure}[!ht]
\centering
\resizebox{\hsize}{!}{\includegraphics[angle=-90]{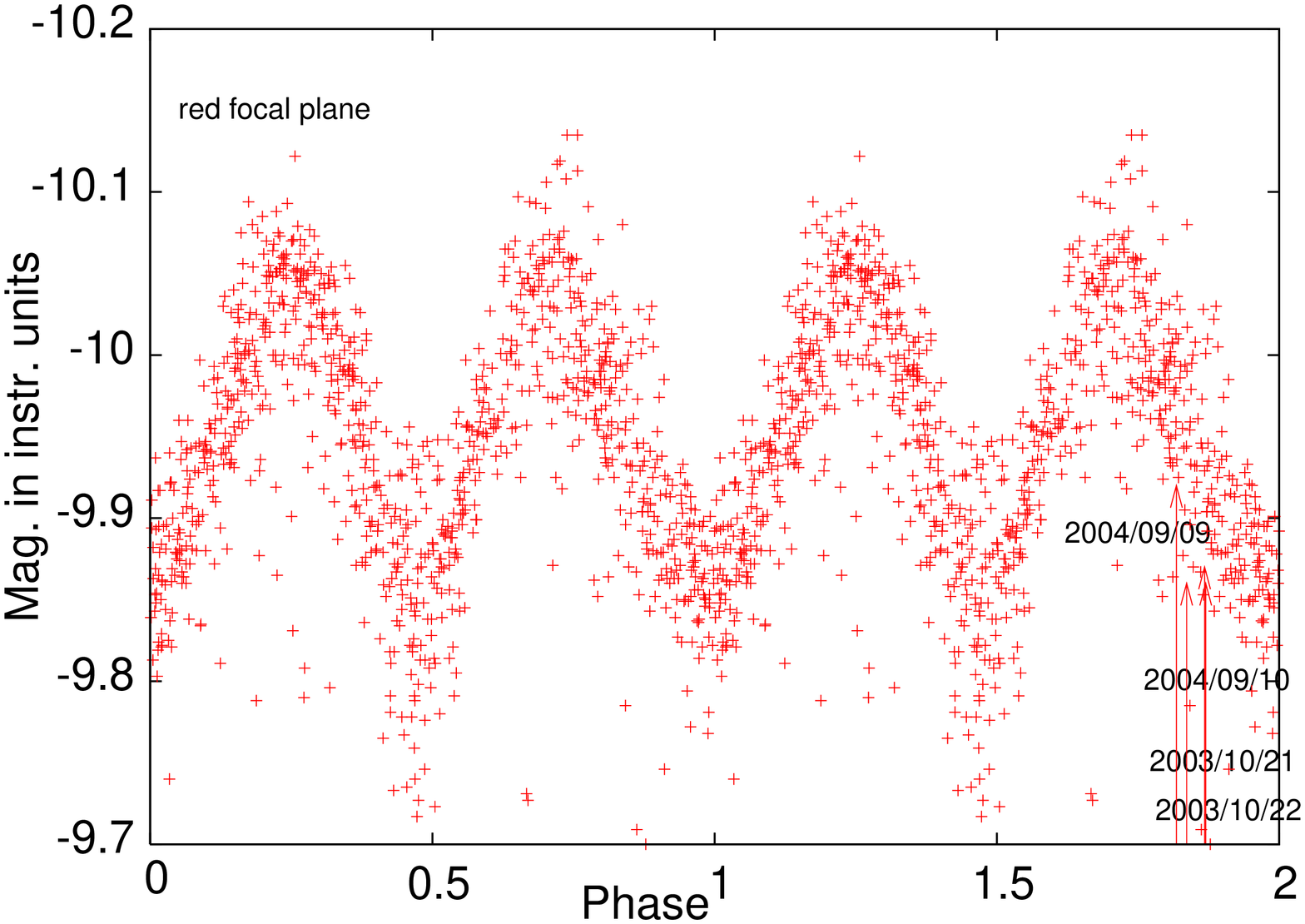}}
 \resizebox{\hsize}{!}{\includegraphics[]{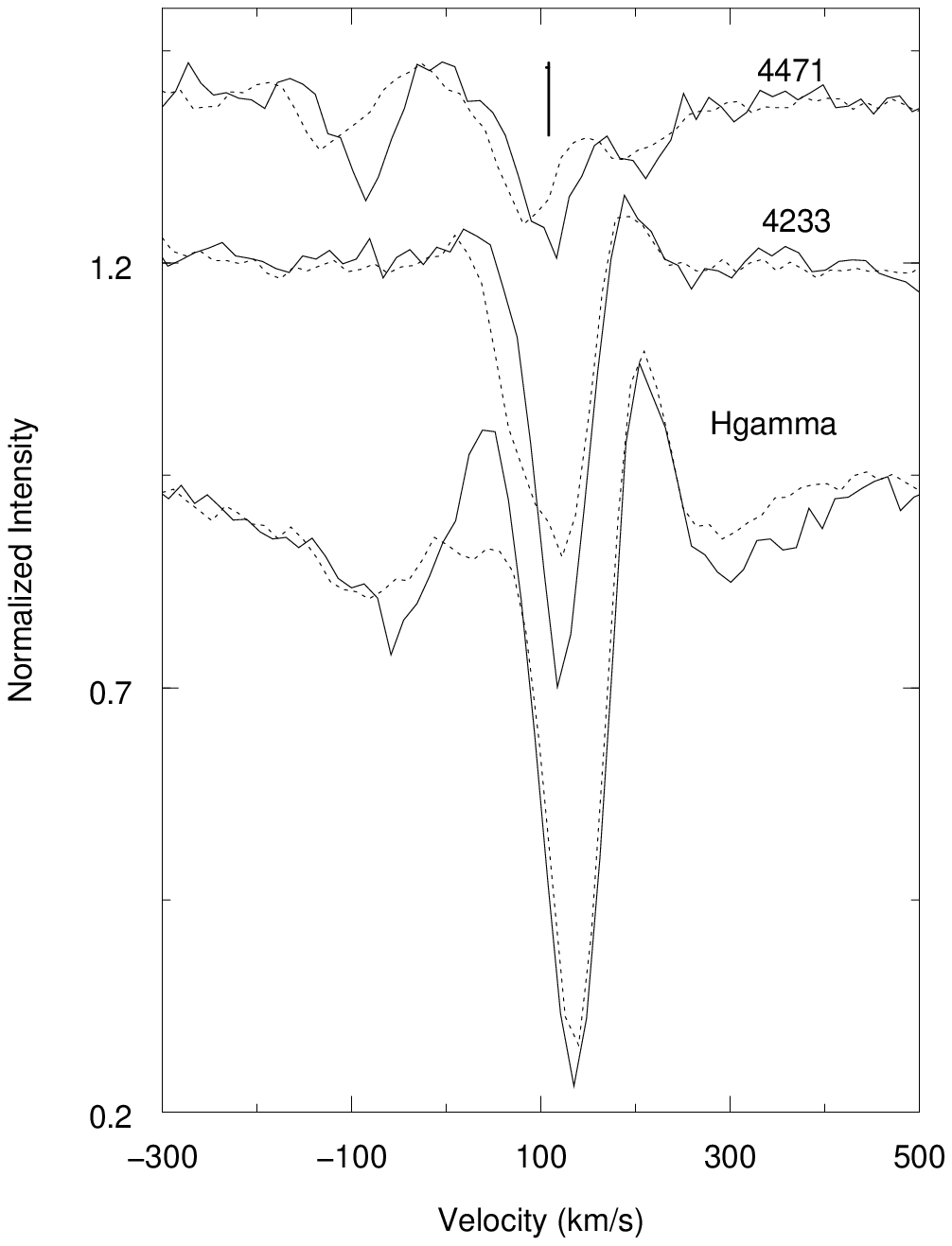}}
\caption{ SMC5\_016461. Top: light curve folded in phase with P=54.337d.
Bottom: Lines of \ion{He}{i} 4471, \ion{Fe}{ii} 4233 and H$\gamma$ for the spectra obtained in 2003 (full line) 
at $\phi$=0.89 and in 2004 (dotted line) at $\phi$=0.82 just before a primary eclipse} 
\label{bin16461}
\end{figure}

\begin{figure}[]
\centering
\centering
\resizebox{\hsize}{!}{\includegraphics[angle=-90]{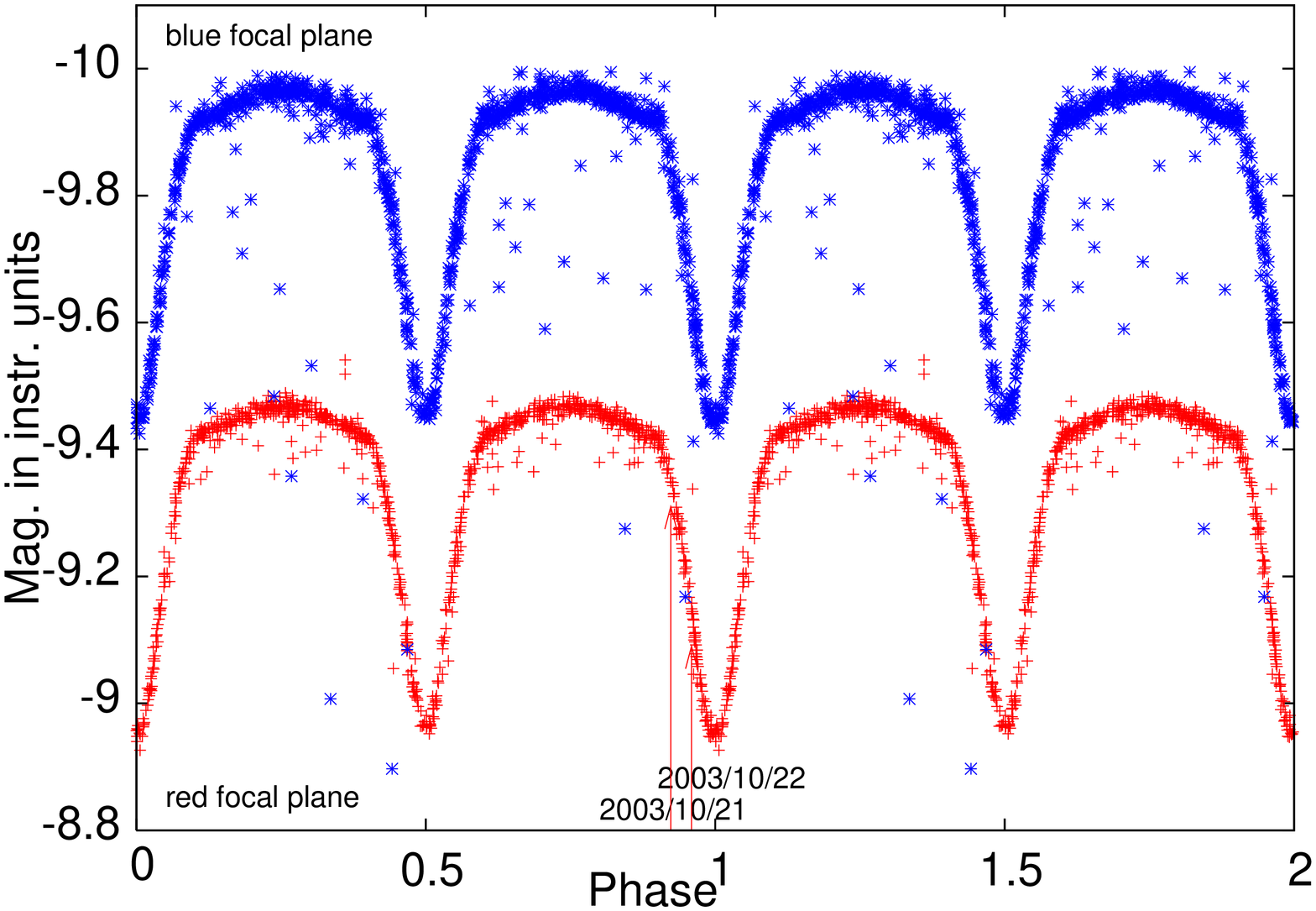}}
\resizebox{\hsize}{!}{\includegraphics[angle=-90]{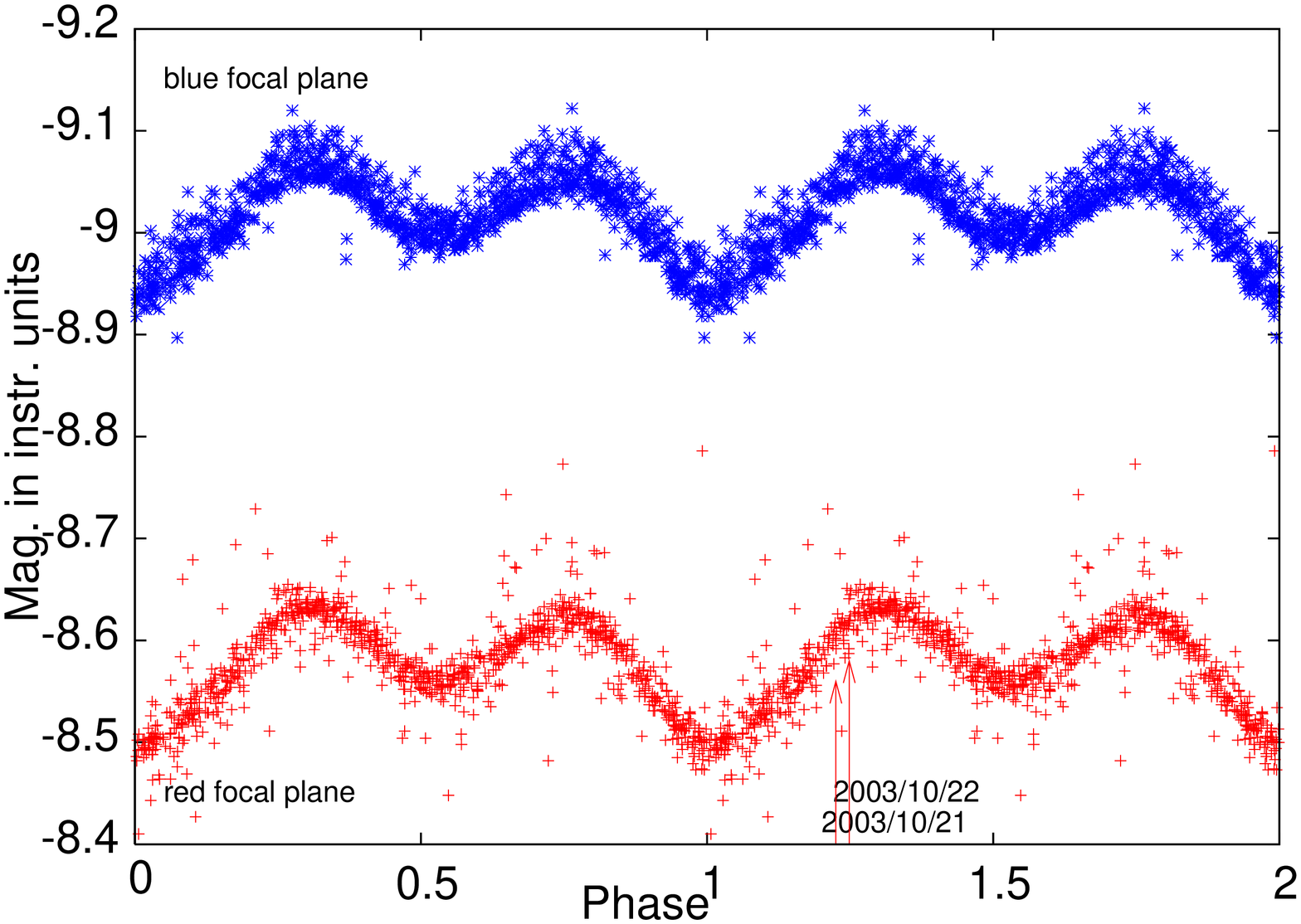}}
\caption{
 Top: SMC5\_020391 folded in phase with P=2.320d.
Bottom: SMC5\_023571 folded in phase with P=3.534d.
}  
\label{bin20391-23571}
\end{figure}

\begin{figure}[!h]
\centering
\resizebox{\hsize}{!}{\includegraphics[angle=-90]{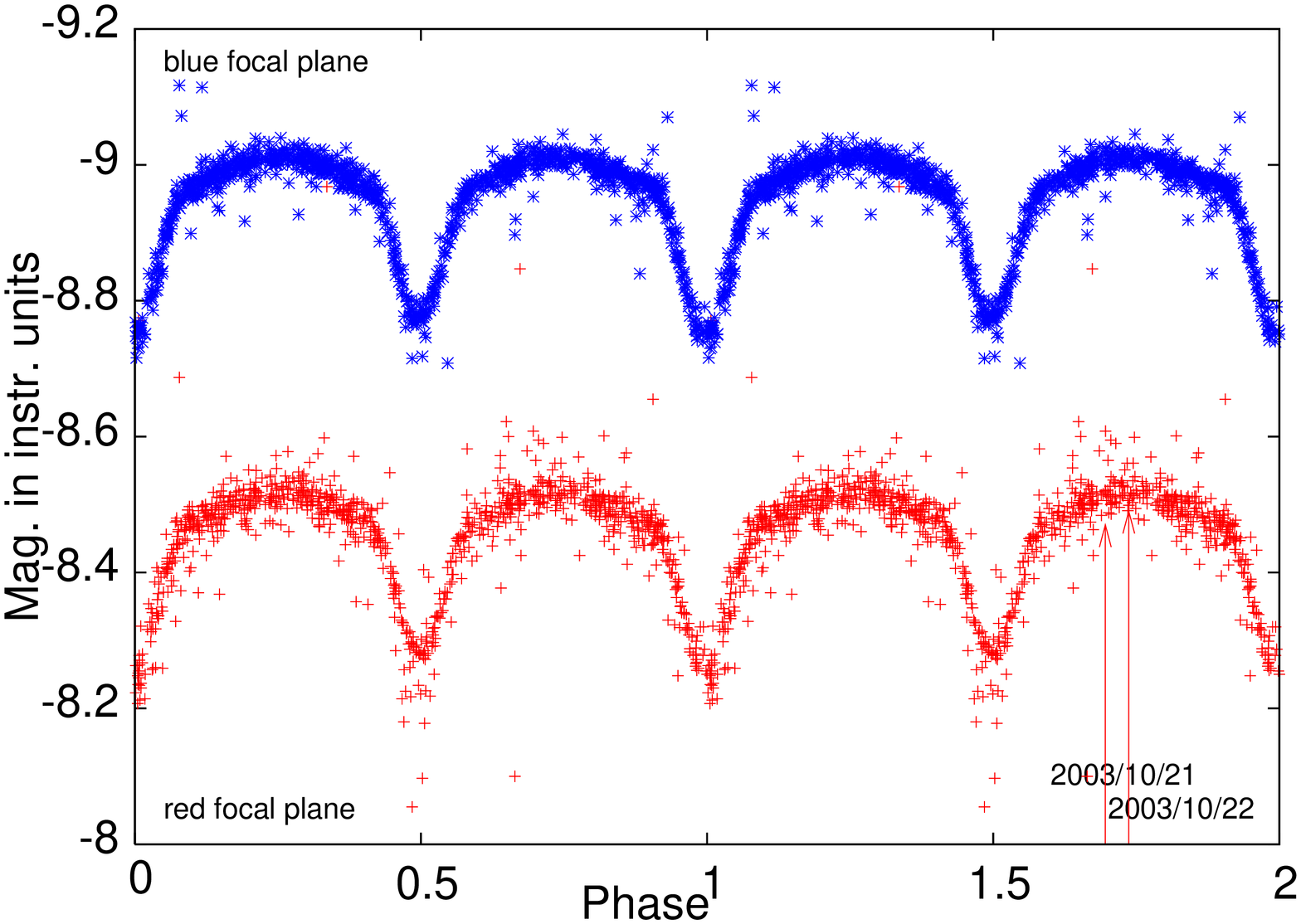}}
 \resizebox{\hsize}{!}{\includegraphics[]{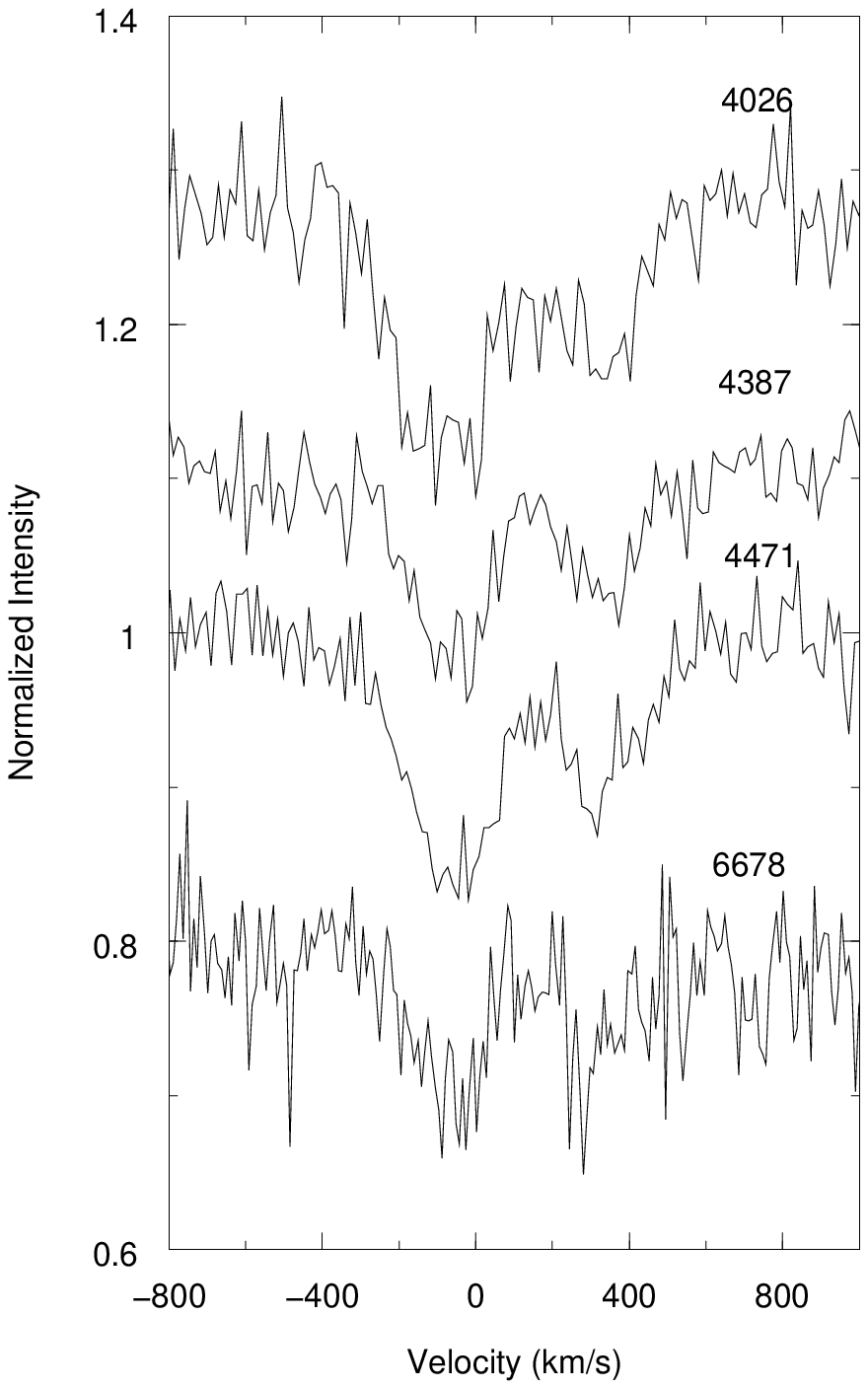}}
\caption{ SMC5\_023641. Top: light curve folded in phase with P=2.010d.
Bottom: Helium lines  taken at $\phi$= 0.70 for 
the blue ones and at $\phi$= 0.74 for the red one.}  
\label{bin23641}
\end{figure}

\begin{figure}[!h]
\centering
 \resizebox{\hsize}{!}{\includegraphics[]{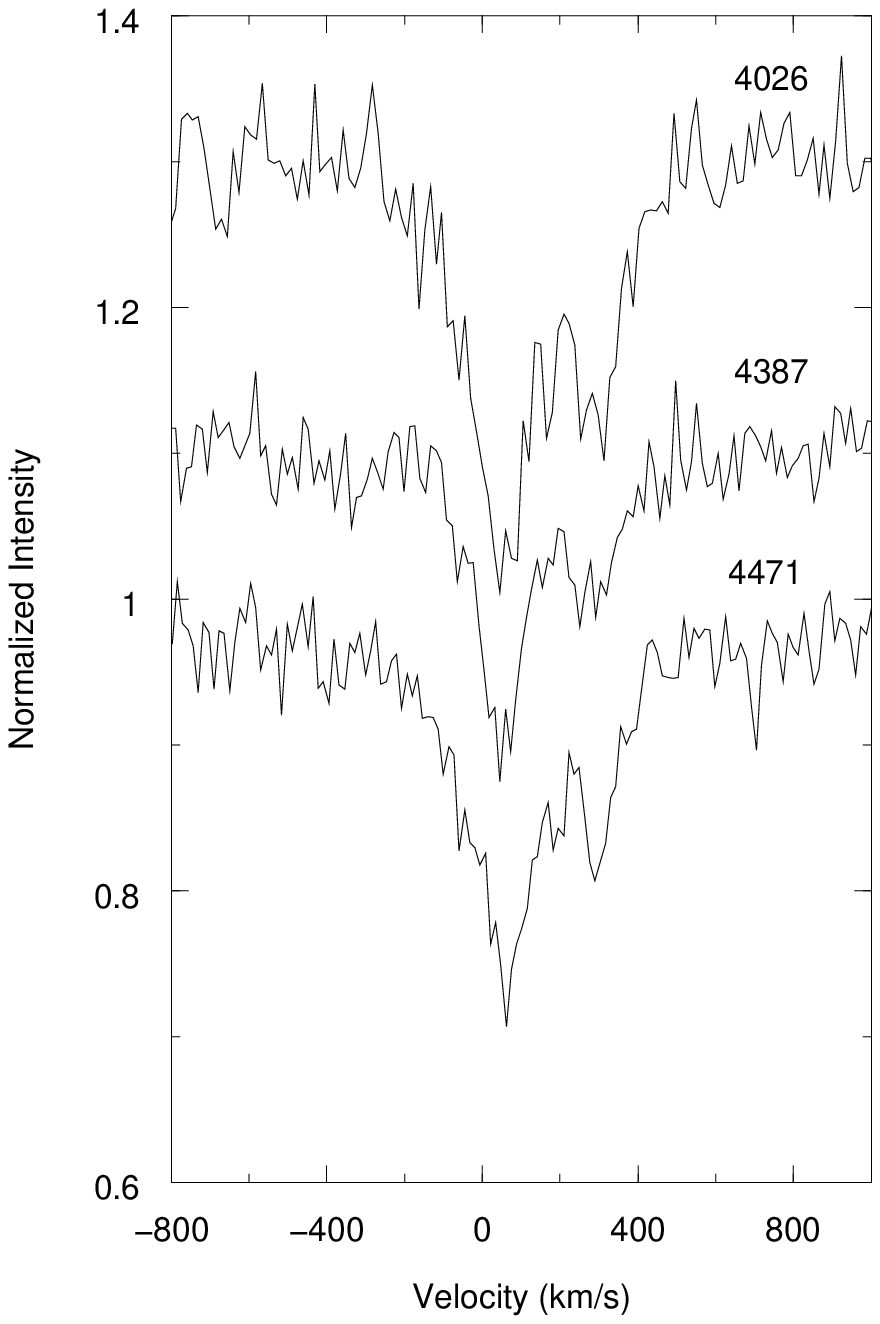}}
\caption{Blue helium lines of SMC5\_052663}  
\label{bin52663}
\end{figure}

\begin{figure}[]
\centering
\resizebox{\hsize}{!}{\includegraphics[angle=-90]{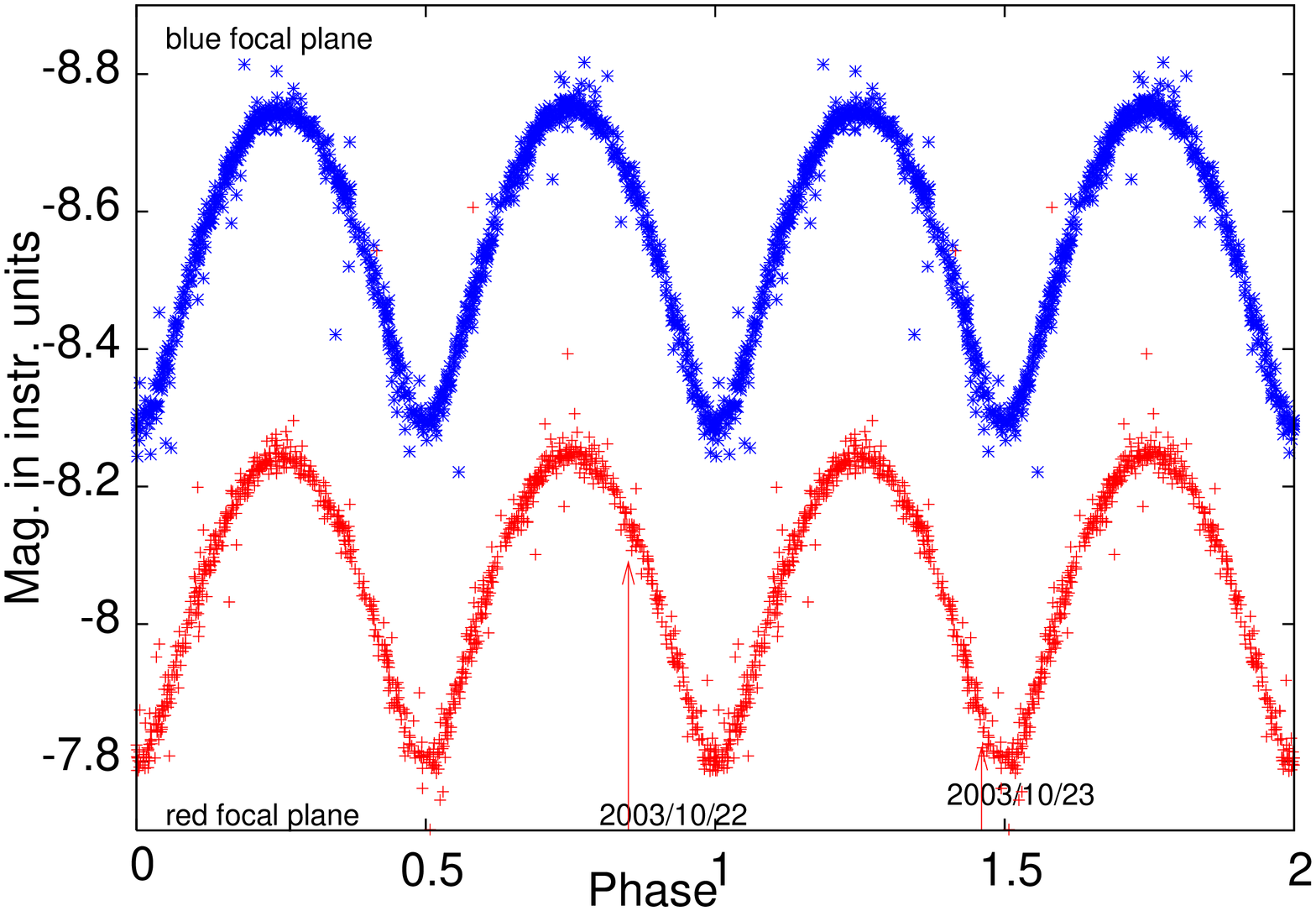}}
\resizebox{\hsize}{!}{\includegraphics[angle=-90]{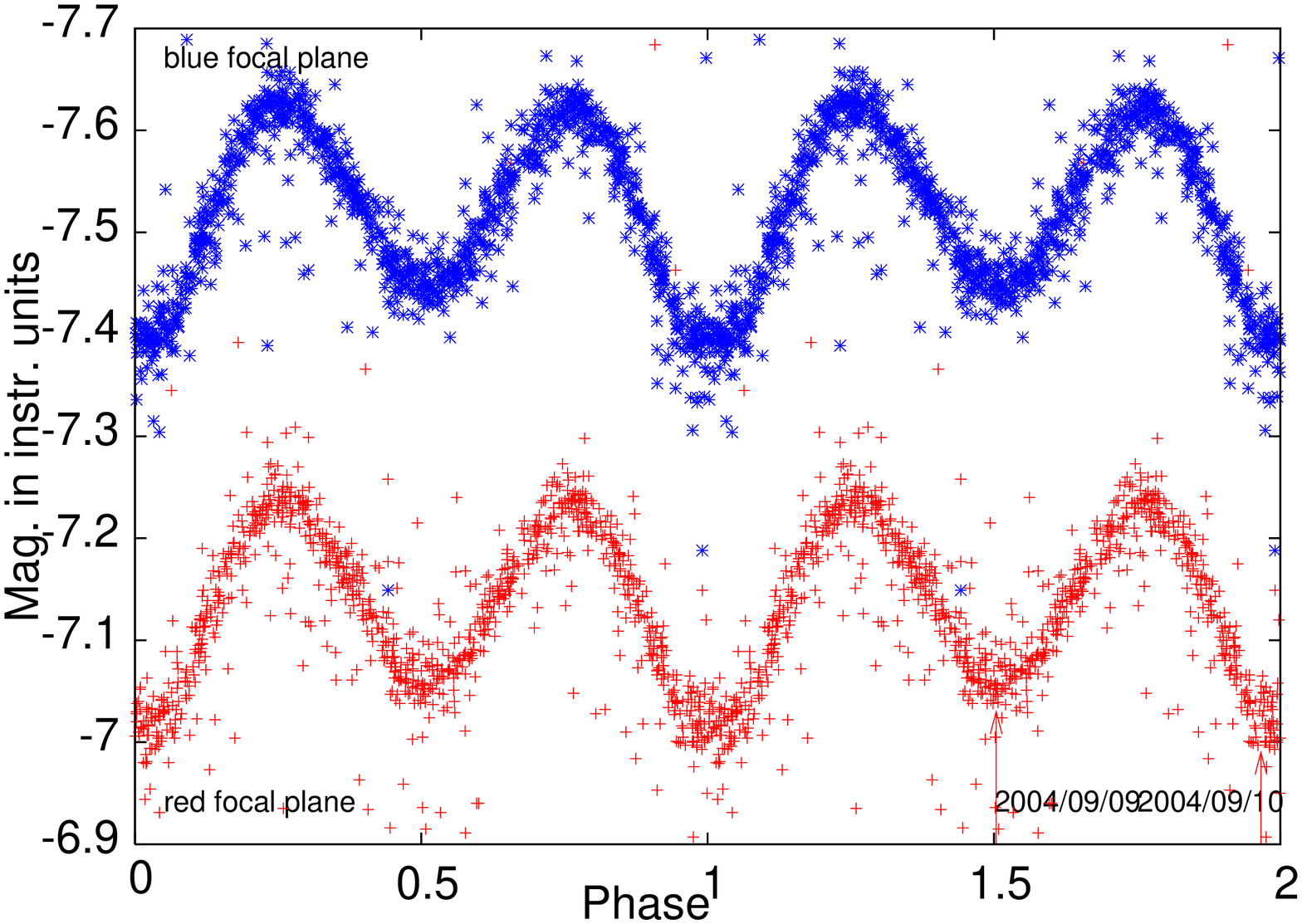}}
\resizebox{\hsize}{!}{\includegraphics[angle=-90]{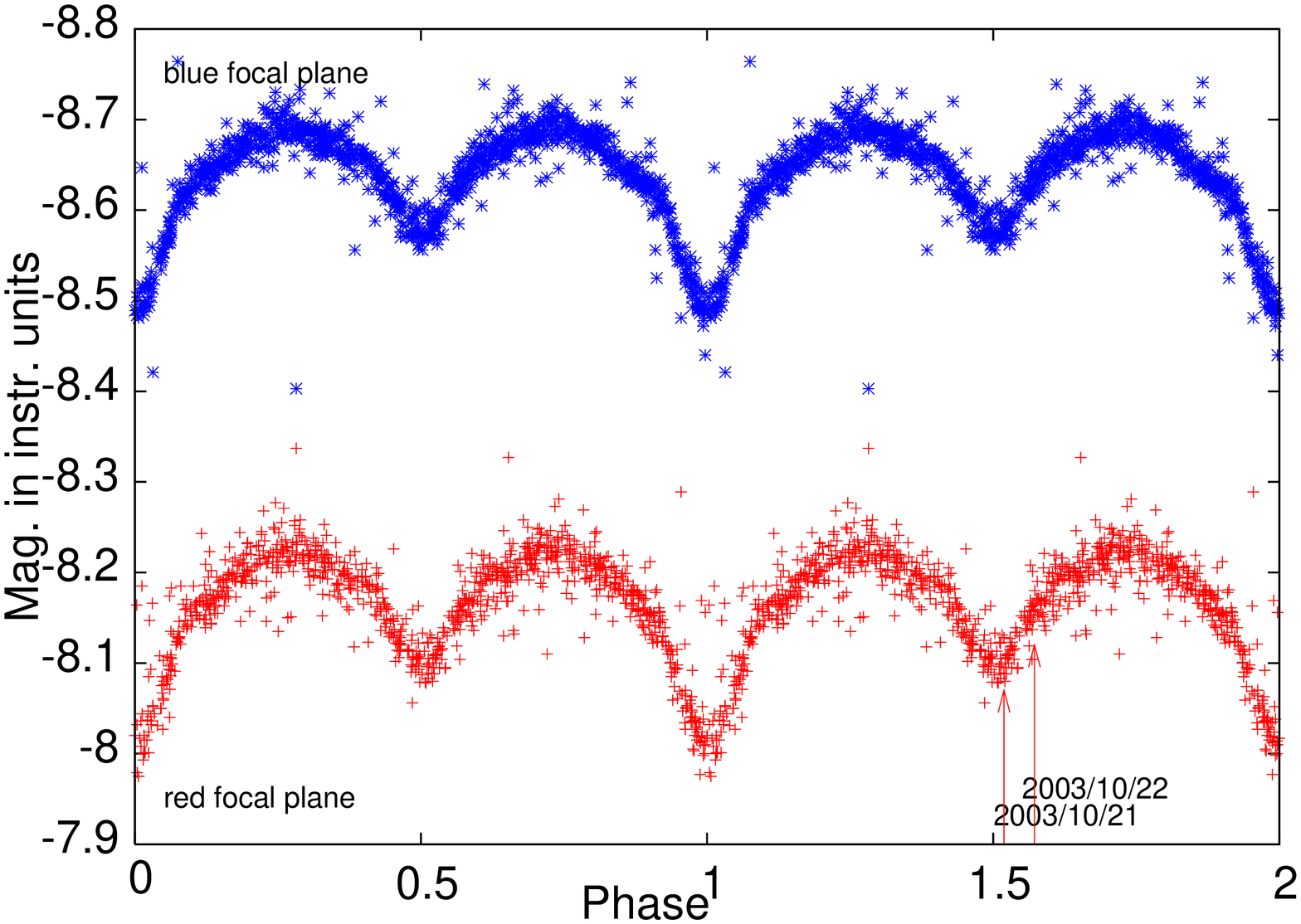}}
\caption{ Top: SMC5\_049816 folded in phase with P=0.664d. 
 Middle: SMC5\_074928 folded in phase with P=2.137d.
Bottom: SMC5\_084353 folded in phase with P=1.557d.}  
\label{bin49816-74928-84353}
\end{figure}

\newpage

\begin{figure}[]
\centering
\resizebox{\hsize}{!}{\includegraphics[width=6cm, angle=0]{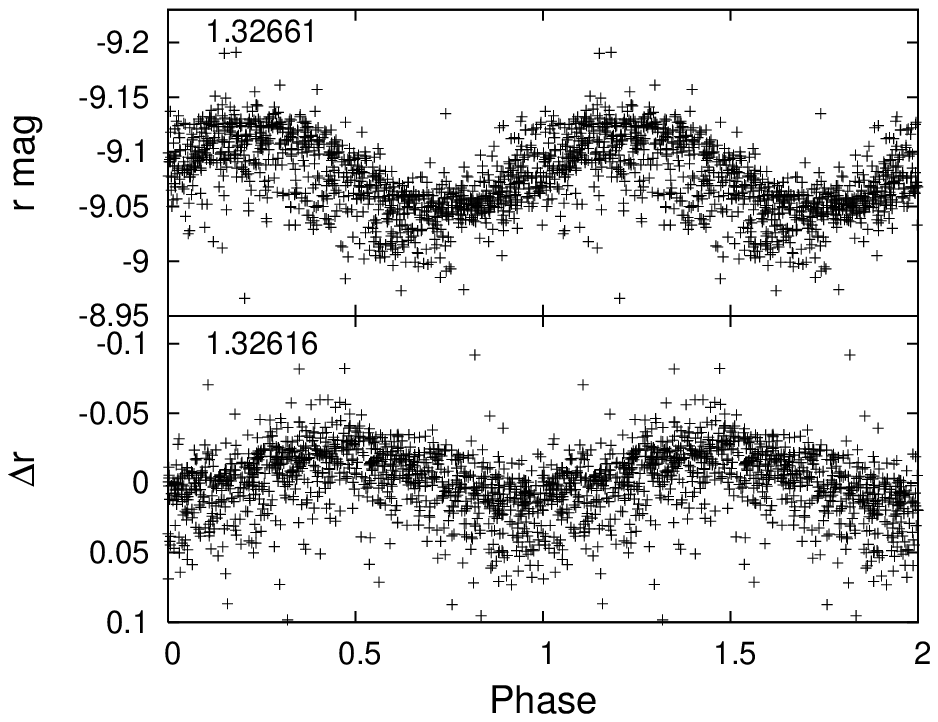}}
\resizebox{\hsize}{!}{\includegraphics[angle=0]{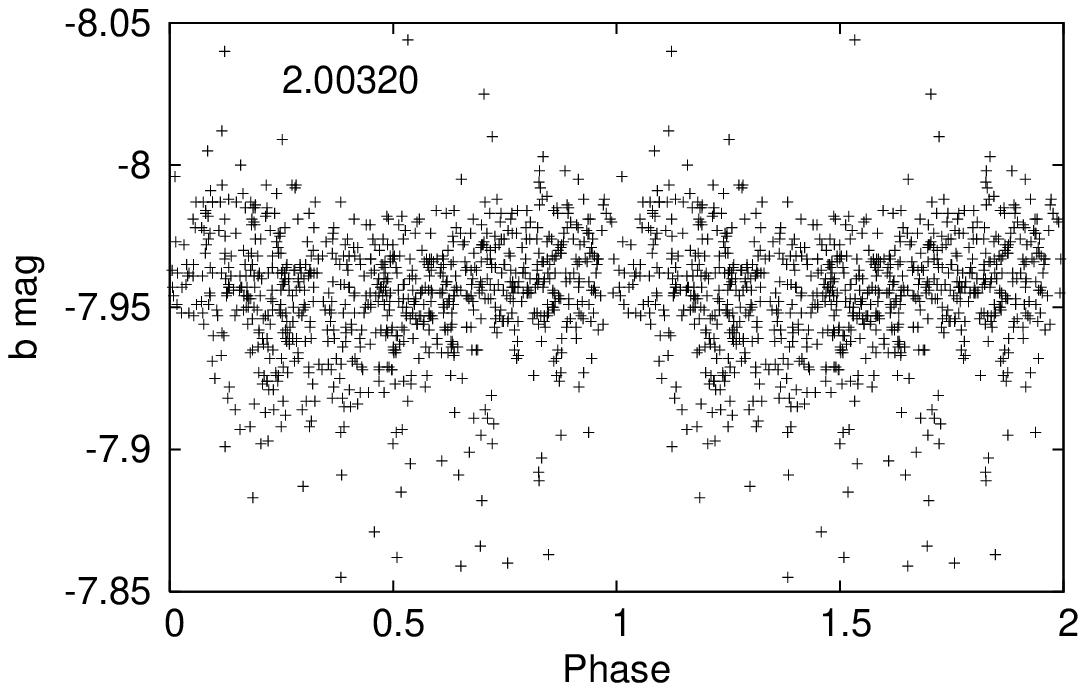}}
\caption{ Be SMC5\_2232 folded in phase: top with f1=1.32661$\mbox{c d}^{-1}$, middle with f2=1.32616$\mbox{c d}^{-1}$ 
after prewhitening for f1. Bottom: Be SMC5\_3296 folded in phase with f=2.00320$\mbox{c d}^{-1}$} 
\label{varbe1}
\end{figure}

\begin{figure}[]
\centering
\resizebox{\hsize}{!}{\includegraphics[angle=0]{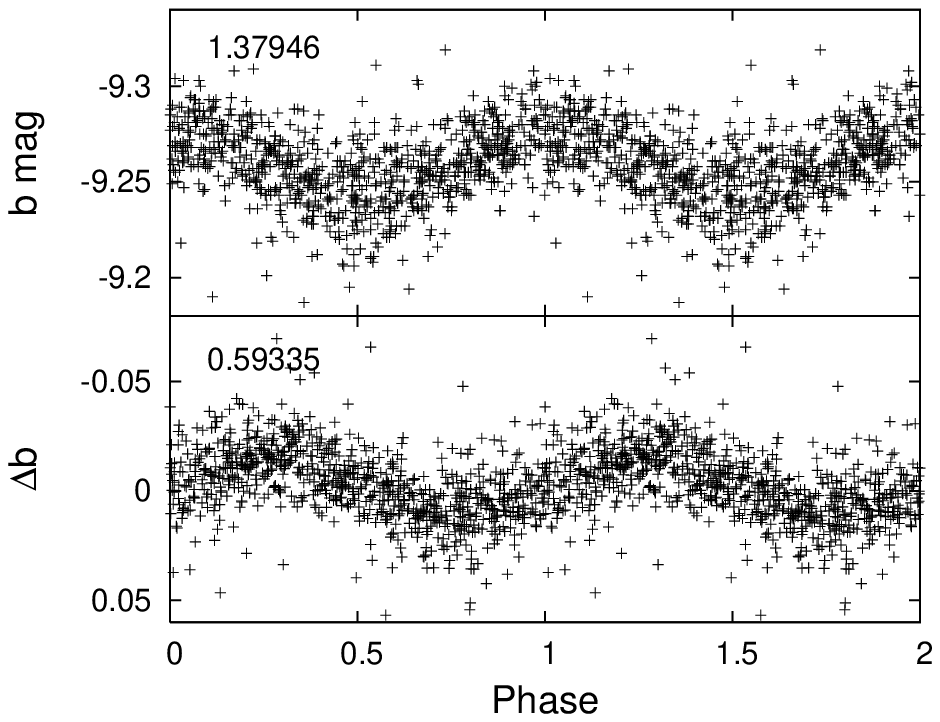}}
\resizebox{\hsize}{!}{\includegraphics[angle=0]{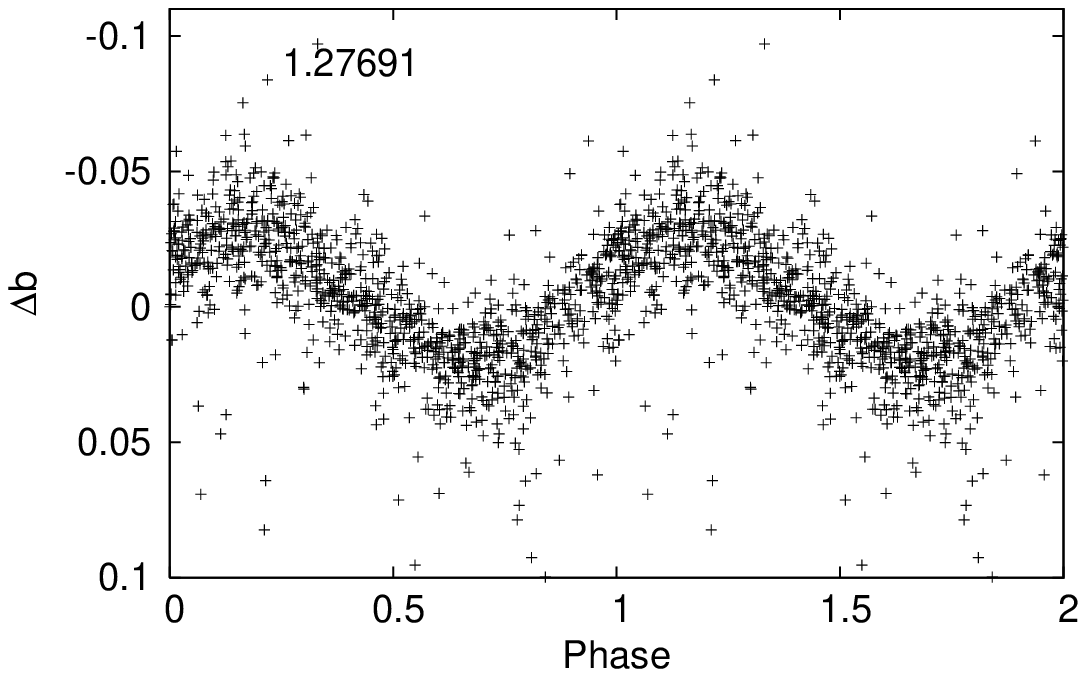}}
\caption{ Be SMC5\_13978 folded in phase: top with f1=1.37946$\mbox{c d}^{-1}$, middle with f2=0.59335 $\mbox{c d}^{-1}$ 
after prewhitening for f1. Bottom: Be SMC5\_14212 folded in phase with f=1.27691$\mbox{c d}^{-1}$.}  
\label{varbe2}
\end{figure}

\begin{figure}[]
\centering
\resizebox{\hsize}{!}{\includegraphics[angle=0]{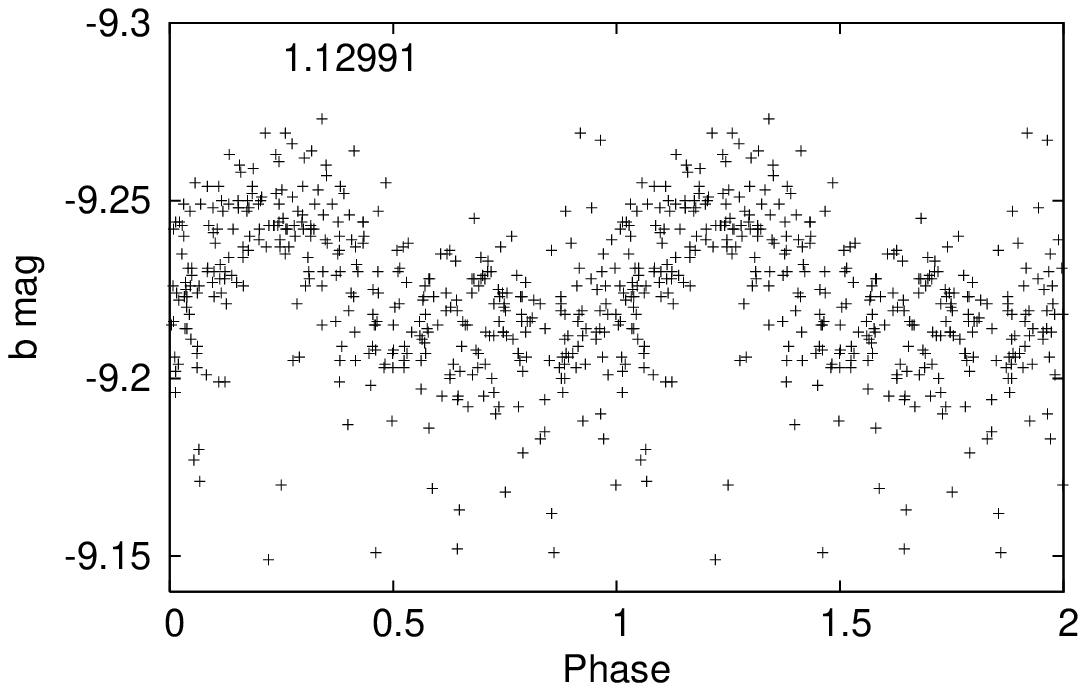}}
\resizebox{\hsize}{!}{\includegraphics[angle=0]{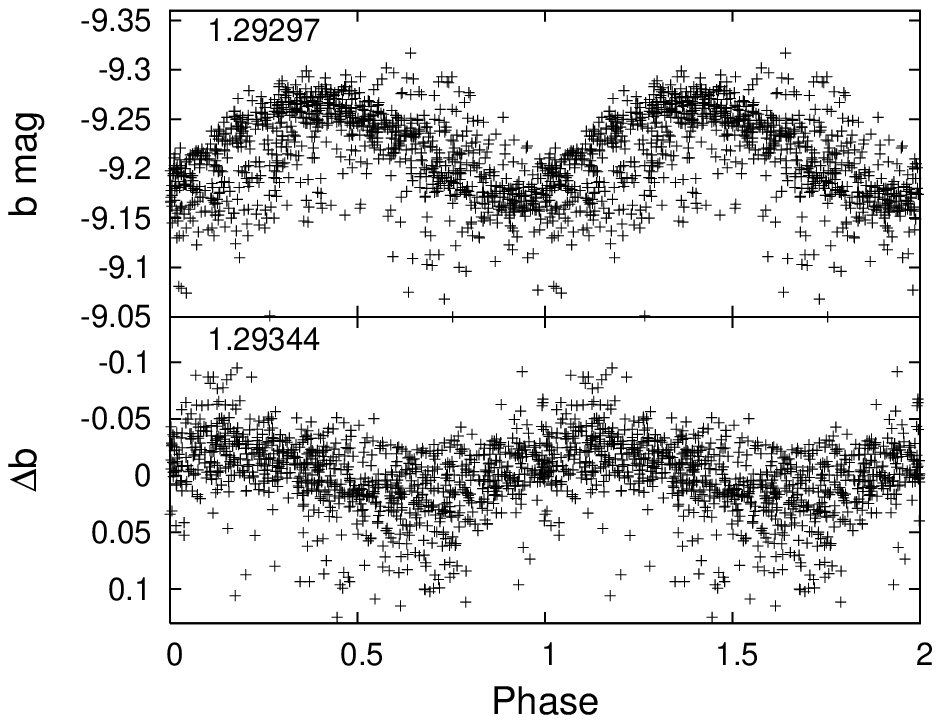}}
\caption{ Top: Be SMC5\_14727 folded in phase with f=1.12991 $\mbox{c d}^{-1}$. Be SMC5\_16523 folded in phase: 
middle with f1=1.29297$\mbox{c d}^{-1}$, bottom with f2=1.29344 $\mbox{c d}^{-1}$ after prewhitening for f1. }  
\label{varbe3}
\end{figure}

\begin{figure}[]
\centering
\resizebox{\hsize}{!}{\includegraphics[angle=0]{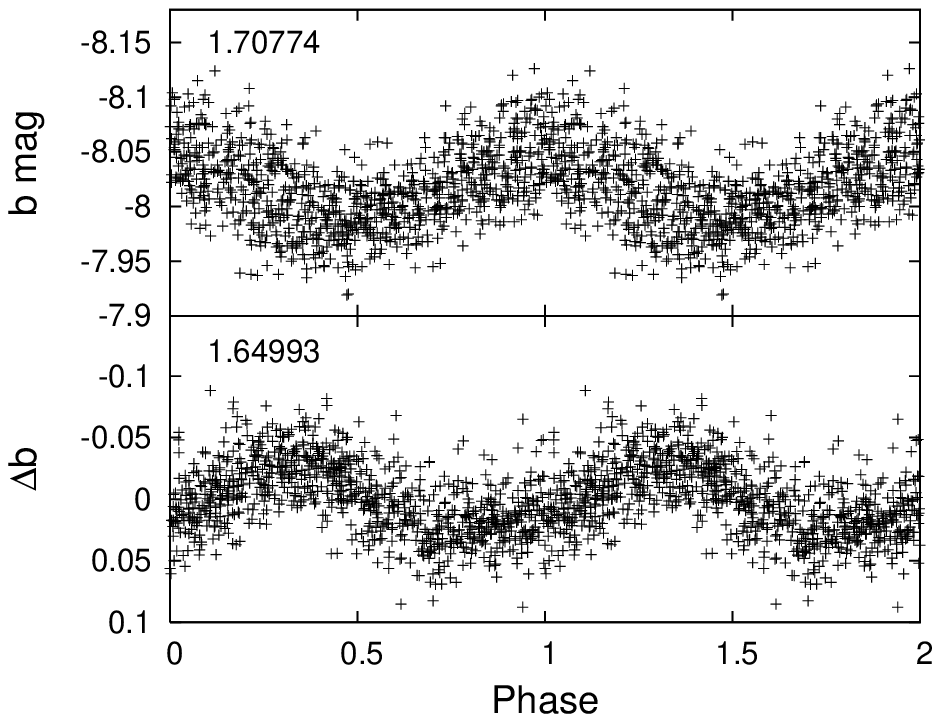}}
\resizebox{\hsize}{!}{\includegraphics[angle=0]{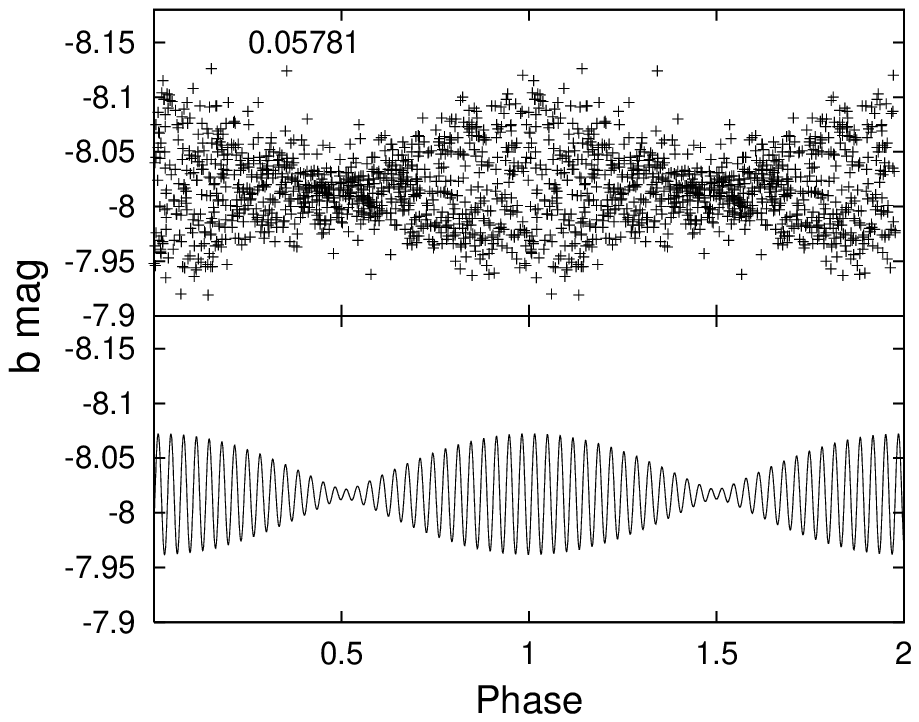}}
\caption{ Be SMC5\_16544 folded in phase: top with f1=1.70774 $\mbox{c d}^{-1}$, middle up with f2=1.64993 $\mbox{c d}^{-1}$ 
after prewhitening for f1. Be SMC5\_16544 beating diagram, middle down:  light-curve folded with the beating frequency 0.05781$\mbox{c d}^{-1}$; 
bottom: sum of the 2 sinusoidal functions.}  
\label{varbe4}
\end{figure}

\begin{figure}[!h]
\centering
\resizebox{\hsize}{!}{\includegraphics[width= 5cm,angle=0]{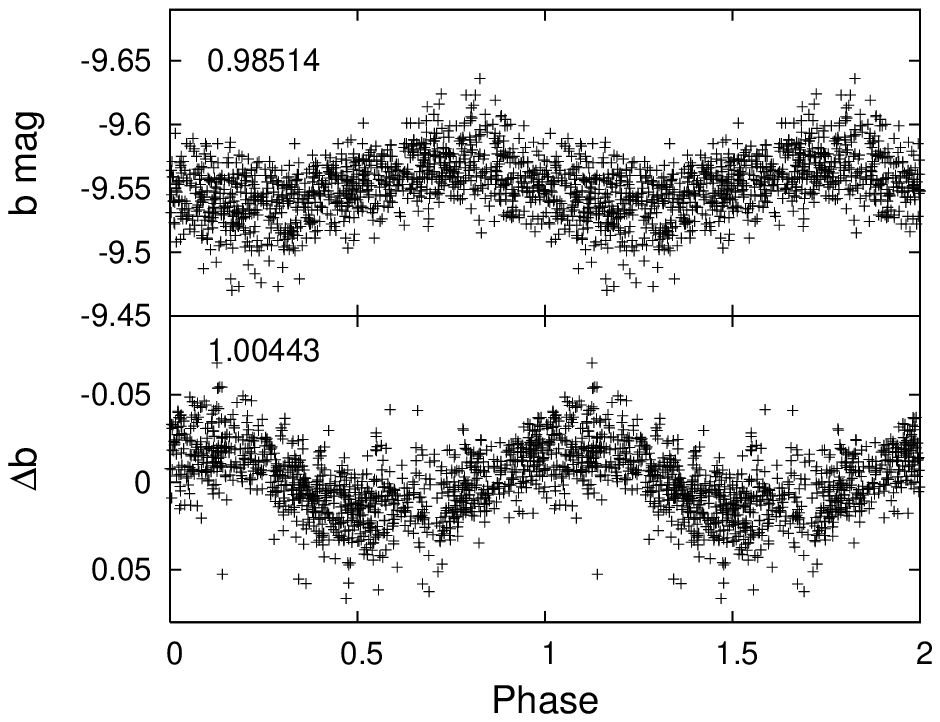}}
\caption{ Be SMC5\_21152 folded in phase: middle down with f1=0.98514$\mbox{c d}^{-1}$, 
bottom with f2=1.00443$\mbox{c d}^{-1}$ after prewhitening for f1.} 
\label{varbe4b}
\end{figure}

\begin{figure}[]
\centering
\resizebox{\hsize}{!}{\includegraphics[angle=0]{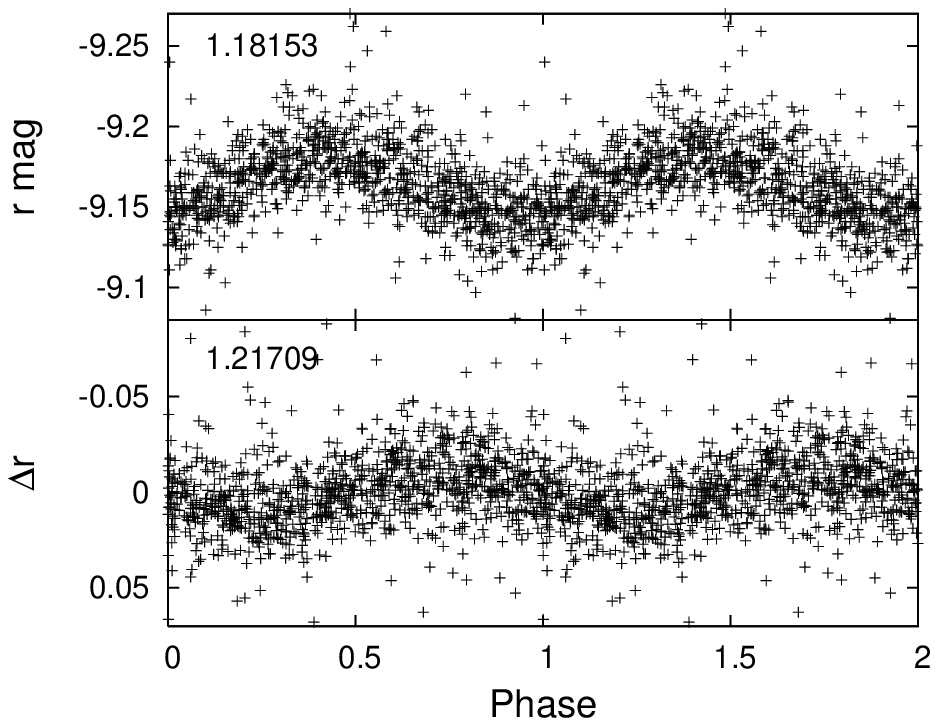}}
\resizebox{\hsize}{!}{\includegraphics[angle=0]{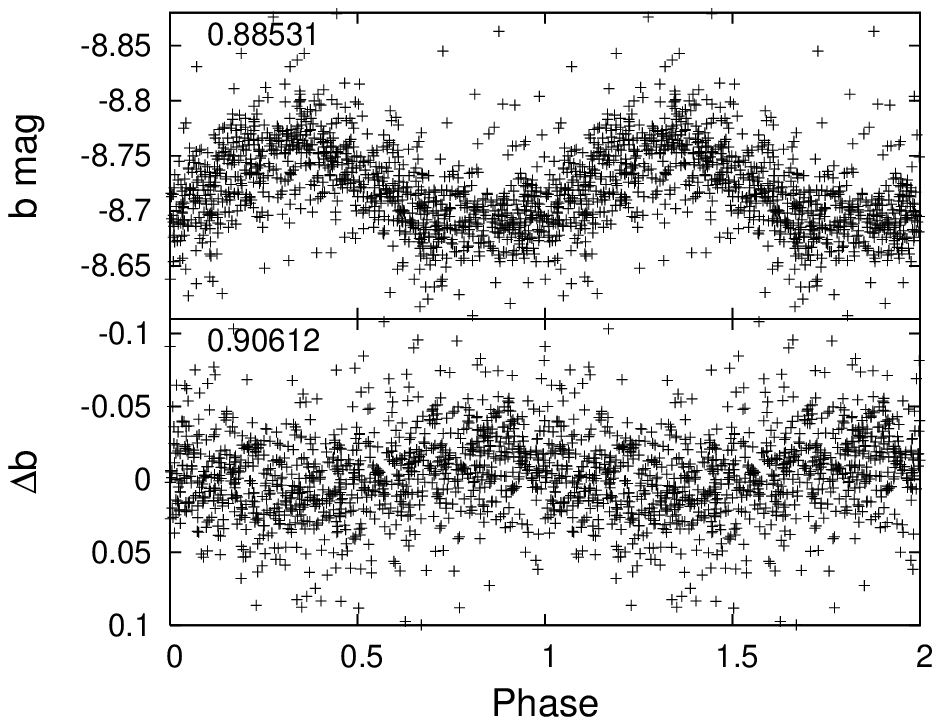}}
\caption{ Be SMC5\_37013 folded in phase: top with f1=1.18153$\mbox{c d}^{-1}$, middle up with f2=1.21709$\mbox{c d}^{-1}$ 
after prewhitening for f1. Be SMC5\_37162 folded in phase: middle down with f1=0.88531$\mbox{c d}^{-1}$, 
bottom with f2=0.90612$\mbox{c d}^{-1}$ after prewhitening for f1.} 
\label{varbe5}
\end{figure}

\begin{figure}[]
\centering
\resizebox{\hsize}{!}{\includegraphics[angle=0]{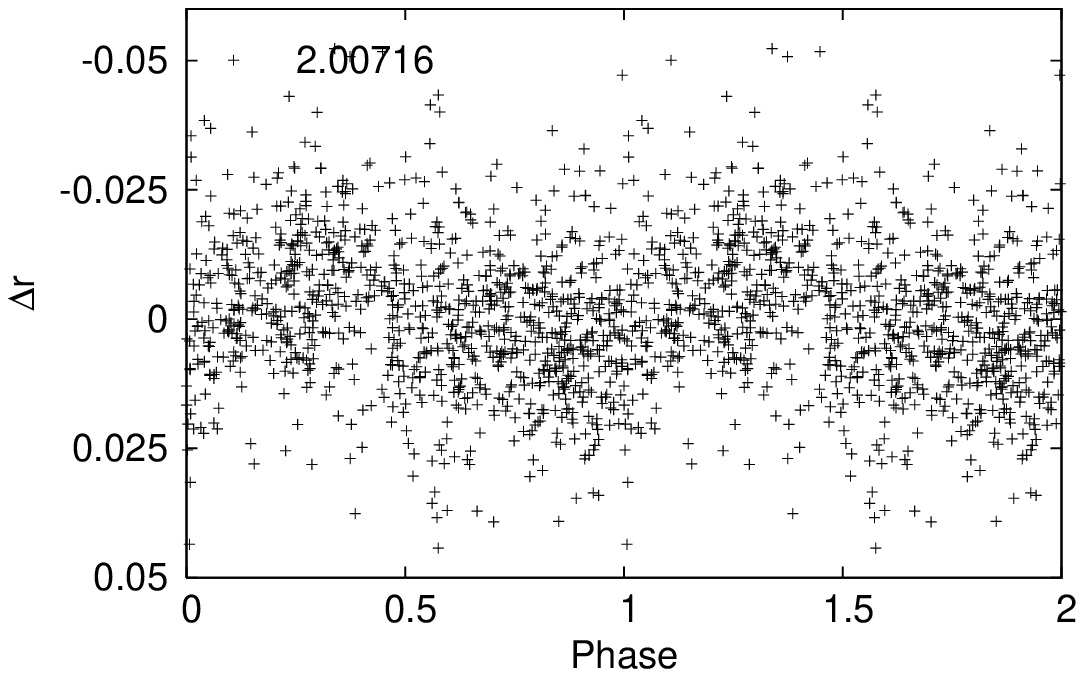}}
\resizebox{\hsize}{!}{\includegraphics[angle=0]{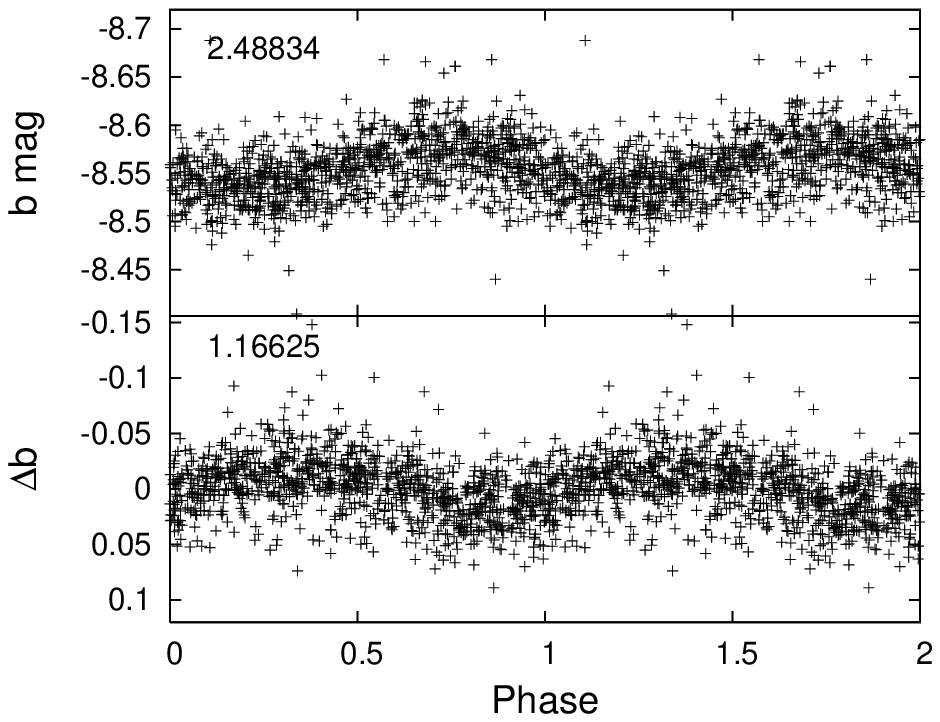}}
\caption{ Top: Be SMC5\_43413  folded in phase with f=2.00716$\mbox{c d}^{-1}$. 
Be SMC5\_82042  folded in phase: middle with f1=2.48834$\mbox{c d}^{-1}$, 
bottom with f2=1.16625$\mbox{c d}^{-1}$ after prewhitening for f1.} 
\label{varbe6}
\end{figure}

\begin{figure}[]
\centering
\resizebox{\hsize}{!}{\includegraphics[angle=0]{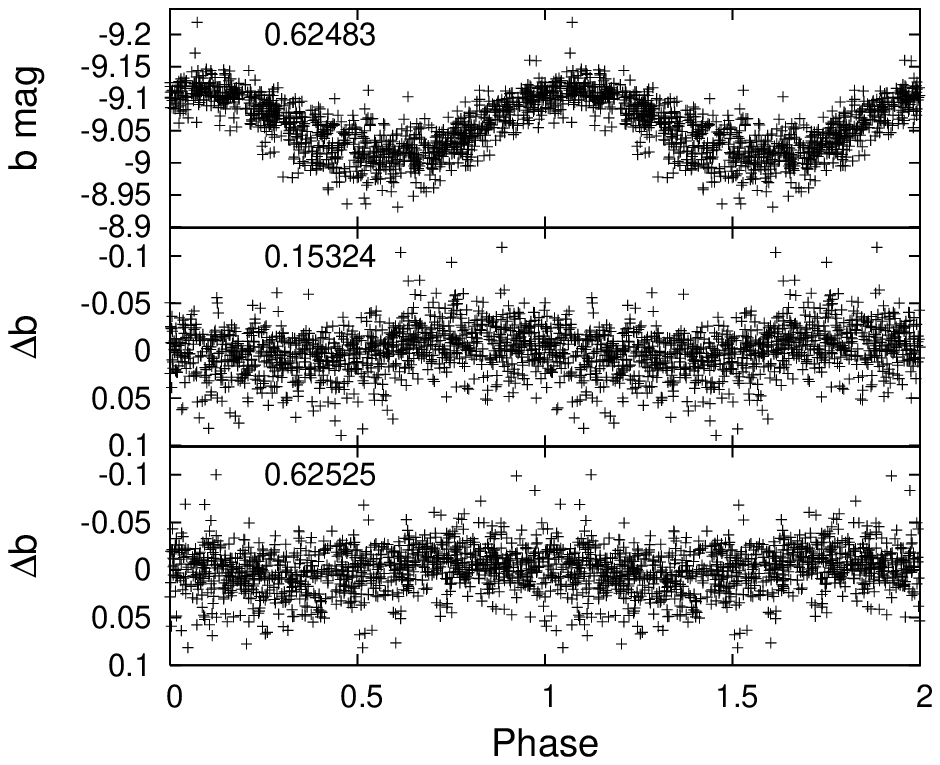}}
\caption{ Be SMC5\_82941 folded in phase: top with f1=0.62483$\mbox{c d}^{-1}$, 
middle with f3=0.15324$\mbox{c d}^{-1}$ after prewhitening for f1, and bottom with f2=0.62525$\mbox{c d}^{-1}$ 
after prewhitening for f1 and f2.}  
\label{varbe7}
\end{figure}

\end{document}